\documentclass[12pt]{article}
\usepackage{amsmath,amssymb,amsfonts,color,graphicx,cite,color,soul,ulem}
\usepackage{rotating}
\pdfoutput=1
\newcount\timecount
\newcount\hours \newcount\minutes  \newcount\temp \newcount\pmhours
\hours = \time
\divide\hours by 60
\temp = \hours
\multiply\temp by 60
\minutes = \time
\advance\minutes by -\temp
\def\hour{\the\hours}
\def\minute{\ifnum\minutes<10 0\the\minutes
            \else\the\minutes\fi}
\def\clock{
\ifnum\hours=0 12:\minute\ AM
\else\ifnum\hours<12 \hour:\minute\ AM
      \else\ifnum\hours=12 12:\minute\ PM
            \else\ifnum\hours>12
                 \pmhours=\hours
                 \advance\pmhours by -12
                 \the\pmhours:\minute\ PM
                 \fi
            \fi
      \fi
\fi
}

\def\monthname{\relax\ifcase\month 0/\or January\or February\or
   March\or April\or May\or June\or July\or August\or September\or
   October\or November\or December\else\number\month/\fi}

\def\bold#1{\setbox0=\hbox{$#1$}%
     \kern-.025em\copy0\kern-\wd0
     \kern.05em\copy0\kern-\wd0
     \kern-.025em\raise.0433em\box0 }

\newcommand{\tb}{\ensuremath{\tan\beta}}

\def\beq{\begin{equation}}
\def\eeq{\end{equation}}

\def\ga{\mathrel{\raise.3ex\hbox{$>$\kern-.75em\lower1ex\hbox{$\sim$}}}}
\def\la{\mathrel{\raise.3ex\hbox{$<$\kern-.75em\lower1ex\hbox{$\sim$}}}}
\def\gyr{{\rm \, G\kern-0.125em yr}}

\def\gappeq{\mathrel{\rlap {\raise.5ex\hbox{$>$}}
{\lower.5ex\hbox{$\sim$}}}}
\def\lappeq{\mathrel{\rlap{\raise.5ex\hbox{$<$}}
{\lower.5ex\hbox{$\sim$}}}}
\def\Toprel#1\over#2{\mathrel{\mathop{#2}\limits^{#1}}}

\def\m12{m_{1\!/2}}

\newcommand{\gsim}{\lower.7ex\hbox{$\;\stackrel{\textstyle>}{\sim}\;$}}
\newcommand{\lsim}{\lower.7ex\hbox{$\;\stackrel{\textstyle<}{\sim}\;$}}

\newcommand{\ETslash}{\ensuremath{/ \hspace{-.7em} E_T}}

\newcommand{\htr}[1]{{#1}}

\graphicspath{{figs/}}

\begin{document}
\begin{titlepage}
\pagestyle{empty}
\begin{flushright}
{\tt KCL-PH-TH/2015-21, LCTS/2015-12, CERN-PH-TH/2015-117 \\
UMN-TH-3437/15, FTPI-MINN-15/25 \\
}
\end{flushright}
\vskip 0.25in
\begin{center}
{\large {\bf Collider Interplay for Supersymmetry, Higgs and Dark Matter}} \\
\end{center}
\begin{center}
\vskip 0.25in
{\bf O.~Buchmueller}$^1$, {\bf M.~Citron}$^1$, {\bf J.~Ellis}$^{2,3}$, {\bf S.~Guha}$^{3,4}$, {\bf J.~Marrouche}$^{3,1}$, \\
\vspace{2mm}
{\bf K.~A.~Olive}$^{5}$, {\bf K.~de~Vries}$^{1}$ and {\bf Jiaming~Zheng}$^{5}$
\vskip 0.2in
{\it
$^1${High Energy Physics Group, Blackett Lab., Imperial College, Prince Consort Road, London SW7 2AZ, UK}\\
$^2${Theoretical Particle Physics and Cosmology Group, Dept.\ of Physics, 
King's College London, London WC2R 2LS, UK}\\
$^3${Physics Department, CERN, CH-1211 Geneva 23, Switzerland}\\
$^4${BITS Pilani Goa Campus, Goa, India}\\
$^5${William\,I.\,Fine\,Theoretical\,Physics\,Institute,\,%
School of Physics and Astronomy, 
Univ.\,of\,Minnesota,\,%
Minneapolis,\,MN\,55455,\,USA}\\}
\vskip 0.3in 
{\bf Abstract}
\vskip 0.2in 
\end{center}
\noindent
We discuss the potential impacts on the CMSSM
of future LHC runs and possible $e^+ e^-$ and higher-energy
proton-proton colliders, considering searches for supersymmetry via
~$\ETslash$ events, precision electroweak physics, 
Higgs measurements and dark matter searches. We validate and present
estimates of the physics reach for exclusion or discovery of supersymmetry via $\ETslash$
searches at the LHC, which should cover the low-mass regions of the CMSSM parameter space
favoured in a recent global analysis. As we illustrate with a low-mass benchmark point, a discovery would make possible accurate
LHC measurements of sparticle masses using the MT2 variable, which could be combined
with cross-section and other measurements to constrain the gluino, squark and stop masses
and hence the soft supersymmetry-breaking parameters $m_0, m_{1/2}$ and $A_0$ of the CMSSM.
Slepton measurements at CLIC would enable $m_0$ and $m_{1/2}$ to be determined with high precision. 
If supersymmetry is indeed discovered in the low-mass region, precision
electroweak and Higgs measurements with a future circular $e^+ e^-$ collider (FCC-ee, also 
known as TLEP) combined with LHC measurements
would provide tests of the CMSSM at the loop level. If supersymmetry is not
discovered at the LHC, is likely to lie somewhere along a focus-point,
stop coannihilation strip or direct-channel $A/H$ resonance funnel. We discuss the prospects for discovering supersymmetry
along these strips at a future circular proton-proton collider such as FCC-hh. Illustrative
benchmark points on these strips indicate that also in this case FCC-ee could provide
tests of the CMSSM at the loop level.

\vfill
\leftline{May 2015}
\end{titlepage}
\baselineskip=16pt

\section{Introduction}

The first run of the LHC at 7 and 8~TeV has framed the agenda for its future runs,
and for possible future colliders. The CMS and ATLAS experiments have discovered 
a Higgs boson~\cite{lhch}, but have found no sign of supersymmetry or any other physics
beyond the Standard Model~\cite{ATLAS20,CMS20}. Present and future studies of the Higgs boson can be
used to constrain scenarios for new physics, as can other high-precision low-energy
measurements and cosmological constraints. We address in this paper the prospects for discovering
supersymmetry during future runs of the LHC at 13/14~TeV in light of the indirect information
currently provided by the Higgs and other measurements, and consider possible
scenarios for discovering or measuring supersymmetry at proposed future linear
and circular colliders, either directly or indirectly, showing how the various colliders
may complement each other.

Our study is within the minimal supersymmetric extension of the Standard Model
with soft supersymmetry-breaking parameters constrained to be universal at a
high input scale, the CMSSM~\cite{funnel,cmssm}. This model is not imposed by top-down
considerations based on string, M- or F-theory, nor is it required by bottom-up
considerations such as limits on flavour-changing neutral interactions. However,
it is the simplest supersymmetric model, so its phenomenology is relatively unambiguous.
As such, it provides a convenient benchmark for considering the interplay between
different high-energy colliders.

One of the most important constraints that we take into account is the bound on the density of cold dark matter,
which provides interesting constraints on the parameters of the CMSSM. In particular,
requiring that the relic density of the lightest supersymmetric particle (LSP), assumed
here to be the lightest neutralino $\chi$~\cite{ehnos}, falls within the range allowed by astrophysics 
and cosmology can be used to provide important constraints, including upper limits, on the soft supersymmetry-breaking
mass parameters in the CMSSM, and hence sparticle masses \cite{cmssmwmap,eo6,ehow+}. 

The LHC measurement of the Higgs mass already provides a significant constraint on the
parameter space of the CMSSM, favouring
sparticle masses that are consistent with the non-observation of
supersymmetric particles at the LHC in Run~1~\cite{eo6,ehow+,mc75,mc8,mc9,post-mh,fp,eoz}.
The starting-point for our analysis
is a recent global fit to the CMSSM model parameters~\cite{mc9},
using these measurements as well as precision electroweak and flavour observables,
as well as direct constraints on the interactions of the LSP with ordinary matter.

In order to evaluate the potential of future LHC runs to probe the CMSSM, we extrapolate
the sensitivities of gluino, squark and stop searches at LHC Run~1 at 7 and 8~TeV to
estimate LHC capabilities with 300 or 3000/fb of data at 13/14~TeV. We find that such data sets
should permit the LHC experiments to discover supersymmetry if it has CMSSM parameters within the low-mass
region favoured by the global fit~\cite{mc9}. Assuming optimistically that they are given by the best current fit
in this low-mass region, we then discuss how accurately the LHC experiments could
measure the gluino, squark and stop masses, and hence the CMSSM soft supersymmetry-breaking parameters $m_0, m_{1/2}$ and $A_0$.

In this optimistic scenario where Nature is described by the CMSSM in the low-mass region,
experiments at the proposed CLIC
$e^+ e^-$ collider at 3~TeV in the centre of mass~\cite{1304.2825, CLIC} would be able to
produce and measure very accurately the masses and other
properties of the sleptons and the lighter gauginos, enabling, for example, high-precision
determinations of the soft supersymmetry-breaking parameters $m_0$ and $m_{1/2}$ of the CMSSM.
An $e^+ e^-$ collider with an energy 1~TeV could also explore parts of the low-mass region, e.g.,
pair-producing the lighter stau slepton at the low-mass best-fit point.
On the other hand, $e^+ e^-$ colliders with energies below 500~GeV in the centre of mass would not
be able to produce and measure sparticles directly.

As we discuss, measurements of $Z$-boson~\cite{LEPEWWG} and Higgs couplings~\cite{ATLASmu, CMSmu, Tevatronmu} do not as yet provide strong supplementary 
constraints on supersymmetric models such as the CMSSM. However, future
higher-precision measurements could be used to constrain the CMSSM
parameters indirectly. In particular, if Nature is indeed described by the CMSSM with
parameters in the low-mass region, measurements of the $Z$ and Higgs boson at
the proposed high-luminosity circular $e^+ e^-$ collider FCC-ee (TLEP)~\cite{TLEP} could be used, in conjunction
with the LHC measurements, to test this supersymmetric model at the
quantum level, as we illustrate in the specific example of the best-fit low-mass point from~\cite{mc9}.
As an aside, we also show how, again in the optimistic low-mass scenario, high-precision
$Z$ measurements at FCC-ee (TLEP) could be used to probe models of supersymmetric
grand unification.

On the other hand, in the pessimistic scenario where the LHC does not discover supersymmetry,
but only establishes 95\% CL lower limits on particle masses, we consider the prospects for
discovering supersymmetry directly at a future higher-energy circular proton-proton collider such as FCC-hh~\cite{FCC-hh,LW}, or
finding indirect evidence for supersymmetry via high-precision $e^+ e^-$ measurements.
Within the CMSSM, high-scale supersymmetric models can be found along narrow
strips where stop-neutralino coannihilation is important~\cite{eoz}, or in the focus-point region~\cite{fp},
and we analyse the prospects of direct and indirect measurements along these strips.
Studies of illustrative benchmark points along these strips indicate that the combination of direct FCC-hh
and indirect FCC-ee measurements could test supersymmetry at the loop level also in this pessimistic case.

The layout of this paper is as follows. In Section~2 we discuss the
extrapolations of current LHC sparticle search sensitivities to future LHC runs.
Then, in Section~3 we discuss possible LHC measurements of particle masses
in the optimistic low-mass best-fit scenario. Section~4 contains our discussion of
$e^+ e^-$ probes of supersymmetry in this optimistic scenarios, including direct
searches at CLIC as well as indirect constraints due to high-precision $Z$ and Higgs
measurements at FCC-ee (TLEP). The pessimistic high-mass scenarios in which
the LHC does not discover supersymmetry are discussed in Section~5, where we
consider the prospects for direct discovery with FCC-hh as well as indirect
measurements with FCC-ee (TLEP). Finally, our conclusions are summarised in Section~6.

\section{Extrapolations of Current LHC Sparticle Search Sensitivities to Higher Energy and Luminosity}

\begin{figure}[ht!]
\centerline{
\includegraphics[height=10cm]{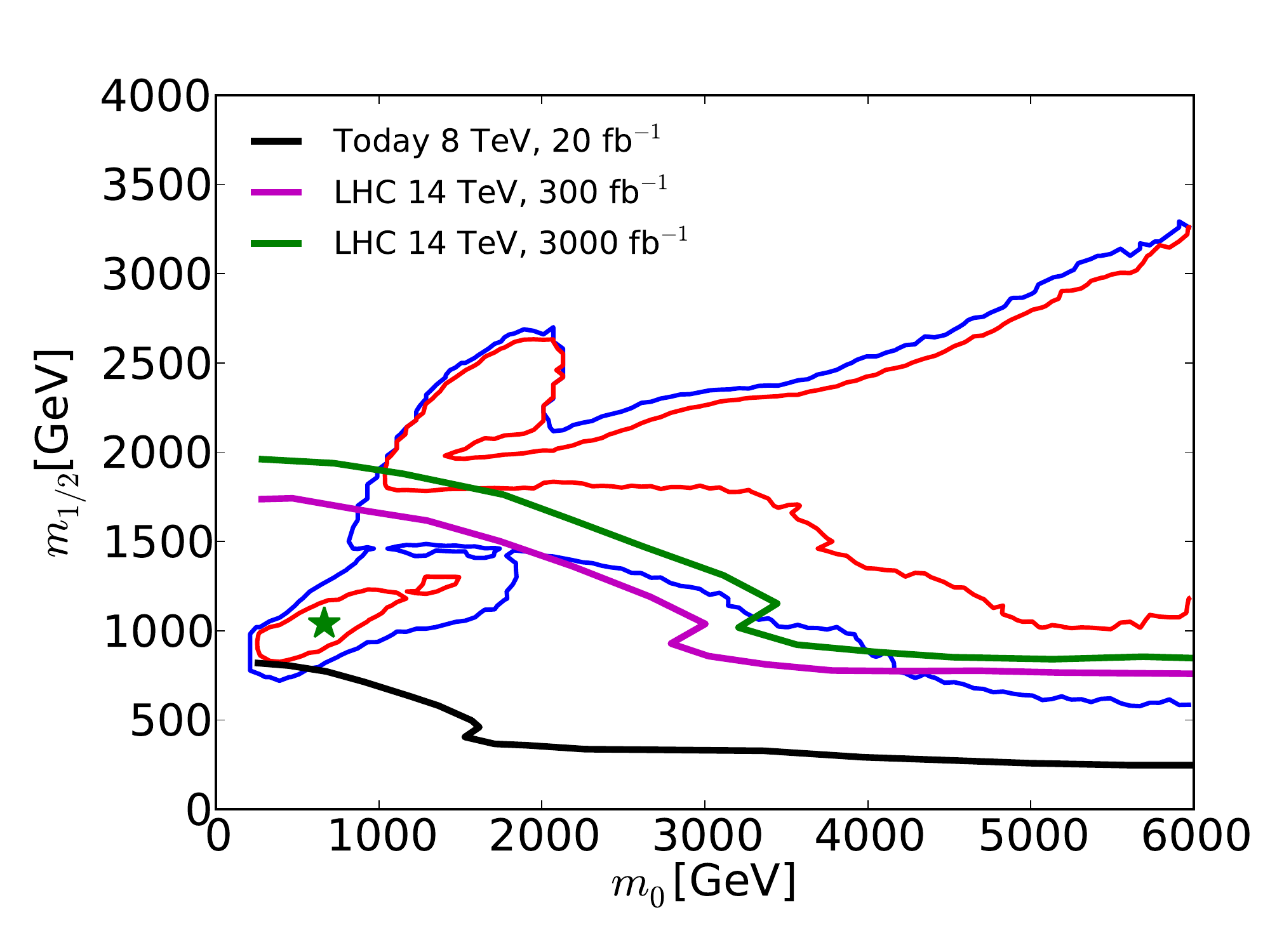}
}
\caption{\label{fig:Kees}\it
The $(m_0, m_{1/2})$ plane in the CMSSM.
The $\Delta \chi^2 =$ 2.30 (68\% CL) and 5.99 (95\% CL) regions 
found in recent global fits are
bounded by solid red and blue lines, respectively. The best-fit point in the low-mass
`Crimea' regions is indicated by a filled green star. Also shown as solid black (purple, green) lines
are the sensitivities of LHC ~$\ETslash$ searches for exclusions at the 95\% CLs with
20/fb of data at 8~TeV (300, 3000/fb of data at 14~TeV). The purple contour is expected
to coincide (within uncertainties) with the 5-$\sigma$ discovery contour at the LHC
with 3000/fb of data at 14~TeV.}
\end{figure}

The baseline for our studies is provided by a recent global fit to the parameters
of the CMSSM~\cite{mc9}~\footnote{This paper also contains a global fit to the NUHM1 \cite{nuhm1,eos}, 
and a fit to the NUHM2 \cite{nuhm2,eos}
can be found in~\cite{mc10}, together with minor updates of these CMSSM and NUHM1 fits. 
A further update of the
CMSSM analysis can be found in~\cite{mc11}. These updates do not impact qualitatively the analyses presented in this paper.}.
In addition to the ATLAS search for jets + $\ETslash$ events with $\sim 20$/fb
of 8~TeV data \cite{ATLAS20,CMS20}, these global fits included the measurement of $m_h$~\cite{lhch,LHCmH} (which was related to
the CMSSM parameters via calculations using {\tt FeynHiggs~2.10.0}~\cite{FeynHiggs,newFH}),
electroweak precision observables and $g_\mu - 2$~\cite{newBNL}, precision flavour observables
including $b \to s \gamma$~\cite{bsgex} and $B_{s,d} \to \mu^+ \mu^-$~\cite{bmm,CMSBsmm,LHCbBsmm,BsmmComb}, 
and dark matter observables
including the direct LUX constraint on dark matter scattering~\cite{LUX} and the total cold dark matter
density~\cite{Planck}. These measurements were combined into a global $\chi^2$ likelihood function, 
whose projection on the $(m_0, m_{1/2})$ plane of the CMSSM is 
displayed in Fig.~\ref{fig:Kees}.
\htr{In this and subsequent figures, we marginalise over the other CMSSM parameters $\tan \beta$ and $A_0$.}
We display in red and blue, respectively, $\Delta \chi^2 = 2.30$ and $5.99$ contours (which we use as proxies for 68\% and 95\% CL
contours). \htr{For each set of $(m_0, m_{1/2})$ values within these contours, there is some choice of $\tan \beta$ and $A_0$
for which $\Delta \chi^2 < 2.30$ or $5.99$, respectively, and outside these contours there are no choices of
$\tan \beta$ and $A_0$ that satisfy these conditions.}
In the figure, a low-mass ``Crimea" region and a high-mass ``Eurasia" region
can be distinguished. \htr{The former consists of points in the stau coannihilation region, and the latter
includes points along rapid $H/A$ annihilation funnels, and along the high-mass focus-point and stop coannihilation
strips we discuss in Section 6.} We also show as a filled green star a representative best-fit point in the low-mass region,
whose parameters are listed in Table~\ref{tab:bestfits}. In the low-mass region, $g_\mu - 2$ makes a significantly smaller contribution to the global
$\chi^2$ function than in the high-mass region, although the CMSSM and related models could 
not by themselves resolve the discrepancy between the experimental measurement
 and the theoretical calculation within the Standard Model~\footnote{This discrepancy can be resolved in a model
 that relaxes the assumption of GUT-scale universality in the soft supersymmetry-breaking parameters~\cite{mc11}.}.
We discuss later characteristics of points in the high-mass  `Eurasia' region: the $\chi^2$ likelihood function is relatively flat
across this region, and there is no well-defined best-fit point that is favoured strongly with respect to other points.

\begin{table*}[!tbh!]
\renewcommand{\arraystretch}{1.5}
\begin{center}
\begin{tabular}{|c|c||c|c|c|c|} \hline
Model & Location & $m_0$ & $m_{1/2}$ & $A_0$ & $\tb$ \\
  &     & (GeV) & (GeV) & (GeV) & \\ 
\hline \hline
CMSSM & Low-mass & 670 &1040 & 3440 & 21 \\
\hline \hline
\end{tabular}
\caption{\it Representative low-mass best-fit point found in  a recent global CMSSM fit~\protect\cite{mc9},
  using the ATLAS jets + $\ETslash$ constraint~\cite{ATLAS20}, and the combination of the
LHCb~\cite{LHCbBsmm} and  CMS~\cite{CMSBsmm} constraints on  $B_{s,d} \to \mu^+ \mu^-$~\cite{BsmmComb},
 and using {\tt FeynHiggs~2.10.0}~\protect\cite{newFH} to calculate $m_h$.
 }
\label{tab:bestfits}
\end{center}
\end{table*}

In  Fig.~\ref{fig:Kees} we also show as a black line
the 95\% CL exclusion contour in the $(m_0, m_{1/2})$ plane
established by ATLAS searches for jets + $\ETslash$ events with
$\sim 20$/fb of data at 8~TeV~\cite{ATLAS20}. This exclusion was derived within the CMSSM with
$\tan \beta = 30$ and (in our sign convention) $A_0 = 2 \, m_0$, but studies have shown that
the limit is relatively insensitive to the values of $\tan \beta$ and $A_0$ \cite{mc8}. 
The ATLAS 95\% CLs contour intersects the 95\% CL contours found in
the global fit, reflecting the importance of other observables in the global fit. For example, as already mentioned,
$g_\mu - 2$ tends to favour relatively low values of $m_0$ and $m_{1/2}$. On the other hand,
the measurement of $m_h$ tends to favour values of $m_0$ and $m_{1/2}$ beyond the
ATLAS $\ETslash$ contour.

We use a simple procedure to estimate the sensitivities of future collider searches exploiting
the $\ETslash$ signature accompanied by jets (possibly $b$-tagged) and/or leptons
at higher centre-of-mass energies and luminosities. We scale the  95\% CL exclusion or 5-$\sigma$ discovery 
contours of the searches at 8 TeV to different luminosity and energy scenarios by
{\it assuming that the signal efficiency and background suppression of the current 8-TeV searches remain unchanged.}
Maintaining the present performance of the searches is motivated by the ATLAS and CMS upgrade programmes, 
and is defined by both experiments as one of the main upgrade goals. The assumption was also used in several studies for 
Snowmass and ECFA (see e.g.~\cite{Snowmass,CMS:2013xfa}) as well as to project collider limits for Dark Matter 
searches~\cite{Malik:2014ggr,Buchmueller:2014yoa}.  It also forms the basis of the Collider Reach~\cite{colliderreach} tool, 
which reports dedicated studies showing good agreement between this extrapolation approach and results obtained from a full simulation.

We caution, however, that various effects could
invalidate our assumption. For example, the signal-to-background ratio could
vary with the centre-of-mass energy and with the number of pile-up events,
which is correlated with the luminosity. Indeed, our extrapolation of the LHC
sensitivities with 300/fb and 3000/fb of integrated luminosity at
14~TeV in the centre of mass is somewhat less conservative than ATLAS estimates of their
exclusion sensitivities~\cite{ATLASHL}. However, we have been able to verify
that our simple assumption gives similar results to Snowmass estimates of the
possible sensitivities of higher-energy colliders based on simplified model searches
at the LHC with $\sim 20$/fb of data at 8~TeV~\cite{Snowmass}, and we consider our assumption a
reasonable objective for future experimental analyses to target. 

We consider in this paper the following LHC sparticle searches: searches for events with jets and missing
transverse energy, $\ETslash$, possibly accompanied by leptons and with some jets $b$-tagged,
dedicated searches for light stop squarks ${\tilde t} \to \chi + c$, and monojet searches.
Using our simple assumption for a number of current LHC searches, 
we calculate cross-sections at higher LHC centre-of-mass energies
with {\sc Pythia~8}~\cite{PYTHIA, py8susy}, using as default the MSTW2008NLO
parton distribution functions~\cite{MSTW}~\footnote{We have verified that very similar results
can be obtained using the NNPDF2.3LO, option 14 parton distribution functions~\cite{NNPDF}.}.
We then require that
the products of the integrated luminosity with the cross-section be the same as for the 8~TeV LHC data.
In this way, we extrapolate current LHC 95\% CLs exclusion limits to higher LHC energies and luminosities, as well
as possible future colliders with 3000/fb at 33 and 100~TeV, as seen in Table~\ref{tab:extrapolations}~\footnote{There are
also extrapolations available for slepton and chargino searches. However, in those cases the interpretations of the searches
are more delicate~\cite{mc11}, and we do not discuss them here.}.

Fig.~\ref{fig:Kees} displays as purple
and green lines, respectively, our extrapolations within the CMSSM 
of the current ATLAS 95\% CLs limit from searches for jets + $\ETslash$ events with
$\sim 20$/fb of data at 8~TeV to LHC searches at 14~TeV (LHC14) with 300/fb and 3000/fb
of integrated luminosity. (We note that the ATLAS study~\cite{ATLASHL} found that the
5-$\sigma$ discovery contour for 3000/fb almost coincides with the 95\% CLs
exclusion contour for 300/fb.) Within the CMSSM, the
ATLAS search for jets + $\ETslash$ events is the most sensitive for $m_0/m_{1/2} \le 2$,
with other searches becoming more important at larger $m_0/m_{1/2}$. We return later
to extrapolations of monojet searches and dedicated searches for light stop squarks,
which are important for our studies of FCC-hh.

\begin{table}
\begin{center}
{\small
\begin{tabular}{ |c||c|c|c||c||c| }
  \hline
  & \multicolumn{3} {|c||} {LHC} & HE-LHC & FCC-hh \\
  \hline
 Search & 8~TeV & 14~TeV & 14~TeV & 33~TeV & 100~TeV \\
signature & 20/fb & 300/fb & 3000/fb & 3000/fb & 3000/fb \\
  \hline
  \hline
 $({\tilde g} \to b {\bar b} \chi)^2$ ($m_{\tilde g}$) & 1300 & 2540 & 2990 & 6080 & 14700 \\
  \hline
 ${\tilde t}{\tilde t^*}$ ($m_{\tilde t}$) & 650 & 1350 & 1740 & 3260 & 7020 \\
      \hline
$({\tilde t} \to c \chi)^2$ ($m_{\tilde t}$) & 240 & 530 & 780 & 1320 & 2510 \\
 \hline \hline
CMSSM & \multicolumn{5} {|c|} {$(m_0, m_{1/2})$} \\
\hline
$m_0 = m_{1/2}$ & (800, 800) & (1610, 1610) & (1860, 1860) & (4080, 4080) &(10800, 10800) \\
  \hline
$m_0 = 2.5 m_{1/2}$ & (1500, 600) & (2950, 1180) & (3390, 1360) & (7310, 2930) &(19000, 7600) \\
  \hline
 \end{tabular}}
\end{center}
\caption{\it Extrapolations of current LHC searches with $\sim 20$/fb of luminosity at 8~TeV
to higher energies and luminosities, assuming sensitivities to the same numbers of signal events.
The first five rows of the Table are possible 95\% CL exclusion sensitivities
derived from searches for specific sparticle pair-production processes,
as indicated, and the numbers correspond to the sparticle masses in GeV. The last two rows are for rays
in the $(m_0, m_{1/2})$ plane, as indicated, and the numbers correspond to the possible 95\% CL
exclusion limits on $m_0$ and $m_{1/2}$.}
\label{tab:extrapolations}
\end{table}

We see that the low-mass
`Crimea' region lies within the purple (95\% CLs exclusion with 300/fb or 5-$\sigma$ discovery with 3000/fb at 14~TeV)
contour where $m_0 \le m_{1/2}$, whereas the high-mass `Eurasia'
region lies largely beyond the purple contour. Based on these comparisons between the extrapolated LHC sensitivity
and current fits within the CMSSM, we have chosen for further study
two scenarios for the outcome of the LHC searches with 3000/fb at 14~TeV.

\begin{itemize}

\item
An `optimistic' scenario in which the LHC discovers supersymmetry in the
`Crimea' region, and for definiteness we assume that its parameters
coincide with those at the the representative low-mass best-fit point in Table~\ref{tab:bestfits}.

\item
A `pessimistic' scenario in which the LHC discovers no evidence for
supersymmetry, in which case the supersymmetry-breaking parameters
must lie somewhere in `Eurasia'.

\end{itemize}

The following sections contain discussions of the interplay between
the various colliders in these scenarios.

\section{LHC Measurements of Supersymmetry in the \\ Optimistic Scenario}

\begin{figure}[ht!]
\centerline{
\includegraphics[height=8cm]{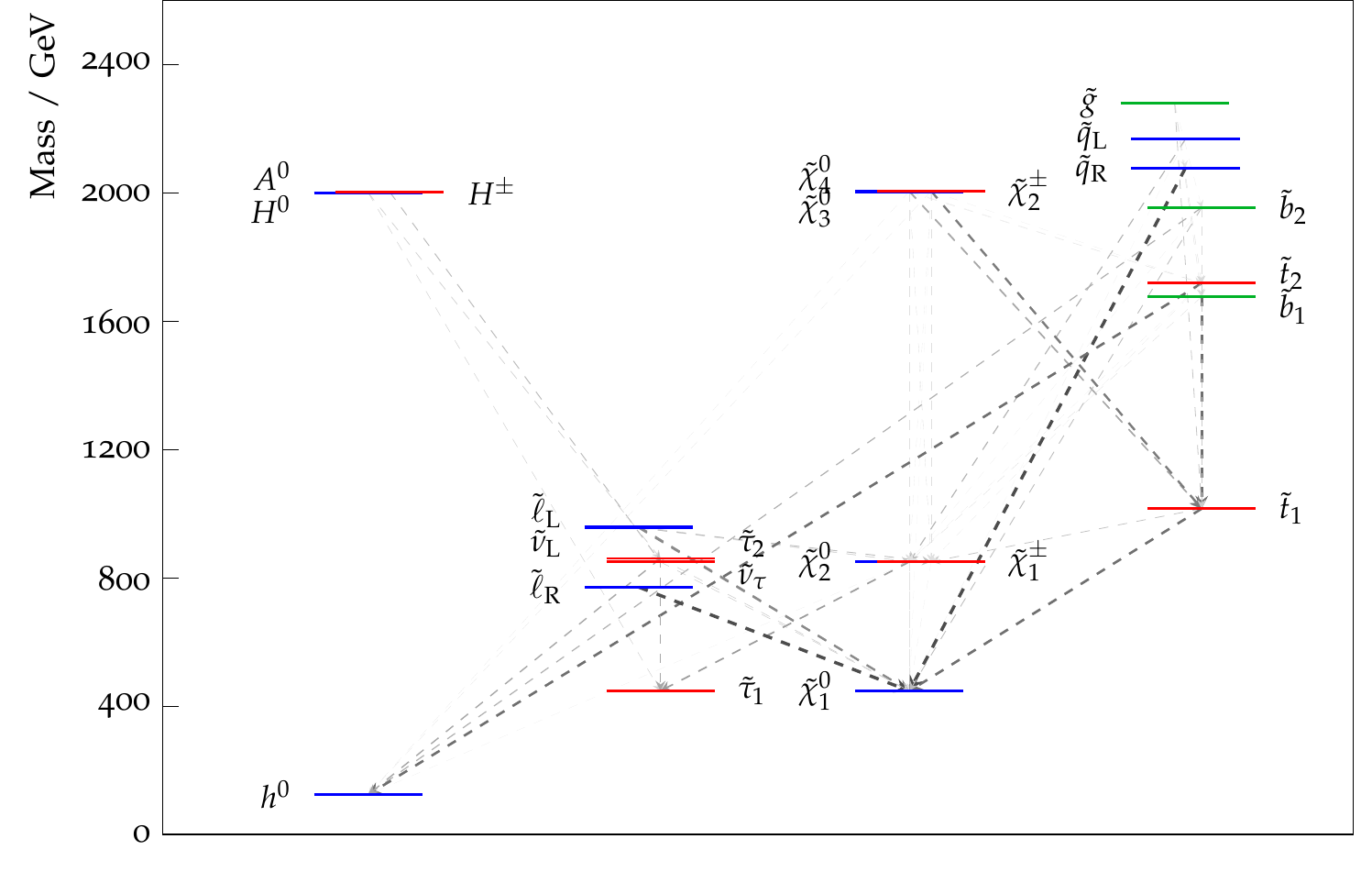}
}
\caption{\label{fig:spectrum}\it
The spectrum at the best-fit point in the CMSSM~\protect\cite{mc9},
whose parameters are listed in Table~\ref{tab:bestfits}. The magnitudes of the branching ratios
for sparticle decays into different final-state particles are represented by the strengths of the dashed lines connecting
them.}
\end{figure}

Assuming that Nature is described by supersymmetry at the CMSSM low-mass best-fit point,
the sparticle mass spectrum is determined, as illustrated in Fig.~\ref{fig:spectrum}.
The most relevant sparticles for searches at the LHC are those with the highest production cross sections,
namely squarks and gluinos. At the best-fit point, the mass of a generic right-handed $u, d, s, c$ or $b$
squark is calculated to be $m_{\tilde q_R} \simeq 2080$~GeV, and the lighter stop squark has
 a mass $m_{\tilde t_1} \simeq 1020$~GeV. Also, $m_{\tilde g} \simeq 2280$~GeV
and the lightest neutralino mass $m_\chi \simeq 450$~GeV. The lighter stau mass $m_{\tilde \tau_1}$
is only very slightly heavier: at the best-fit point and the rest of the low-mass region stau-$\chi$
coannihilation is responsible for bringing the relic density into the range allowed by cosmology. In the following we consider
the possible LHC measurements of generic right-handed squarks ${\tilde q_R}$, gluinos ${\tilde g}$
and the lighter stop ${\tilde t_1}$, with either 300 or 3000/fb of luminosity at LHC14.
We assume that experiments at the LHC discover supersymmetry with the mass spectrum
characteristic of the best-fit point shown in Fig.~\ref{fig:spectrum}, and ask how
accurately its parameters can be measured.

\subsection{Gluinos and Squarks}

\begin{figure}[ht!]
\centerline{
\includegraphics[height=6cm]{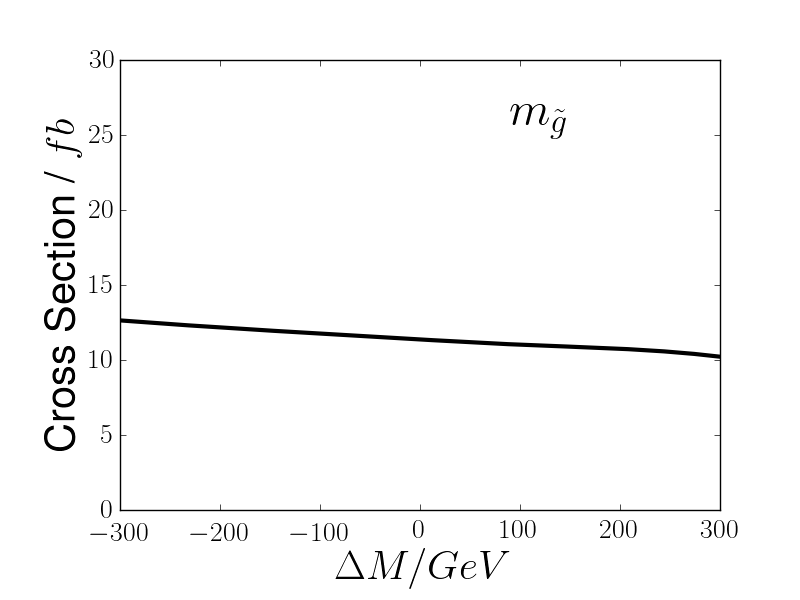}
\includegraphics[height=6cm]{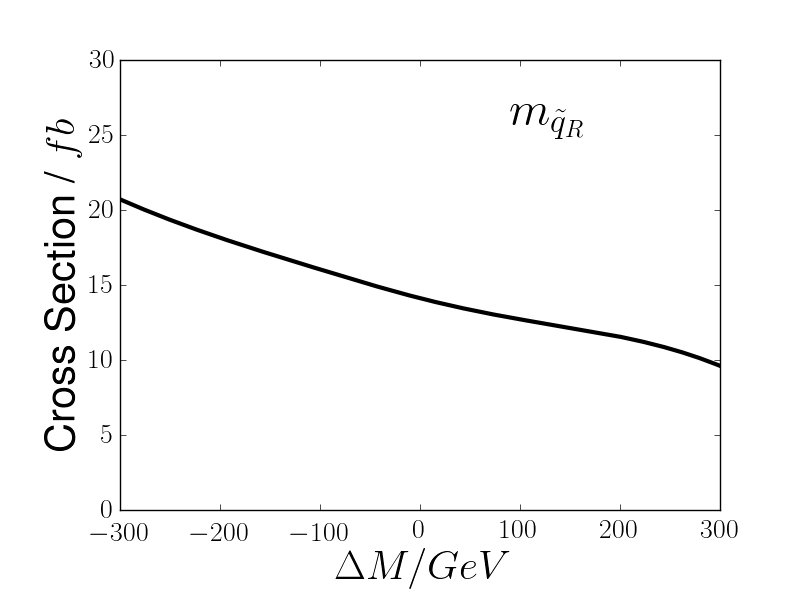}
}
\caption{\label{fig:sigmas}\it
The sensitivities of the total sparticle cross-section to $m_{\tilde g}$ (left panel) and
$m_{\tilde q_R}$ (right panel), expressed as functions of the mass differences $\Delta M$ 
relative to the low-mass best-fit values in the CMSSM.}
\end{figure}

We estimate first the potential resolution with which the gluino and squark masses could be measured.
For this purpose, we consider three contributions to the determination of these model parameters: measurements of
the total cross-section, the distribution in the MT2 variable~\cite{MT2}, and the spectator
jet energies in ${\tilde g} \to q + {\tilde q_R}$ decay.
Fig.~\ref{fig:sigmas} shows how the total cross-section for strongly-interacting
particle production at LHC14 obtained from {\sc Pythia} depends on the gluino 
mass $m_{\tilde g}$ (left panel) and the squark mass $m_{\tilde q_R}$ (right panel),
expressed as functions of the mass differences $\Delta M$ relative to the low-mass best-fit values in the CMSSM~\footnote{\htr{We note that the relative fractions of the gluon-gluon, gluon-squark and squark-squark final states
vary continously in these plots.}}.
We see that the dependence of the cross-section on $m_{\tilde g}$
is much weaker than that on $m_{\tilde q_R}$. In the following we combine the information
that can be derived the cross-section with that obtainable from an analysis using the
MT2 variable.

In order to assess how MT2 measurements could contribute to constraining the gluino and squark masses,
we construct a set of MT2 templates for various values of these and the neutralino mass,
and fit these templates to a simulation of the prospective MT2 distribution for the central best-fit
values of the masses. \htr{For this analysis, we first matched the reconstructed jets from the {\sc Pythia} output
to the squarks, gluinos and neutralinos at the generator level. We then applied the same logic to construct MT2 
as in experimental papers, treating the neutralinos as $\ETslash$ and the decay products of the 
squarks and gluinos as the jets. Thus, this approach does not consider combinatoric effects
as could arise in a study that used a full detector simulation.}  We use the same kinematic specifications for the search regions
as in the published 8~TeV search~\cite{ATLAS20}, and assume that the sensitivity remains the same for 14 TeV~\footnote{We
consider this to be a conservative assumption, as the signal-to-background ratio is likely to improve.}.

\begin{figure}[ht!]
\begin{center}
\includegraphics[height=6cm]{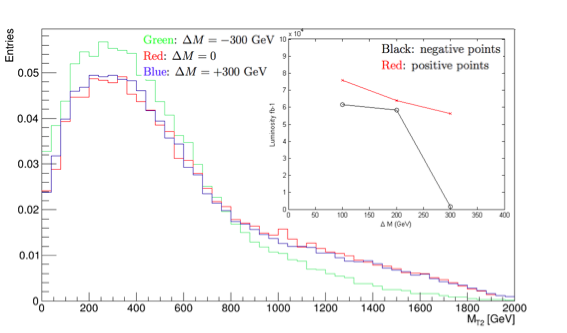}\\
\includegraphics[height=6cm]{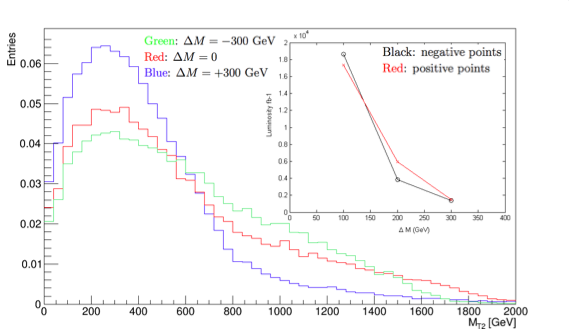}
\end{center}
\caption{\label{fig:MT2}\it
Simulations for 14-TeV collisions, \htr{using {\sc Pythia~8}~\protect\cite{PYTHIA, py8susy} and
including Standard Model backgrounds}, of
the distributions in the MT2 variable for (upper panel) the nominal value of the gluino mass at the low-mass CMSSM best-fit
point, $m_{\tilde g} \simeq 2280$~GeV (blue histogram), and gluino masses differing by
$\pm 300$~GeV (green and blue histograms), and
similarly for (lower panel) the nominal value of the squark mass $m_{\tilde q} \simeq 2080$~GeV and
values $\pm 300$~GeV. In both cases, we fix the other
sparticle masses to their nominal best-fit values, assuming in particular that the LSP mass
$m_\chi = 450$~GeV. The inserts show the integrated luminosities at 14~TeV that
would be required to distinguish at the 3-$\sigma$ level
between the best fit and other models with the indicated mass shifts $\Delta M$.}
\end{figure}

Fig.~\ref{fig:MT2} displays prospective histograms of the MT2 distributions obtained from
simulations using {\sc Pythia~8}~\cite{PYTHIA, py8susy} and the MSTW2008NLO
parton distribution functions~\cite{MSTW} for different values of $m_{\tilde g}$ (upper panel),
the right-handed squark mass $m_{\tilde q}$ (lower panel)~\footnote{\htr{Here and subsequently,
we include in our simulations the Standard Model backgrounds from $W^\pm$, $Z^0$, ${\bar t} t$
and single-$t$ production. However, we do not embark on
a simulation of either ATLAS or CMS, since the purpose of our exploratory study is to give a
first feeling for what might be possible in future LHC runs, and suitable detector simulations for the
high-luminosity LHC are not available.}}.
In both cases, we compare the distribution for the nominal mass at the best-fit point with the corresponding
distributions for values of the mass deviating from the nominal value by $\pm 300$~GeV,
keeping the other sparticle masses fixed. In the gluino case,
we see that the MT2 histogram for the nominal value $m_{\tilde g} = 2280$~GeV (in red) is very similar
to that for the $- 300$~GeV choice (in blue), whereas the histogram for the $+ 300$~GeV choice is
less similar. The reverse is true for the squark case (middle panel): here the nominal histogram
for $m_{\tilde q_R} \simeq 2080$~GeV (red) is more
similar to that for the $+ 300$~GeV choice (green), and less similar to that for the $-300$~GeV case (blue).

\htr{The plots in Fig.~\ref{fig:MT2} were obtained by recalculating the full {\sc Pythia} output
as the squark and gluino masses were varied around the best-fit CMSSM point.
In some cases, the variation changed the ordering of the squark and gluino masses, 
leading to substantial changes in the MT2 distribution, e.g., in the $\Delta M = -300$~GeV 
case in the upper panel of Fig.~\ref{fig:MT2} (green histogram) and in the $\Delta M = +300$~GeV case in the lower panel
(blue histogram). The changes in the shapes of the MT2 distributions were less important when the mass ordering
stayed the same. In addition, the variations in the shape of the MT2 distribution include the effects of changes 
in the relative production rates of ${\tilde g} {\tilde g}$, ${\tilde g} {\tilde q}$, ${\tilde q} {\tilde q}$ and ${\tilde q} {\tilde {\bar q}}$
final states arising from the mass variation.}

\begin{figure}[ht!]
\centerline{
\includegraphics[height=6cm]{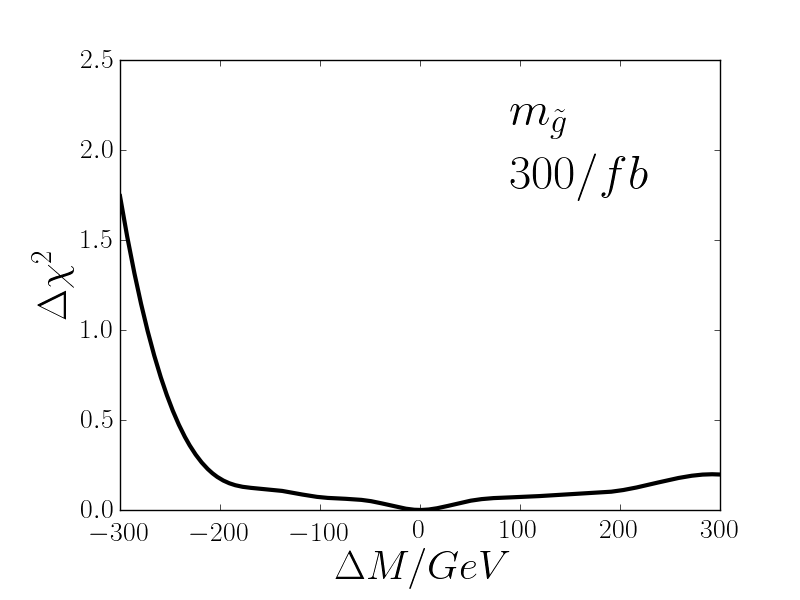}
\includegraphics[height=6cm]{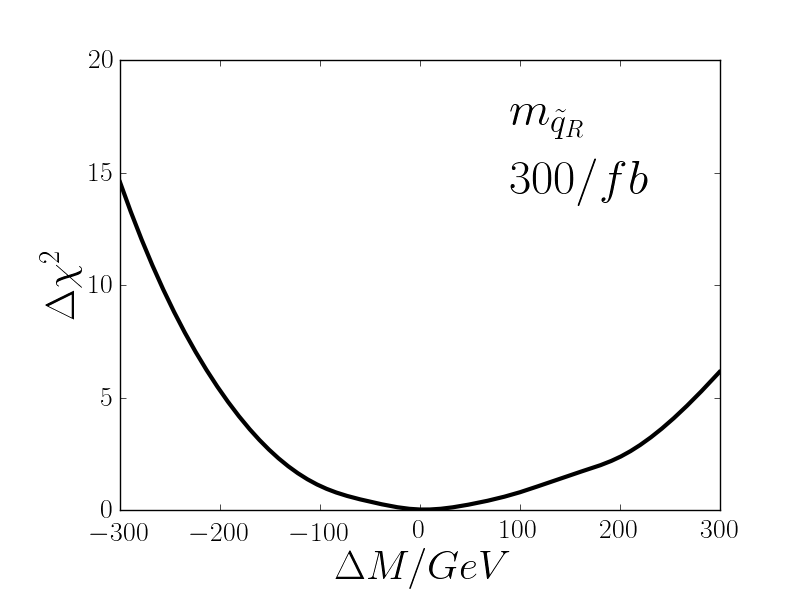}
}
\centerline{
\includegraphics[height=6cm]{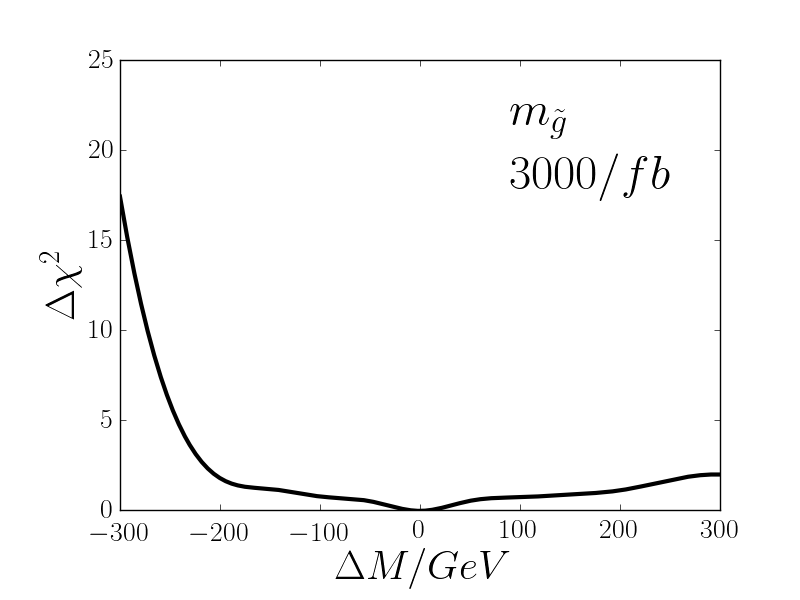}
\includegraphics[height=6cm]{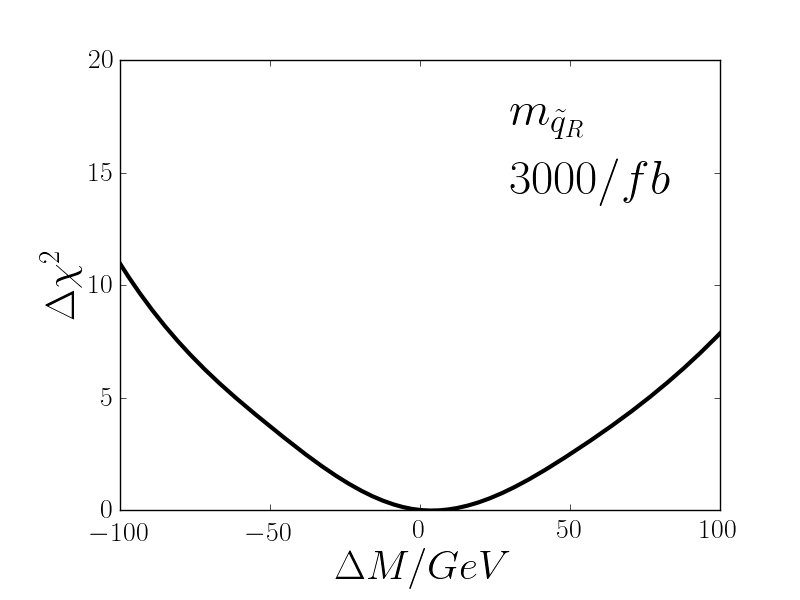}
}
\caption{\label{fig:masschi2}\it
The $\chi^2$ functions for $m_{\tilde g}$ (left panels) and
$m_{\tilde q_R}$ (right panels), as estimated from cross-section and MT2
measurements with 300/fb (upper panels) and 3000/fb (lower panels).}
\end{figure}

In order to estimate the uncertainties in measurements of sparticle masses that could be possible at the LHC,
we have performed fits to the simulated data for varying amounts of integrated luminosity. We use these to
estimate the 68\% CL ranges of mass estimates obtainable with either 300 or 3000/fb of integrated luminosity.
As seen in Fig.~\ref{fig:MT2}, the changes in the MT2 distributions for gluino and squark mass
changes of $\pm 300$~GeV are quite different, so we do not expect symmetric Gaussian uncertainties, and we
note that the same is true for the projected cross-section measurements shown in Fig.~\ref{fig:sigmas}.
Combining these with the MT2 measurements, we find the $\chi^2$ distributions as functions of
$m_{\tilde g}$ and $m_{\tilde q_R}$ shown in Fig.~\ref{fig:masschi2}  in the
left and right panels, respectively. The $\chi^2$ functions are evaluated as
\begin{equation}
\chi^2 (m) \; = \; \frac{1}{n} \sum_{i = 1}^n \left( \frac{(N_i (m) - N_i (\hat m))^2}{\sigma_i^2} \right) \, ,
\label{constructchi2}
\end{equation}
where $\hat m$ is the nominal mass, the $N_i$ are numbers of events in the simulation and $\sigma_i$ is
the statistical error in each bin for the assumed luminosity,
and the sum over $i = 1, ... n$ includes all the bins in the histograms added in quadrature.
The upper row of panels is for 300/fb of integrated luminosity,
and the lower row is for 3000/fb of integrated luminosity.
On the basis of this analysis, we estimate the following fit uncertainties with 300/fb of data at 14~TeV:
\begin{eqnarray}
300/{\rm fb}: \; \; \Delta m_{\tilde g} & = & (-270, + \dots)~{\rm GeV} \, , \nonumber \\
\Delta m_{\tilde q_R} & = & (-100, +110)~{\rm GeV} \, .
\label{300uncertainties}
\end{eqnarray}
where the $\dots$ indicate that these measurements provide no useful upper limit on $m_{\tilde g}$,
and with 3000/fb:
\begin{eqnarray}
3000/{\rm fb}: \; \; \Delta m_{\tilde g} & = & (-110, +150)~{\rm GeV} \, , \nonumber \\
\Delta m_{\tilde q_R} & = & (-30, +35)~{\rm GeV} \, .
\label{3000uncertainties}
\end{eqnarray}
These uncertainties do not include a potential systematic effect from jet energy scale uncertainties.
However, as we expect these to be at the level of 10\% or below, their overall impact is expected to be sub-dominant.

\begin{figure}[ht!]
\centerline{
\begin{turn}{90}
{\small ~~ $p_T$ in ${\tilde g} \to {\tilde q_R} + {\bar q}$ decay}
\end{turn}
\includegraphics[height=5cm]{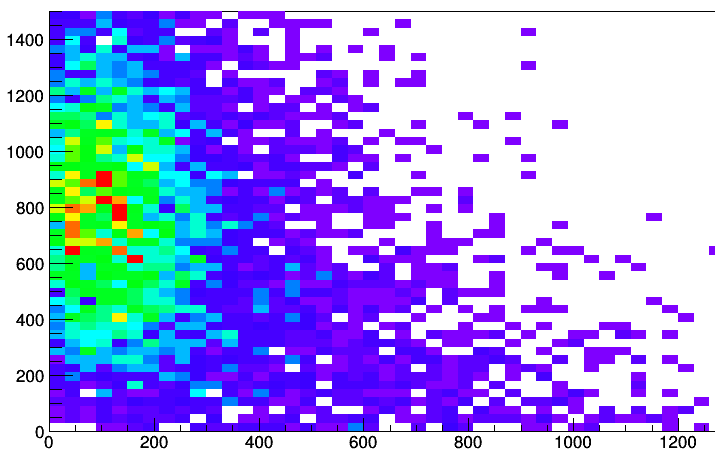}
\hspace{0.5cm}
\begin{turn}{90}
{\small ~~ $p_T$ in ${\tilde g} \to {\tilde q_R} + {\bar q}$ decay}
\end{turn}
\includegraphics[height=5cm]{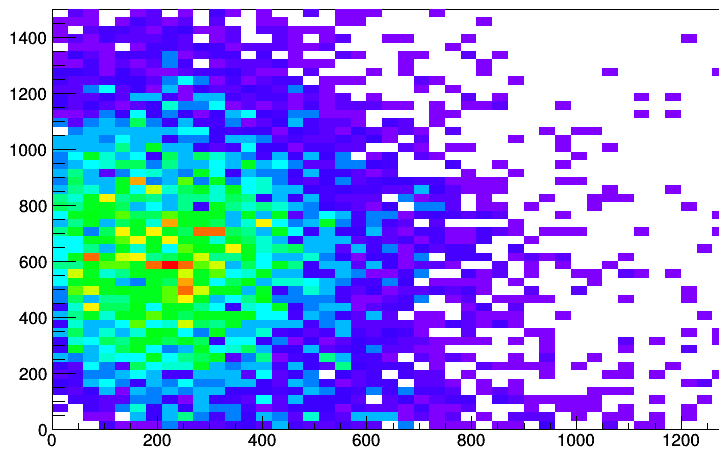}}
~~~~~~~~~~~~~~~~ {\small $p_T$ in $q_R \to q + \chi$ decay} 
~~~~~~~~~~~~~~~~~~~~~~~~~~~~~~~~~ {\small $p_T$ in $q_R \to q + \chi$ decay} 
\caption{\label{fig:squarkscatter}\it
Scatter plots of the $p_T$ (in GeV) of the jet emitted in ${\tilde q_R} \to q + \chi$ decay
(horizontal axis) and the jet emitted in ${\tilde g} \to {\tilde q_R} + {\bar q}$ decay
(vertical axis) resulting from a simulation of gluino 
pair-production at the LHC at 14~TeV. Left panel: for the best-fit ${\tilde g}$ and ${\tilde q_R}$ masses.
Right panel: for the best-fit values of $m_{\tilde g}$ but with $m_{\tilde q_R}$ reduced by 300~GeV.}
\end{figure}

The upper and lower panels of Fig.~\ref{fig:MT2} show that the mass difference
$m_{\tilde g} - m_{\tilde q_R}$ is poorly constrained by the MT2 measurement, and
this is reflected in the asymmetric $\chi^2$ distributions seen in Fig.~\ref{fig:masschi2}. However,
there are many other possible measurements at the LHC. In particular, we have
considered the extra information that could be obtained from measurements of the
(relatively) soft jet emitted in the decay ${\tilde g} \to {\tilde q_R} + {\bar q}$, which would be
monochromatic in the gluino rest frame. Fig.~\ref{fig:squarkscatter} displays scatter plots of
the $p_T$ of the jet emitted in ${\tilde q_R} \to q + \chi$ decay
(horizontal axis) and the jet emitted in ${\tilde g} \to {\tilde q_R} + {\bar q}$ decay
(vertical axis) based on simulations of ${\tilde g}$ pair-production.
The left panel is for the best-fit values of $m_{\tilde g}$ and $m_{\tilde q_R}$,
and the right panel is for the same value of $m_{\tilde g}$ but with $m_{\tilde q_R}$
reduced by 300~GeV: the plots are clearly distinct.

\begin{figure}[ht!]
\centerline{
\begin{turn}{90}
{\small Distribution (arbitrary units)}
\end{turn}
\includegraphics[height=5cm]{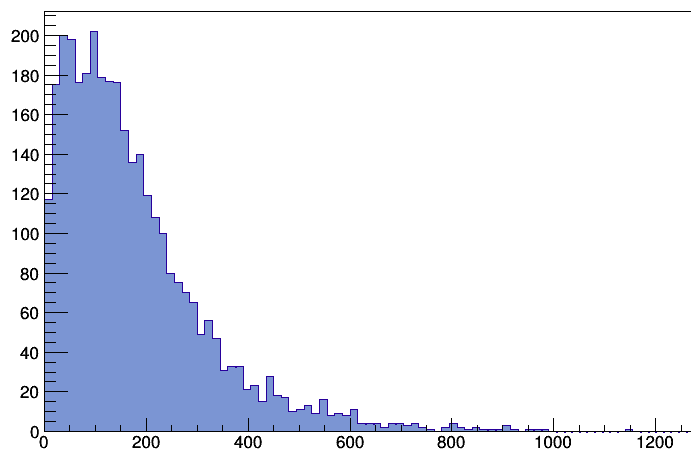}
\hspace{0.5cm}
\begin{turn}{90}
{\small Distribution (arbitrary units)}
\end{turn}
\includegraphics[height=5cm]{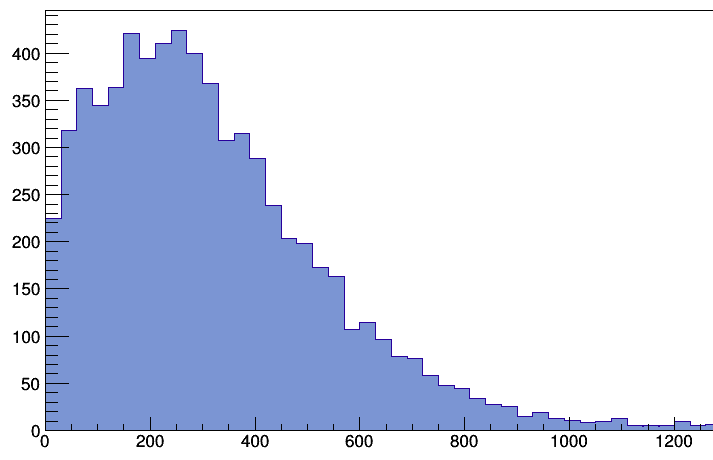}}
~~~~~~~~~~~~~~~~ {\small $p_T$ in ${\tilde g} \to {\tilde q_R} + q$ decay} 
~~~~~~~~~~~~~~~~~~~~~~~~~~~~~~~~~ {\small $p_T$ in ${\tilde g} \to {\tilde q_R} + q$ decay} 
\caption{\label{fig:ptxsquark}\it
Simulations of of the distributions of the quark $p_T$ (in GeV) from ${\tilde g}$ 
pair-production at the LHC at 14~TeV followed by ${\tilde g} \to {\tilde q_R} + q$
decays. Left panel: for the best-fit ${\tilde g}$ and ${\tilde q_R}$ masses,
and (right panel) for the same value of $m_{\tilde g}$ but with  $m_{\tilde q_R}$
reduced by 300~GeV.}
\end{figure}

Fig.~\ref{fig:ptxsquark} displays the spectrum of the `soft' jet in the same two
cases: in the left panel with the best-fit ${\tilde g}$ and ${\tilde q_R}$, and in the right
panel with the same value of $m_{\tilde g}$ but with  $m_{\tilde q_R}$ reduced by 300~GeV.
These can clearly be distinguished with a high degree of confidence. We do not display
the corresponding distribution with $m_{\tilde q_R}$ increased by 300~GeV, since in this
case the ${\tilde q_R}$ is heavier than the ${\tilde g}$ and there is no `monochromatic'
supplementary jet in gluino decay.

\begin{figure}[ht!]
\centerline{
\includegraphics[height=6.2cm]{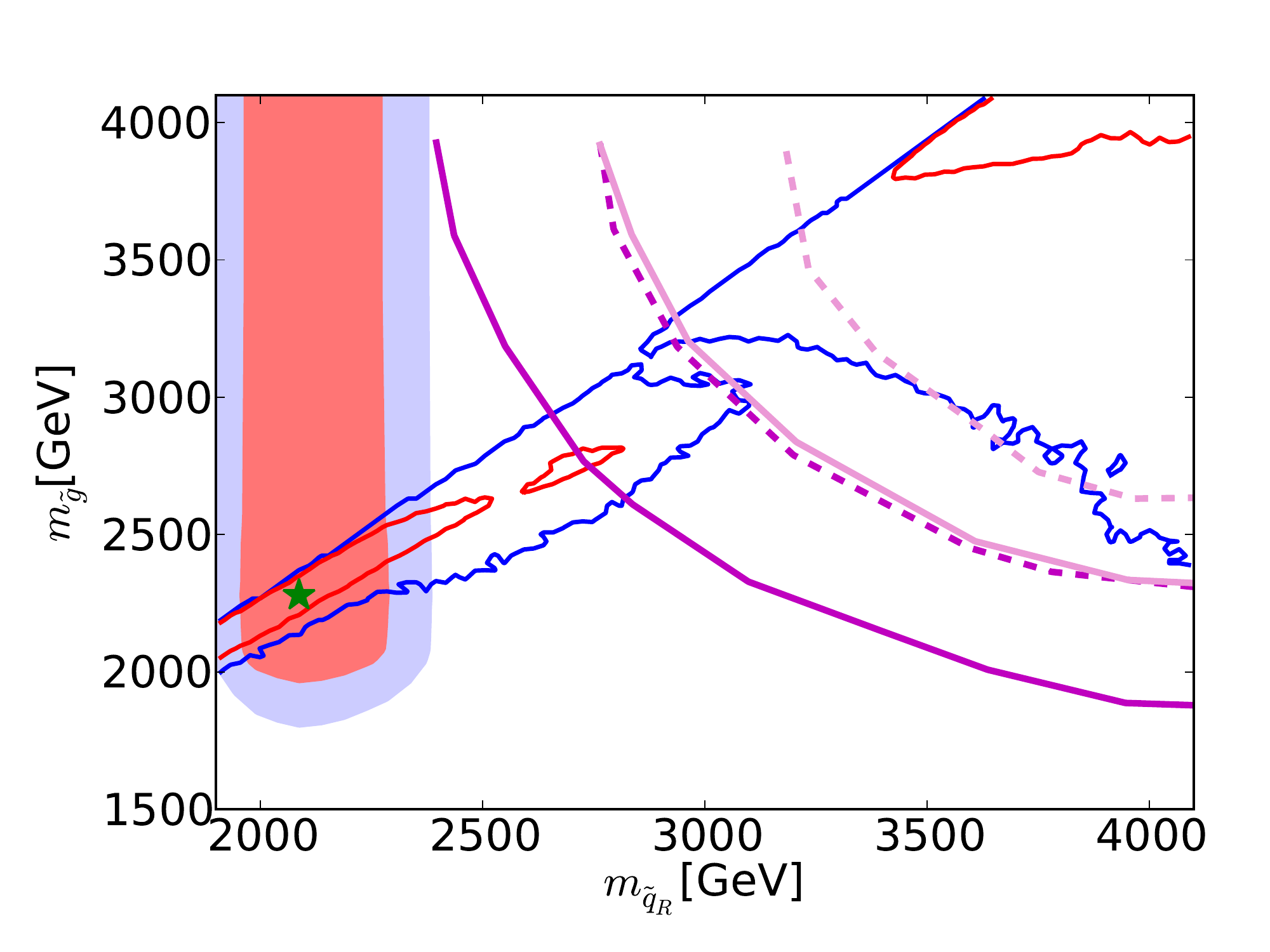}
\includegraphics[height=6.2cm]{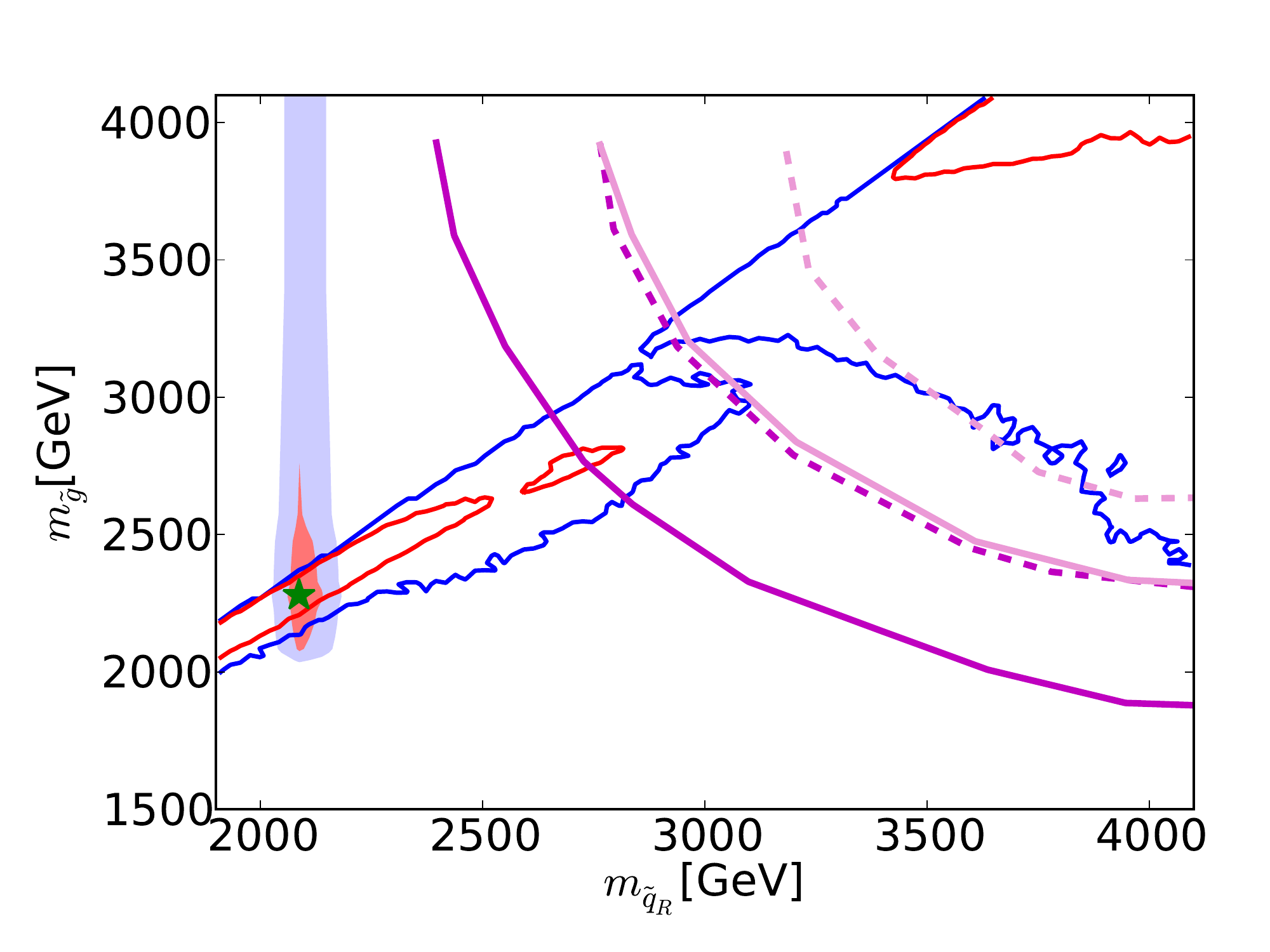} 
}
\centerline{
\includegraphics[height=6.2cm]{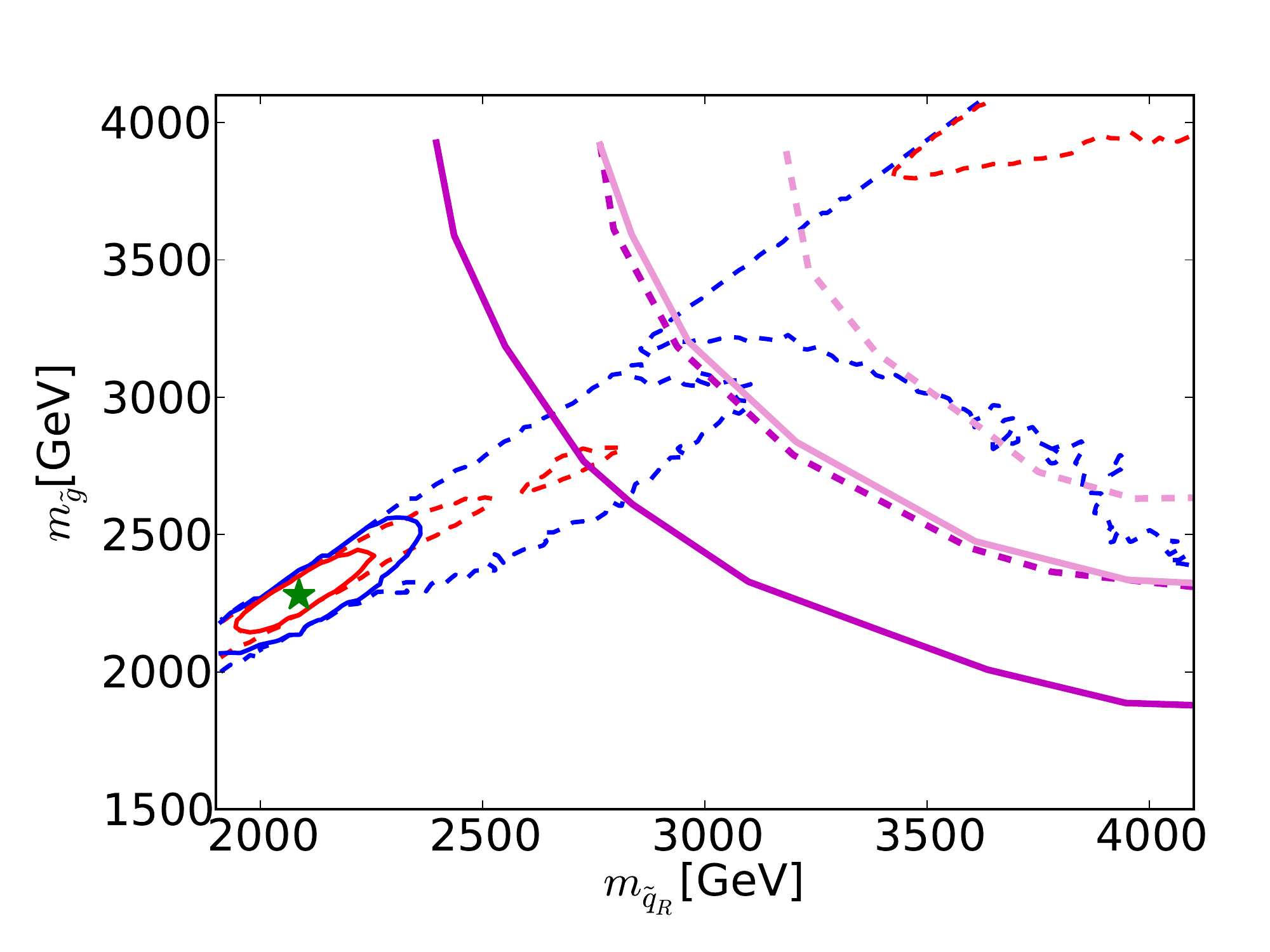}
\includegraphics[height=6.2cm]{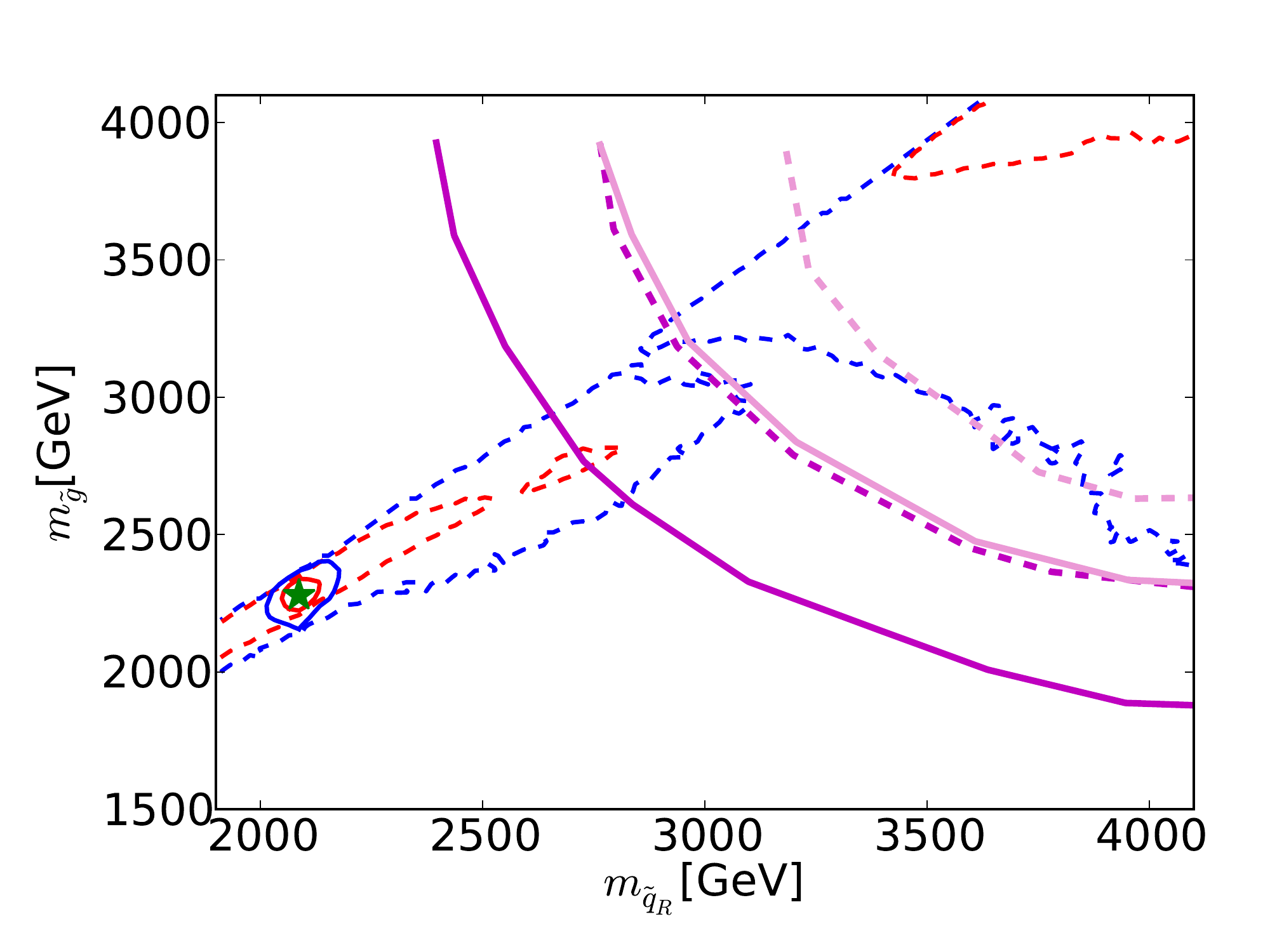} 
}
\caption{\label{fig:msqmgl}\it
The upper panels show the 68 and 95\% CL regions (shaded pink and blue, respectively) in the $(m_{\tilde q_R}, m_{\tilde g})$ planes
obtained from cross-section, MT2 and supplementary jet measurements
at LHC14 with 300/fb (left panel) and 3000/fb (right panel). These regions
are superposed on the best-fit point (green star) and the 68 and 95\% CL regions found in a recent global
fit to the CMSSM~\protect\cite{mc9} (solid red and blue lines), and the magenta lines show the prospective capabilities of the
LHC to exclude at the 95\% CL (dashed) supersymmetry or discover it
at the 5-$\sigma$ level (solid) with 300/fb or 3000/fb (darker and lighter lines).
In the lower panels we show as solid red and blue lines the results of fits combining the LHC measurements
with the recent global fit (here shown as dashed lines).
}
\end{figure}

We assume that the jet energy in ${\tilde g} \to {\tilde q_R} + {\bar q}$ decay
can be measured with an accuracy of 50~GeV. This information can then be
combined with the cross-section and MT2 distribution discussed earlier to
estimate 68 and 95\% CL regions in the $(m_{\tilde q_R}, m_{\tilde g})$ plane. These are shown shaded pink and blue, respectively,
in Fig.~\ref{fig:msqmgl} for 300/fb of integrated luminosity (upper left panel)
and for 3000/fb of integrated luminosity (upper right panel). As in Fig.~\ref{fig:Kees},
the low-mass portions of the solid red and blue contours outline the Crimea region 
and the high-mass portions correspond to the Eurasia region Finally, the solid [dashed] magenta lines
(darker and lighter) show the 5-$\sigma$ discovery (95\% CL exclusion) reaches of the LHC with 300 (3000)/fb.
The lower panels of Fig.~\ref{fig:msqmgl} show as solid red and blue lines the 68 and 95\% CL contours from 
fits combining the prospective LHC measurements with the recent global fit~\cite{mc9} (whose CL contours are
displayed as dashed lines in these panels).

\begin{figure}[ht!]
\centerline{
\includegraphics[height=6.2cm]{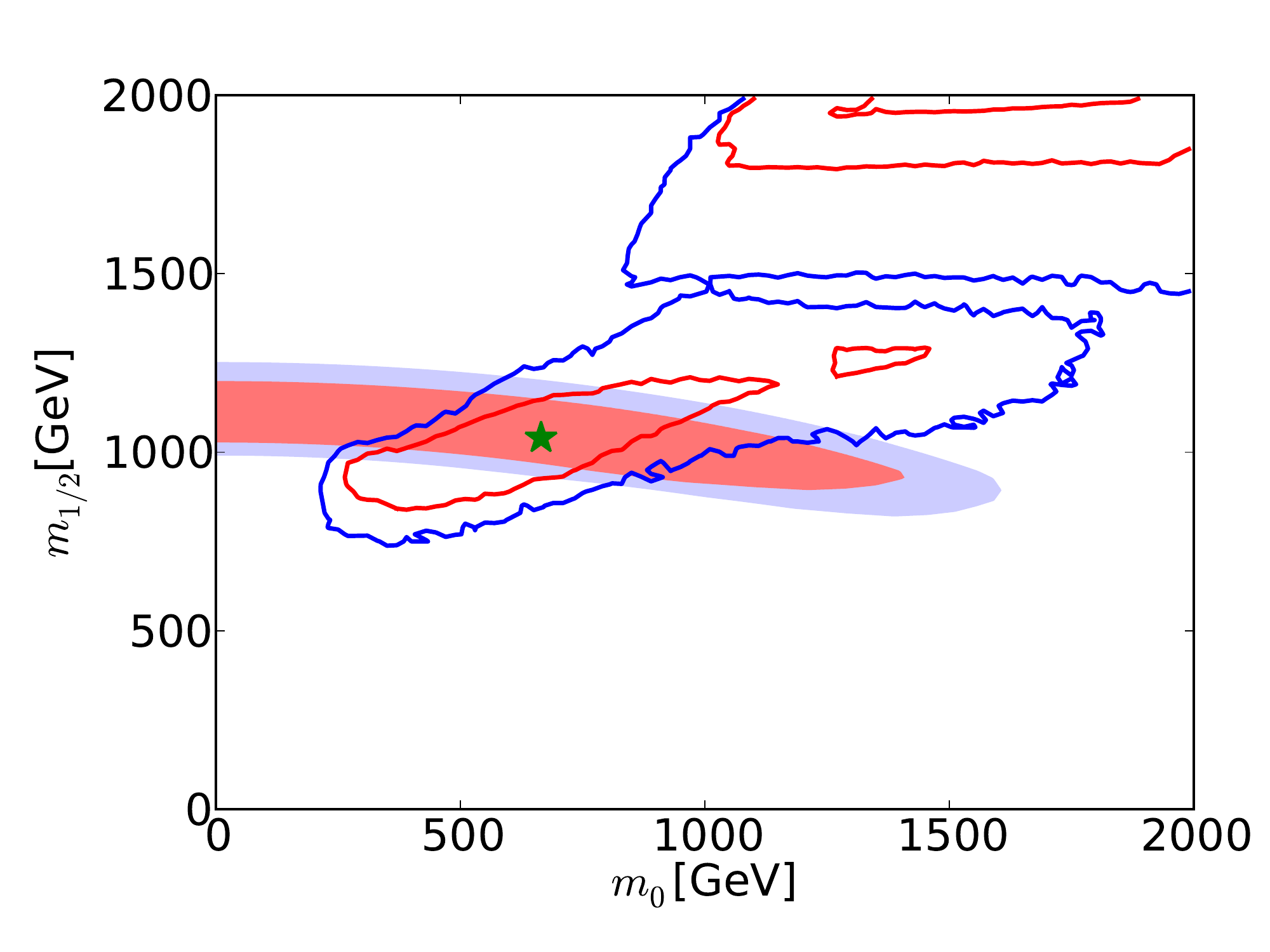}
\includegraphics[height=6.2cm]{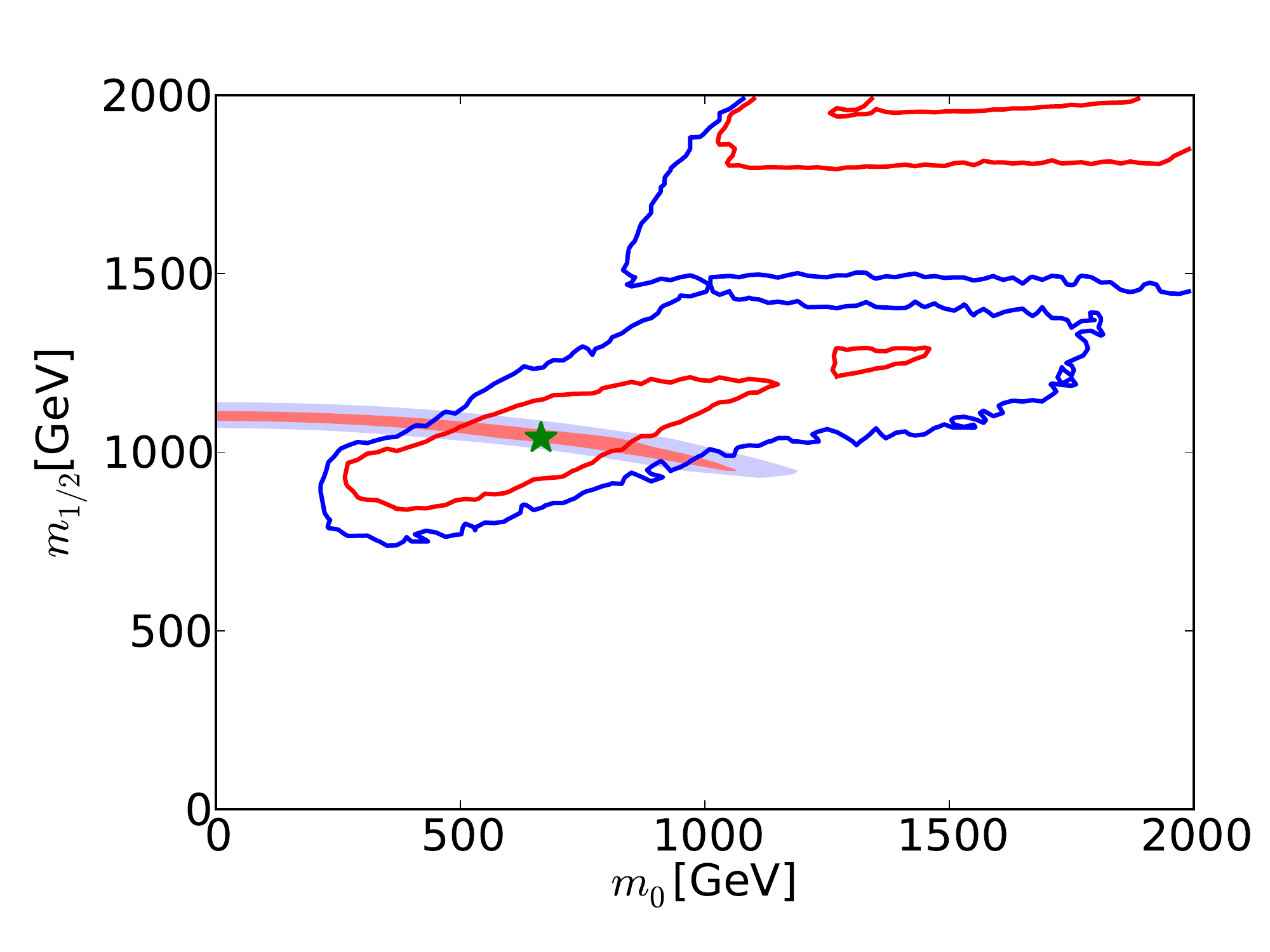}
}
\centerline{
\includegraphics[height=6.2cm]{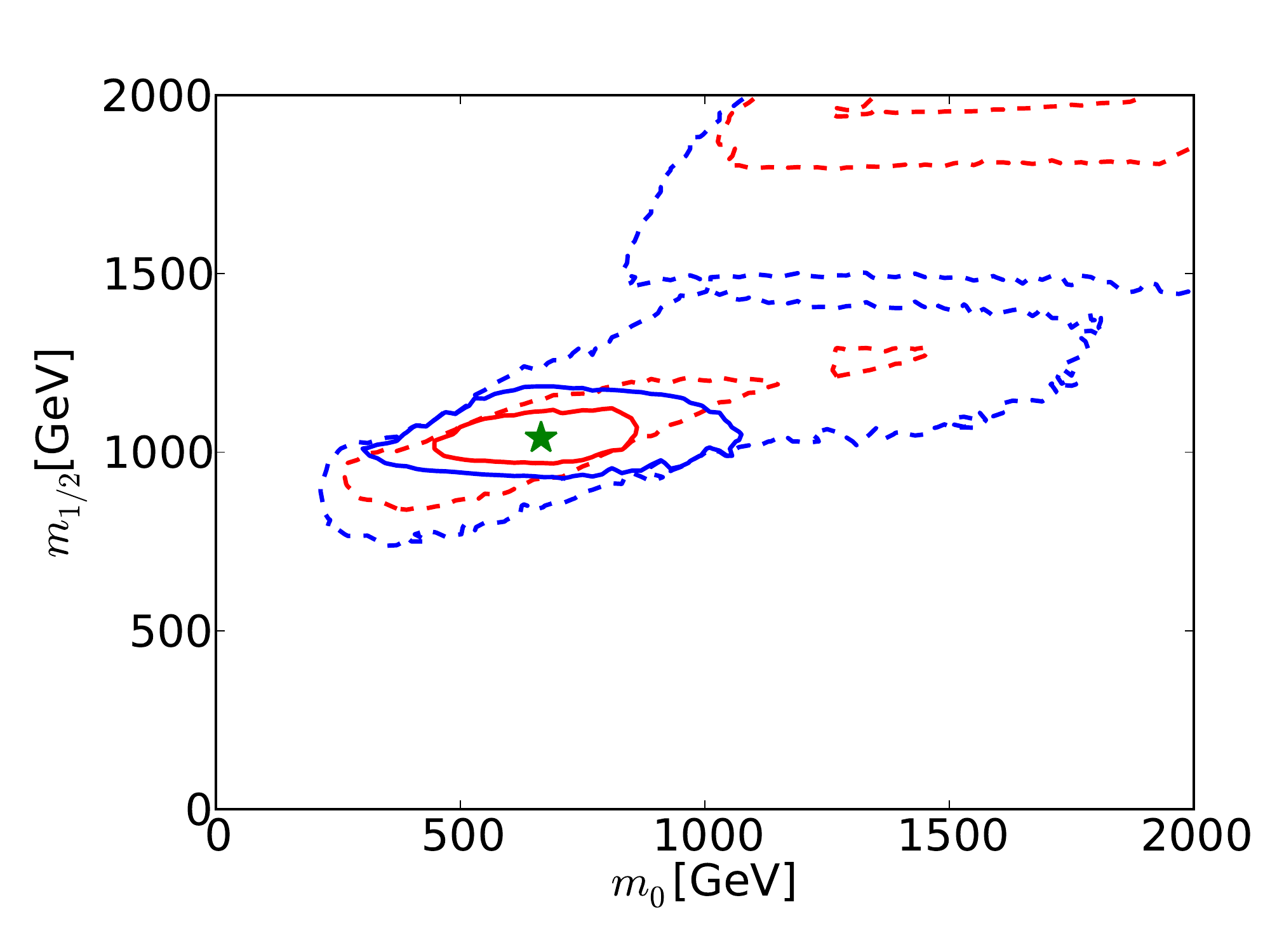}
\includegraphics[height=6.2cm]{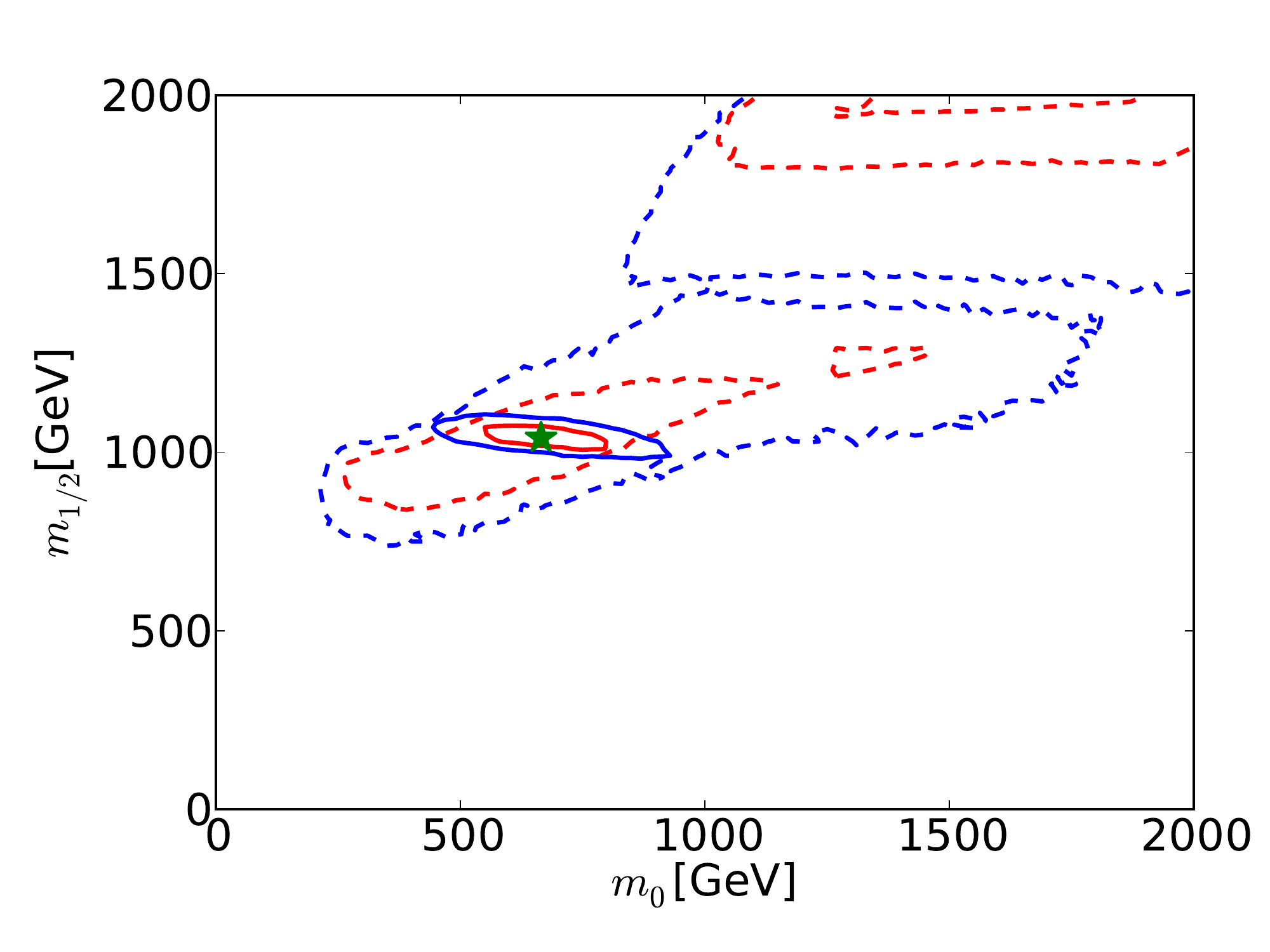}
}
\caption{\label{fig:m0m12Jad}\it
The upper panels show the 68 and 95\% CL regions (shaded pink and blue, respectively) in the $(m_0, m_{1/2})$ planes
obtained from cross-section, MT2 and supplementary jet measurements
at LHC14 with 300/fb (left panel) and 3000/fb (right panel). These regions
are superposed on the 68 and 95\% CL regions found in a recent global
fit to the CMSSM~\protect\cite{mc9} (red and blue lines).
In the lower panels we show as solid lines the results of fits combining the LHC measurements
with this global fit (here shown as dashed lines).}
\end{figure}

These prospective measurements can be projected onto the
$(m_0, m_{1/2})$ plane of the CMSSM, as seen in the upper panels of Fig.~\ref{fig:m0m12Jad},
also for 300/fb of integrated luminosity (upper left panel) and 3000/fb of integrated luminosity (upper right panel).
Formally, the corresponding numerical 68\% CL uncertainties are
\begin{eqnarray}   
300/{\rm fb}: \; \; \Delta m_0 & = & (-670, +620)~{\rm GeV} \, , \nonumber \\    
3000/{\rm fb}: \; \; \Delta m_0 & = & (-670, +220)~{\rm GeV} \, ,
\label{m0uncertainties}
\end{eqnarray}     
and
\begin{eqnarray}
300/{\rm fb}: \; \; \Delta m_{1/2} & = & (-140, +100)~{\rm GeV} \, , \nonumber \\
3000/{\rm fb}: \; \; \Delta m_{1/2} & = & (-90, +20)~{\rm GeV} \, .
\label{m12uncertainties}
\end{eqnarray}
The upper panels of Fig.~\ref{fig:m0m12Jad} also show that these numbers imply non-trivial
correlations between $m_0$ and $m_{1/2}$.
They also show, as solid red and blue lines, respectively, the boundaries of the 68 and 95\%
CL regions found in the recent global analysis of current data~\cite{mc9}. We see that the prospective future LHC measurements could
provide information that would be complementary to this global fit~\cite{mc9}~\footnote{The main effect of the measurement of the 
supplementary jet in ${\tilde g} \to {\tilde q_R} + q$ decay is to truncate the
preferred strip in the 300/fb case: it has no visible effect in the 3000/fb case.}.

The lower panels of Fig.~\ref{fig:m0m12Jad} show the results (solid red and blue lines) of combining these LHC measurements with
the recent global fit~\cite{mc9} (dashed red and blue lines). We see that the LHC measurements would
reduce substantially the sizes of the 68 and 95\% CL regions already with 300/fb, and that the
prospective 3000/fb measurements would be particularly powerful in this regard.

\subsection{Stop Measurements}

\begin{figure}[ht!]
\centerline{
\hspace{-0.5cm}
\includegraphics[height=6cm]{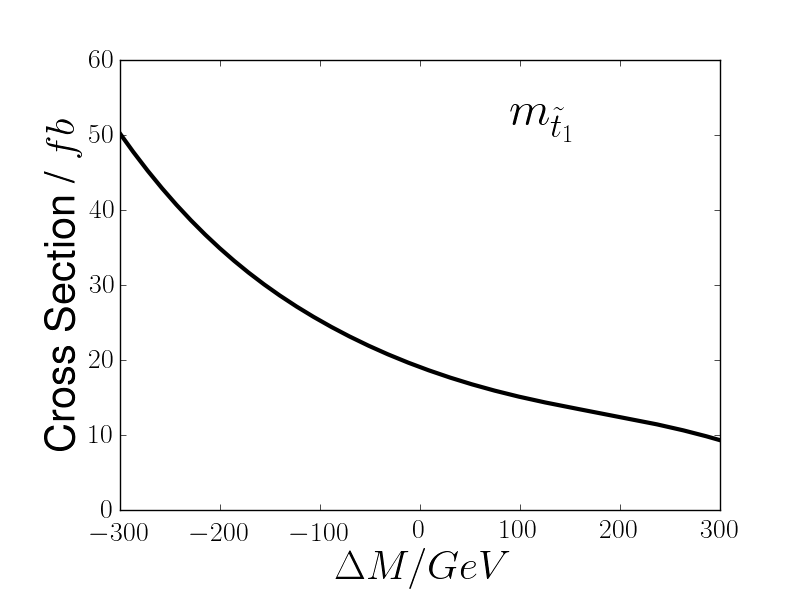}
\hspace{-0.5cm}
\includegraphics[height=6cm]{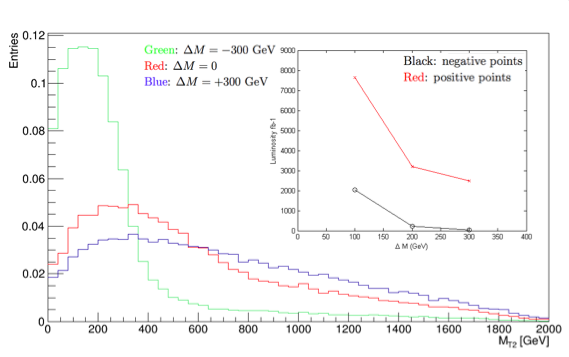}
}
\caption{\label{fig:mst1}\it
Left panel: The sensitivity of the total stop pair-production cross-section to $m_{\tilde t_1}$.
Right panel: Simulations for 14-TeV collisions of
the distributions in the MT2 variable for the nominal value of the lighter
stop mass $m_{\tilde t_1} = 1020$~GeV and values $\pm 300$~GeV, with the other
sparticle masses fixed to their nominal best-fit values. The insert shows the integrated luminosities at 14~TeV that
would be required to distinguish at the 3-$\sigma$ level
between the best fit and other models with the indicated mass shifts $\Delta M$ relative to the value at the
low-mass best-fit point.}
\end{figure}

We have also considered the possible accuracy in measuring $m_{\tilde t_1}$ via
${\tilde t_1} + \overline{\tilde t_1}$ production at the LHC at 14~TeV.
The left panel of Fig.~\ref{fig:mst1} shows the sensitivity of the
total stop pair-production cross-section to $m_{\tilde t_1}$: we see that
over the displayed range it
is greater than those to the $m_{\tilde g}$ and $m_{\tilde q_R}$, that were
shown in Fig.~\ref{fig:sigmas}. The right panel of Fig.~\ref{fig:mst1} shows
histograms of MT2 for the nominal mass $m_{\tilde t_1} \simeq 1020$~GeV
at the representative low-mass best-fit point
and for choices differing by $\pm 300$~GeV. These cases are quite distinct,
as is also seen in the inset, which displays the luminosities required for
3-$\sigma$ discrimination between the nominal value of $m_{\tilde t_1}$
and selected larger or smaller values.

\begin{figure}[ht!]
\centerline{
\begin{turn}{90}
{\small ~~~ Distribution (arbitrary units)}
\end{turn}
\includegraphics[height=5.5cm]{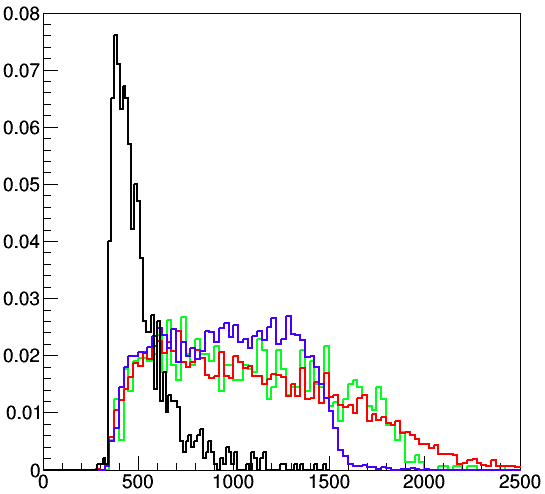}
\begin{turn}{90}
{\small ~~~ Distribution (arbitrary units)}
\end{turn}\includegraphics[height=5.5cm]{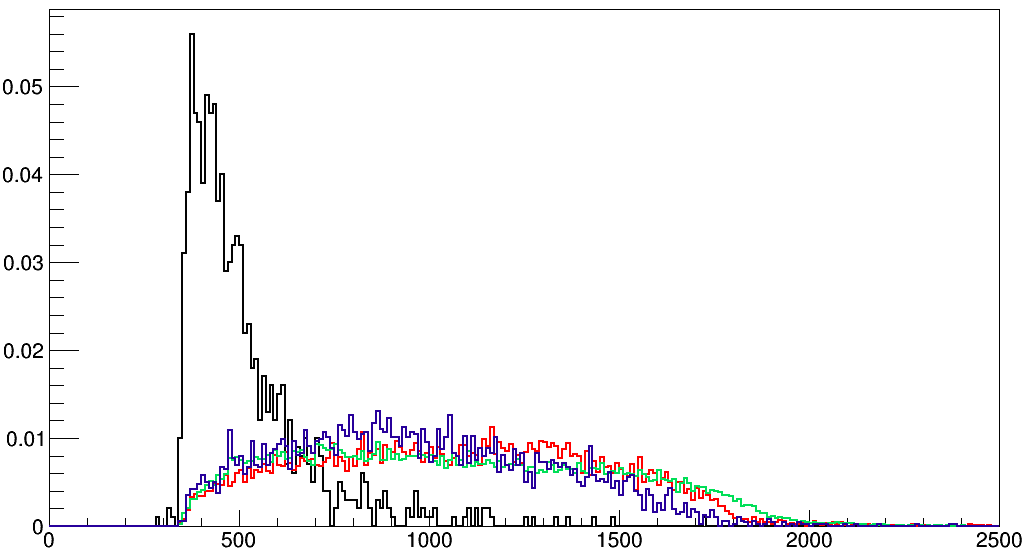}}
~~~~~~~ {\small ${\bar t} t$ invariant mass (in GeV)} 
~~~~~~~~~~~~~~~~~~~~~~~~~~~~~ {\small ${\bar t} t$ invariant mass (in GeV)} 
\caption{\label{fig:topquark_invmass}\it
The unit-normalised $t {\bar t}$ invariant-mass distribution resulting from a simulation of ${\tilde t_1} + \overline{\tilde t_1}$ 
production at the LHC at 14~TeV. Left panel: for the best-fit ${\tilde g}$ and ${\tilde t_1}$ masses
of 2280 and 1020~GeV
(green histogram), compared with the Standard Model background (black histogram) 
and simulations with ${\tilde g}$ masses 300~GeV above (red histogram) and
300~GeV below (blue histogram) the nominal value of $m_{\tilde g}$.
Right panel: similarly for the best-fit ${\tilde g}$ and ${\tilde t_1}$ masses
(green histogram), compared with the Standard Model background (black histogram) 
and simulations with ${\tilde t_1}$ masses 300~GeV above (red histogram) and
300~GeV below (blue histogram) the nominal value of $m_{\tilde t_1}$.}
\end{figure}

The left panel of Fig.~\ref{fig:topquark_invmass}
displays the \htr{shape of the unit-normalised} $t {\bar t}$ invariant mass distribution resulting from a simulation of such events
using {\sc Pythia~8}~\cite{PYTHIA, py8susy} and the MSTW2008NLO
parton distribution functions~\cite{MSTW},
produced with the nominal CMSSM best-fit values of $m_{\tilde t_1} = 1020$~GeV and $m_{\tilde g} = 2280$~GeV (green histogram),
compared with the Standard Model background (black histogram), which is sharply
peaked at low invariant masses close to the $t {\bar t}$ threshold. Also shown in Fig.~\ref{fig:topquark_invmass} are the
invariant-mass distributions for ${\tilde g}$ masses 300~GeV above (red histogram) and
300~GeV below (blue histogram) the nominal value of $m_{\tilde t_1}$. As expected, the
higher (lower) mass gives a longer (shorter) tail in the invariant-mass distribution.
On the other hand, as we see in the right panel of Fig.~\ref{fig:topquark_invmass}
that the invariant ${\tilde t_1} + \overline{\tilde t_1}$ mass distribution in ${\tilde g}$
decays is almost independent of $m_{\tilde t_1}$ for fixed $m_{\tilde g}$.

\begin{figure}[ht!]
\centerline{
\includegraphics[height=6cm]{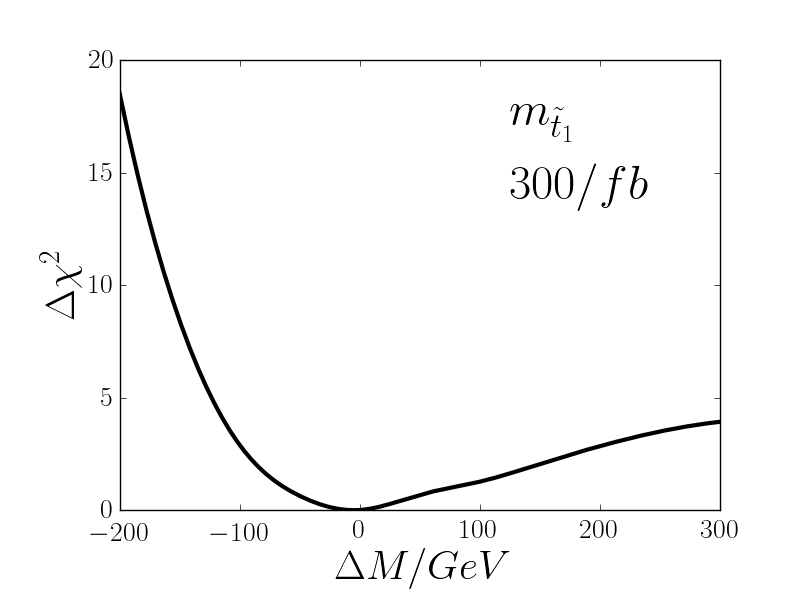}
\includegraphics[height=6cm]{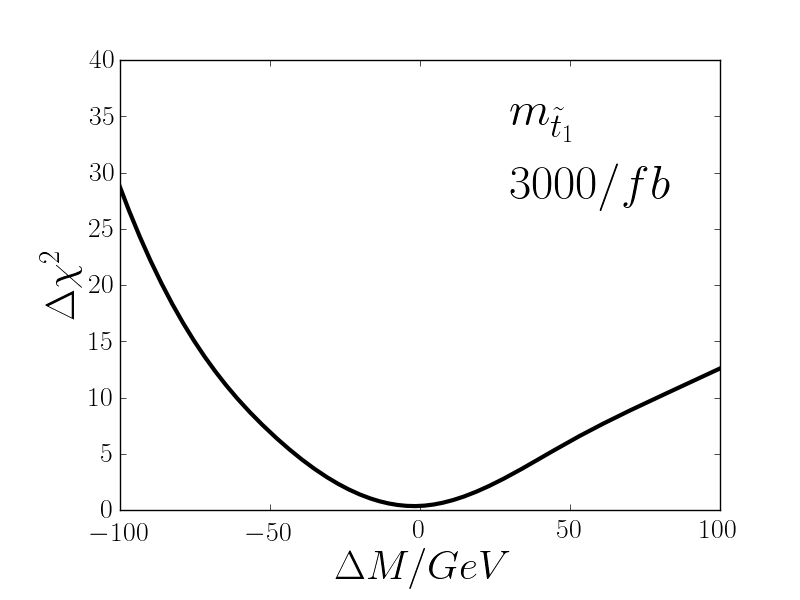}}
\caption{\label{fig:stopmass}\it
The $\chi^2$ functions for $m_{\tilde t_1}$ estimated from LHC14
measurements with 300/fb (left panel) and 3000/fb (right panel).}
\end{figure}

Combining the cross-section, MT2 and $t {\bar t}$ invariant-mass measurements, 
we find the $\chi^2$ distributions as functions of
$m_{\tilde t_1}$ shown in Fig.~\ref{fig:stopmass}. The left panel is for 300/fb of integrated luminosity,
and the right panel is for 3000/fb of integrated luminosity.
We find the following fit uncertainties with 300/fb or 3000/fb of data at 14~TeV:
\begin{eqnarray}
300/{\rm fb}: \; \; \Delta m_{\tilde t_1} & = & (-30, +50)~{\rm GeV} \, , \nonumber \\
3000/{\rm fb}: \; \; \Delta m_{\tilde t_1} & = & (-10, +15)~{\rm GeV} \, .
\label{stopuncertainties}
\end{eqnarray}
As in the previous cases, these uncertainties should be convoluted with a systematic jet energy scale
uncertainty of $\sim 10$\%.

The uncertainties (\ref{stopuncertainties}) may be used to estimate the corresponding
uncertainties in the trilinear soft supersymmetry-breaking parameter $A_0$, by comparing
the stop mass (which is very sensitive to $A_0$) with the squark and gluino masses
(which are insensitive to $A_0$). The effect of marginalising over the latter masses
can be incorporated by assuming that $m_0$ and $m_{1/2}$ have their best-fit values,
as is also the case for $\tan \beta$.
In estimating the uncertainty in $A_0$, we incorporate the correlation between $A_0$
and $\mu$ that is imposed by the electroweak vacuum conditions within the CMSSM, finding
\begin{eqnarray}
300/{\rm fb}: \; \; \Delta A_0 & = & (+80, -150)~{\rm GeV} \, , \nonumber \\
3000/{\rm fb}: \; \; \Delta A_0 & = & (+30, -40)~{\rm GeV} \, .
\label{A0uncertainties}
\end{eqnarray}
We emphasise again that these uncertainties do not take into account the jet energy scale
uncertainty, which we expect to be subdominant.

As a final point in this Section, we comment on the magnitudes of some of the branching ratios
for sparticle decays that are represented by dashed lines in Fig.~\ref{fig:spectrum}.
The earlier analysis of ${\tilde g} \to {\tilde q} + {\bar q}$ decays exploits the fact that this
decay mode is dominant if $m_{\tilde g} > m_{\tilde q_R}$, as reflected in the boldness
of the dashed line connecting the ${\tilde g}$ and ${\tilde q_R}$ states. We draw attention
to the decay ${\tilde t_2} \to {\tilde t_1} + h$, which is also dominant, having a branching
ratio of 77\% represented also by a bold dashed line. This implies that about 50\% of
${\tilde t_2} \overline{\tilde t_2}$ events would contain, in addition to a $t {\bar t}$ pair,
a pair of high-$p_T$ Higgs bosons and substantial missing transverse energy. Typical
boost factors for the Higgs bosons would be $\sim 5$. A detailed exploration of this
experimental signature lies beyond the scope of this paper.

\section{$e^+ e^-$ Probes of Supersymmetry in the Optimistic Scenario}

In the low-mass `optimistic' CMSSM scenario there would be interesting opportunities
for both direct and indirect precision probes of supersymmetry at an $e^+ e^-$ collider,
which we now explore.

\subsection{Direct Sparticle-Pair Production}

\begin{figure}[hbtp!]
\centerline{
\includegraphics[height=8cm]{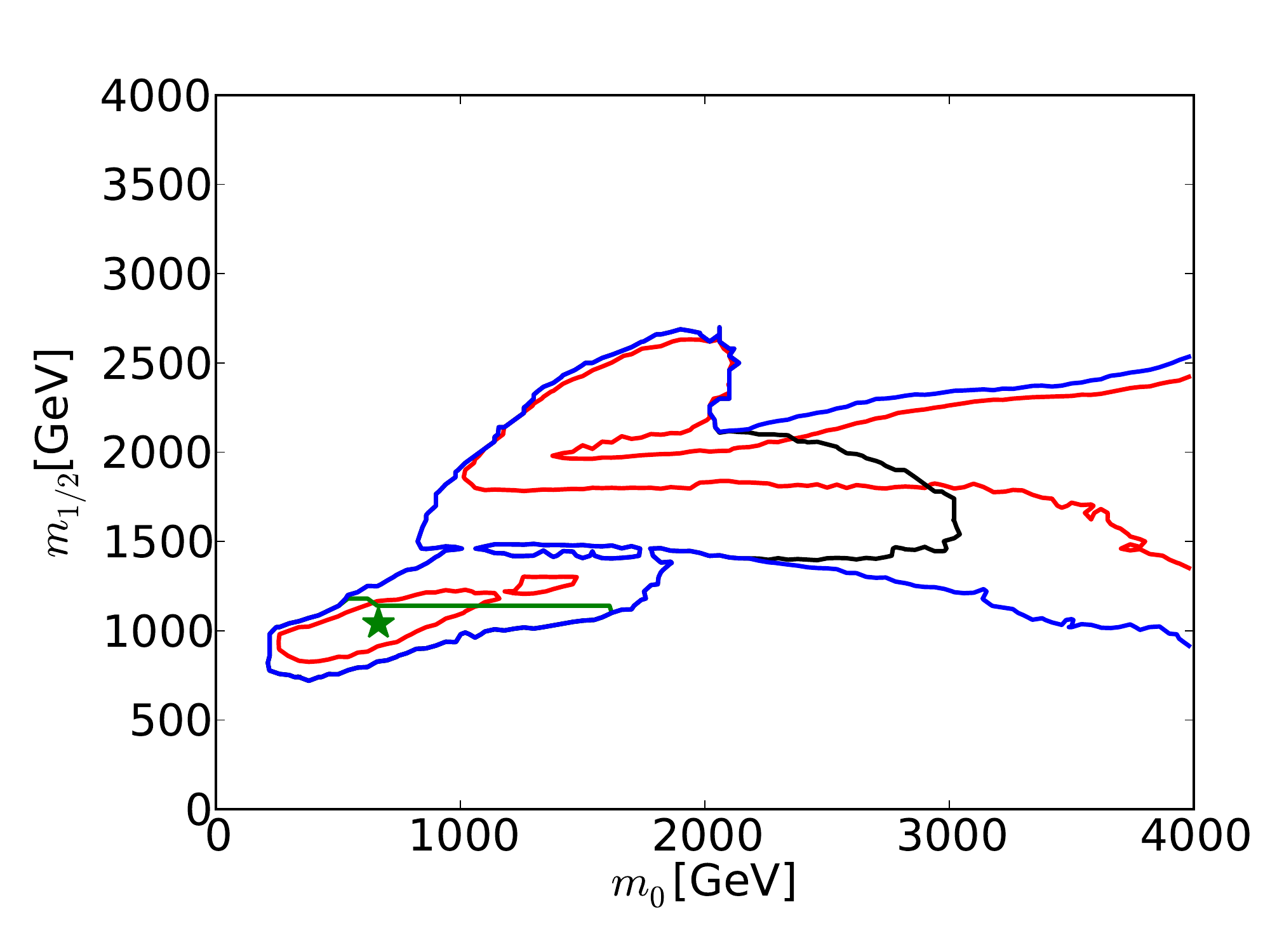}
}
\caption{\label{fig:stau}\it
Contours where it is possible to attain at the 95\% CL $m_{\tilde \tau_1} = 500 \, (1500)$~GeV, indicated by solid
green (black) lines, are overlaid on the
$(m_0, m_{1/2})$ plane in the CMSSM, 
with the same CL contours and best-fit point from a global fit~\protect\cite{mc9} as
displayed previously in Fig.~\protect\ref{fig:Kees}.}
\end{figure}

The most direct possibility would be pair-production and measurement of
electroweakly-interacting sparticles. The next-to-lightest supersymmetric
particle (NLSP) is expected, in generic regions of the CMSSM
parameter space, to be the lighter stau slepton ${\tilde \tau_1}$. Accordingly, 
Fig.~\ref{fig:stau} displays, superimposed on the same CMSSM
$(m_0, m_{1/2})$ plane discussed previously, contours showing where it is possible at the 95\% CL to attain $m_{\tilde \tau_1}
= 500$~GeV (green), the largest mass that could be pair-produced with an $E_{CM} = 1$~TeV
linear collider, and 1500~GeV (black), the largest mass that could be 
pair-produced with an $E_{CM} = 3$-TeV linear collider such as CLIC. These contours are restricted to the
regions within the 68 and 95\% CL regions found in the recent global fit~\cite{mc9},
where the CMSSM parameter space is well sampled.
We see that the $m_{\tilde \tau_1} = 500$~GeV line crosses the `Crimea' region, whereas the
$m_{\tilde \tau_1} = 1500$~GeV lines reach deep into the `Eurasia' region.
In particular, the low-mass best-fit point in the CMSSM lies within the $m_{\tilde \tau_1} 
= 500$~GeV reach of a 1-TeV $e^+ e^-$ collider. In the low-mass `Crimea' region, the cold dark
matter density is brought into the range acceptable to cosmology by coannihilation with the stau,
so the $m_{\tilde \tau_1} \le 500$~GeV contour has $m_{1/2}$ almost constant. On the other hand,
in the `Eurasia' region other mechanisms such as neutralino annihilation via direct-channel heavy
Higgs poles come into play, and the $m_{\tilde \tau_1} \le 1500$~GeV contour has a more complicated
shape.

Within the specific CMSSM model studied, a 500-GeV $e^+ e^-$ collider
would very likely be unable to detect directly any supersymmetric particles. This is because
the contour for $m_{\tilde \tau_1} = 250$~GeV, the largest mass that could be pair-produced with an $E_{CM} = 500$-GeV
linear collider, would lie at $m_{1/2} \lesssim 600$~GeV, which is outside the 95\% CL contour in the $(m_0, m_{1/2})$ plane
shown in Fig.~\ref{fig:stau}. 
A similar conclusion could be drawn from the lower right panels of Figs.~5 and 13 of~\cite{mc9},
where we see that $\Delta \chi^2 > 9$ for $m_{\tilde \tau_1} \le 250$~GeV~\footnote{On the other hand,
this would be possible within the pMSSM10 analysis discussed in~\cite{mc11}.}.

If slepton-antislepton pair-production is accessible at an $e^+ e^-$ collider,
many very precise direct measurements become possible. Two benchmark supersymmetric
scenarios were analysed in~\cite{1304.2825}, and the prospective accuracies for
sparticle mass measurements were assessed. In one of these scenarios (P1), the
slepton mass spectrum was very similar to that at the low-mass CMSSM best-fit
point (see Table~1 and Fig.~\ref{fig:spectrum}), with masses between 1000 and
1100~GeV. The ${\tilde \chi_1^\pm}$ and ${\tilde \chi_1^0}$ masses in scenario P1
were somewhat lower than in the low-mass best-fit CMSSM spectrum, whereas
the ${\tilde \chi_2^0}$ mass was again very similar. Based on simulations of
2/ab of CLIC data at 3~TeV, the following uncertainties in sparticle masses were
estimated:
\begin{equation}
\Delta m_{\tilde e_R} \; = \; 2.9 \, {\rm GeV}, \; \Delta m_{\tilde \chi_1^0} \; = \; 4.6 \, {\rm GeV}, \;
\Delta m_{\tilde \chi_1^\pm} \; = \; 3.6 \, {\rm GeV} \, .
\label{CLICmasserrors}
\end{equation}
The measurement uncertainty in $m_{\tilde \chi_1^0}$ can be converted directly
into the corresponding uncertainty in $m_{1/2}$:
\begin{equation}
\Delta m_{1/2} \; = \; 11 \, {\rm GeV} \, .
\label{CLICm12}
\end{equation}
Combining this uncertainty with the uncertainty in the $m_{\tilde e_R}$ measurement, one finds
\begin{equation}
\Delta m_0 \; = \; 4 \, {\rm GeV} \, .
\label{CLICm0}
\end{equation}
As discussed in~\cite{1304.2825} and~\cite{CLIC}, many precision measurements of supersymmetric
particle masses and other properties would be possible at CLIC point, including
tests of the universality hypotheses of the CMSSM.
However, it is already clear that CLIC could provide exceptional precision in the
determination of CMSSM model parameters, if Nature is described by a model
in the Crimea region. Moreover, the comparison between the CLIC determinations
of the CMSSM parameters with those from the LHC discussed earlier would enable
non-trivial checks to be made of the consistency of the CMSSM assumption of
universal input soft supersymmetry-breaking parameters.

\subsection{Electroweak Precision Observables}

It is also possible to obtain indirect information about supersymmetric models
from electroweak precision observables (EWPOs), similar in principle to the information about $m_t$ and $m_H$
that were obtained previously from precision measurements at LEP and the SLC~\cite{LEPEWWG}.

\begin{figure}[hbt]
\begin{center}
\begin{tabular}{c c}
\includegraphics[height=10cm]{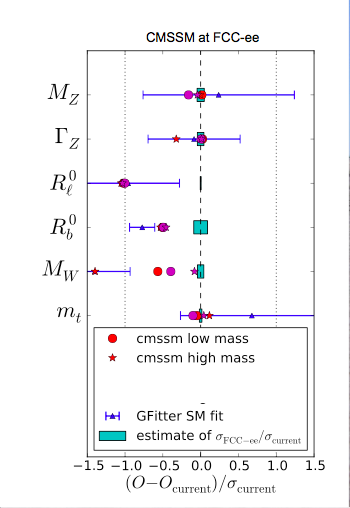} &
\includegraphics[height=10cm]{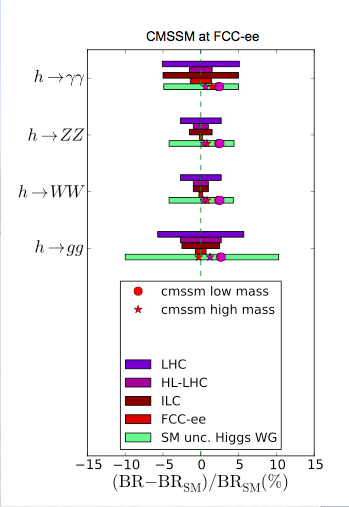} \\
\end{tabular}
\end{center}   
\caption{\label{fig:precision}\it
The left panel compares current measurements of electroweak precision
observables (EWPOs) taken from a Standard Model fit~\protect\cite{GFitter} (blue, with error bars), predictions at low- and high-mass
best-fit points in the CMSSM~\protect\cite{mc9} (red and purple symbols) and prospective FCC-ee (TLEP)
experimental errors~\protect\cite{TLEP} (turquoise bars). The right panel compares prospective
measurements of Higgs branching ratios at future colliders,  low- and high-mass CMSSM
predictions (red and purple symbols) and the current uncertainties
within the Standard Model (turquoise bars).}
\end{figure}

\begin{figure}
\vspace{-2cm}
\begin{center}
\begin{tabular}{c c}
\includegraphics[height=6cm]{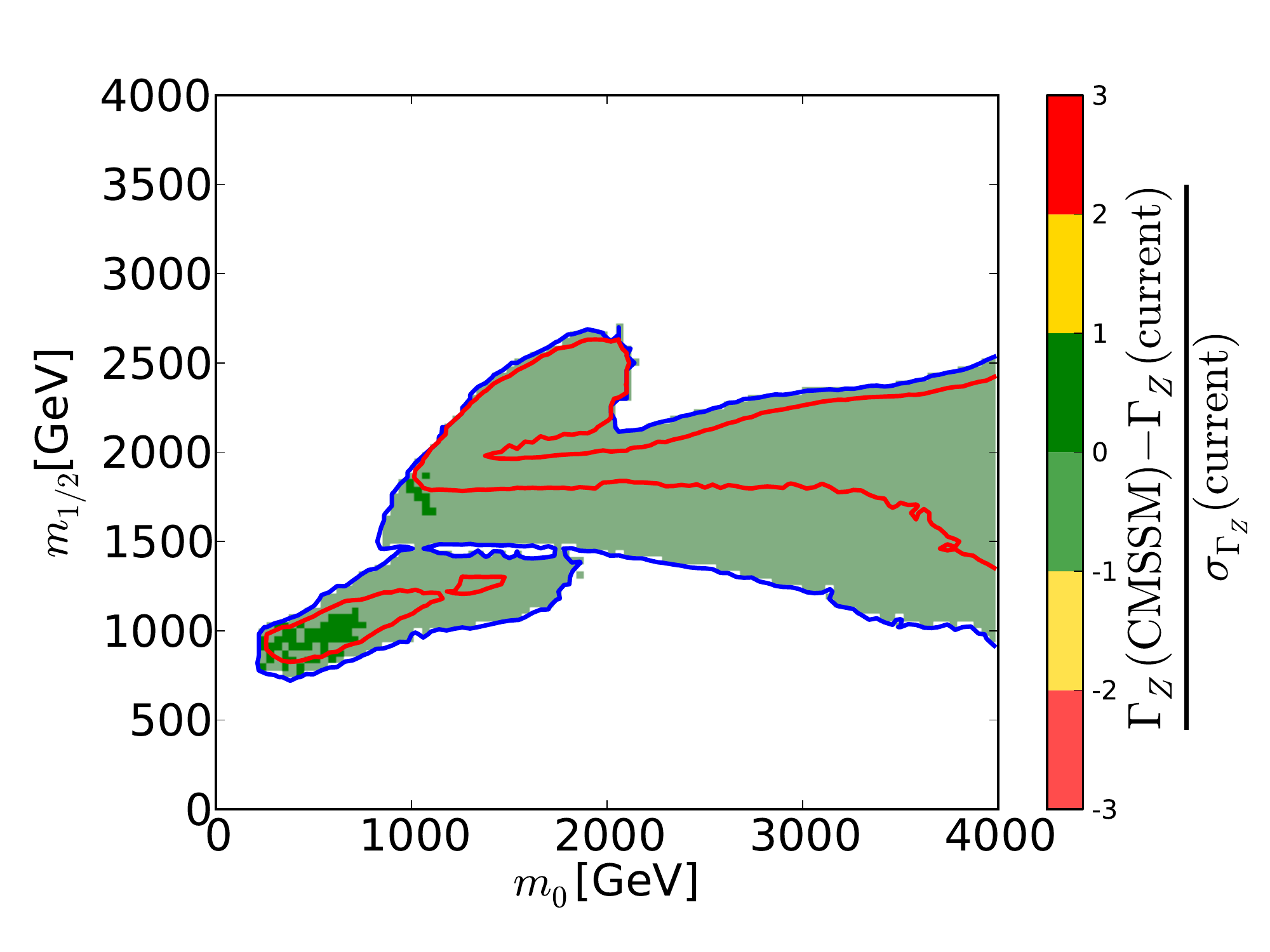} & 
\includegraphics[height=6cm]{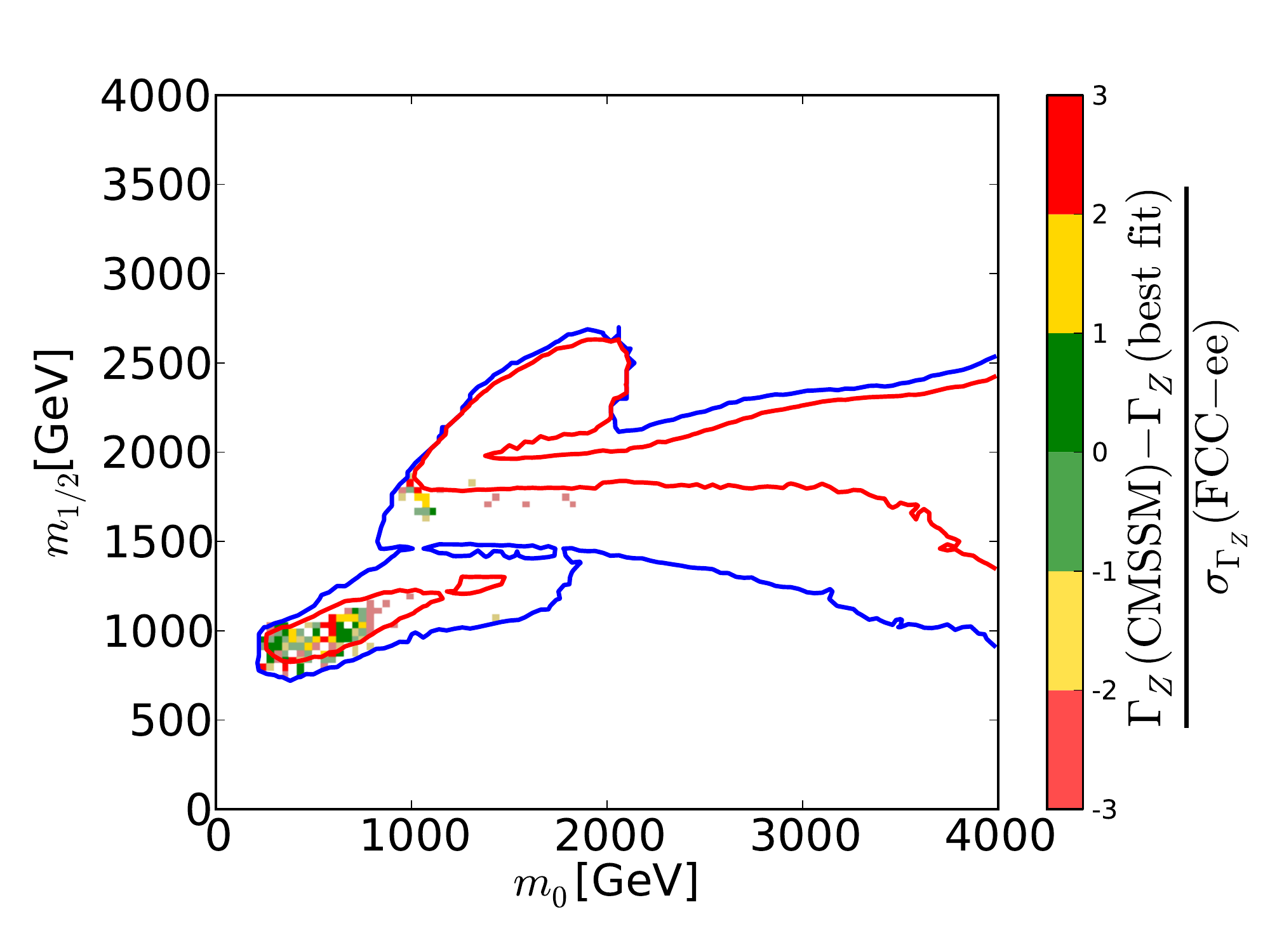} \\
\end{tabular}
\end{center}   
\begin{center}
\begin{tabular}{c c}
\includegraphics[height=6cm]{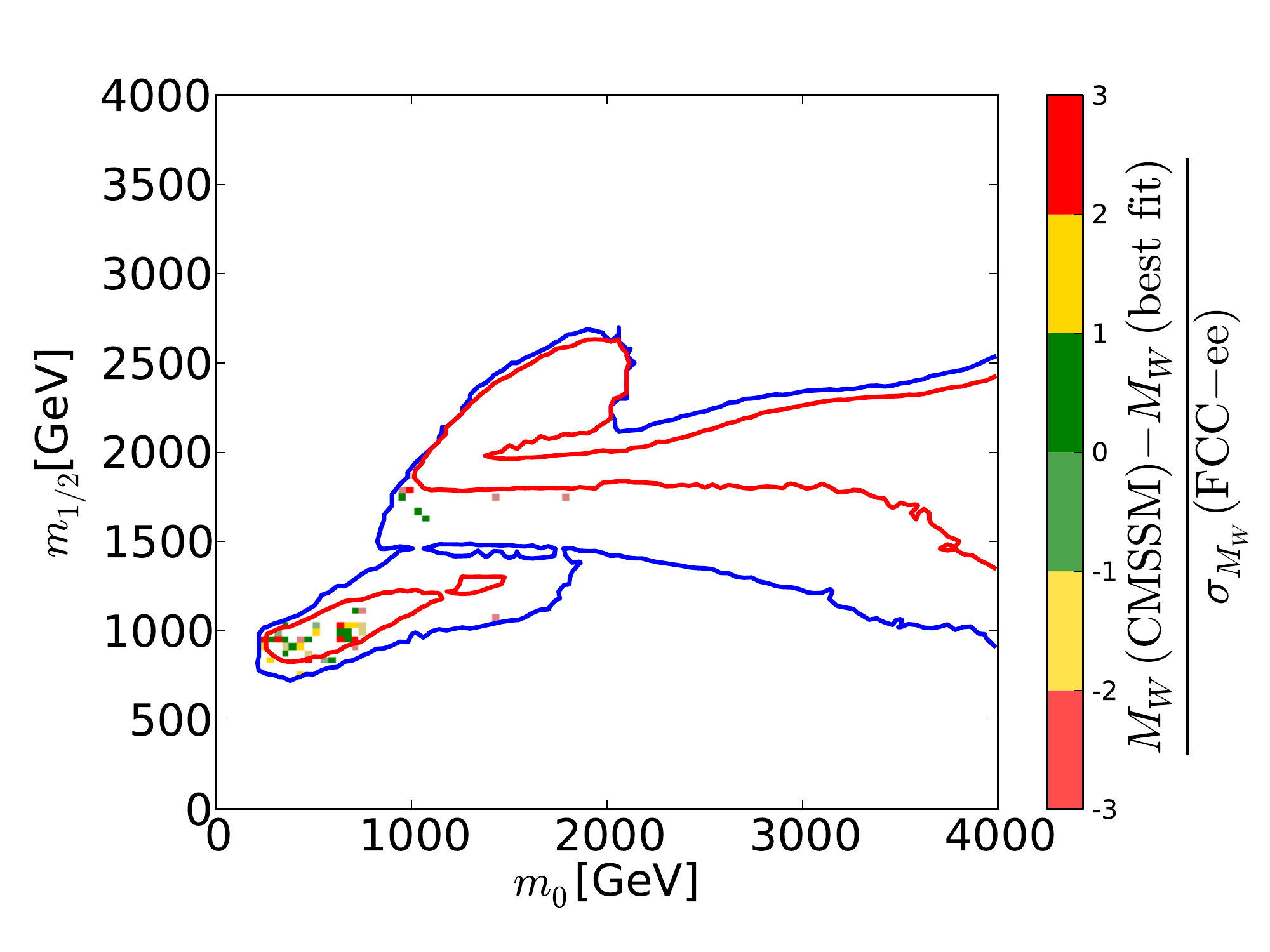} & 
\includegraphics[height=6cm]{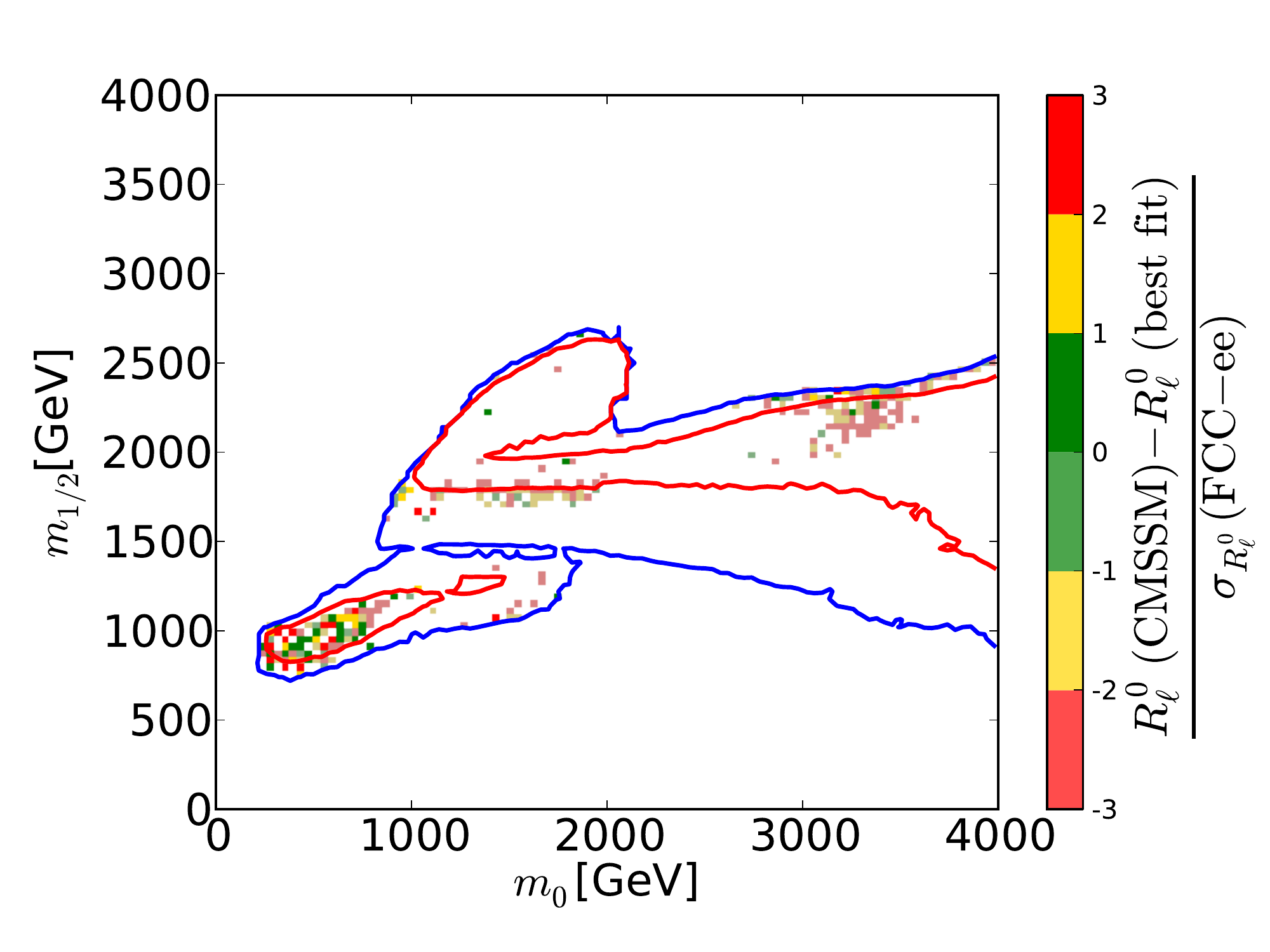} \\
\end{tabular}
\end{center}   
\caption{\label{fig:PrecisionZ}\it
The present measurement of $\Gamma_Z$ (upper left panel)~\protect\cite{LEPEWWG}, and prospective
FCC-ee (TLEP) measurements~\protect\cite{TLEP} of $\Gamma_Z$ (upper right), $M_W$ (lower
left) and $R_\ell$ (lower right) are superposed on the preferred region of the $(m_0, m_{1/2})$ plane
in the CMSSM~\protect\cite{mc9} shown previously in Fig.~\protect\ref{fig:Kees}.
The colours represent deviations from the present central value in units
of the present LHC experimental error (upper left panel), and the deviations from the
values at the low-mass best-fit point of the values at other points in the $(m_0, m_{1/2})$ plane
in units of the estimated future FCC-ee (TLEP) experimental errors (other panels).
}
\end{figure}

The left panel of Fig.~\ref{fig:precision} 
displays as blue points with error bars the central values and 1-$\sigma$ uncertainties of several
such observables, as calculated in a recent global fit~\cite{GFitter}, compared with their values
and current individual experimental uncertainties in the Standard Model. 
Also shown (without theoretical uncertainties) are the values of these observables calculated
at the representative low- and high-mass best-fit points in the CMSSM found in~\cite{mc9}. As is apparent from
the left panel of Fig.~\ref{fig:precision} and the upper left panel of Fig.~\ref{fig:PrecisionZ},
the current experimental error in the measurement of $\Gamma_Z$ is too large
to provide much information about supersymmetric model parameters. The entire
region of the CMSSM $(m_0, m_{1/2})$ plane currently allowed at the 95\% CL
according to the global fit~\cite{mc9} is compatible with the current measurement of $\Gamma_Z$
at the 1-$\sigma$ level~\cite{LEPEWWG}. However, also shown in the left panel of Fig.~\ref{fig:precision},
as turquoise bars, are the prospective experimental errors in measurements at
FCC-ee (TLEP) (neglecting theoretical uncertainties)~\cite{TLEP},
normalized relative to the current experimental errors. It is clear that, for $\Gamma_Z$ and many other
electroweak precision observables, the prospective FCC-ee (TLEP) uncertainties are sufficiently small
to be very sensitive to deviations from their Standard Model values and capable
of constraining supersymmetric scenarios.

The right panel of Fig.~\ref{fig:precision}
makes a similar point for precision Higgs observables, by comparing the estimated precisions
of measurements at the LHC, the ILC and FCC-ee (TLEP)~\cite{ICFA,TLEP}
(shown as colour-coded horizontal bars) with the deviations of the observables from their
Standard Model values that are calculated for the low- and high-mass CMSSM
best-fit points~\cite{mc9}. It is clear that FCC-ee (TLEP) has the best ability to distinguish
these models from the Standard Model, as we discuss in more detail later.

As a first example of the possible utility of the precision electroweak
measurements possible with FCC-ee (TLEP), we consider the
optimistic scenario in which supersymmetry is within the LHC discovery range and
assume, for definiteness, that the model parameters correspond
to the current best-fit low-mass point in the CMSSM. We see in the upper
left panel of Fig.~\ref{fig:PrecisionZ} that, as already commented, the current experimental
uncertainty in $\Gamma_Z$, namely $\Delta \Gamma_Z = 2.3$~MeV~\cite{LEPEWWG}, is too
large to provide significant information about CMSSM model parameters
within the 95\% CL regions displayed in Fig.~\ref{fig:Kees}. \htr{For this reason, all of the
95\% CL region in the upper left panel of Fig.~\ref{fig:Kees} is shaded green, since it lies
within one current standard deviation of the present measurement}. On the other hand, we see in the
upper right panel of Fig.~\ref{fig:PrecisionZ} that the prospective experimental
uncertainty at FCC-ee (TLEP), namely $\Delta \Gamma_Z = 0.1$~MeV~\cite{TLEP}, is
far smaller than the variation in $\Gamma_Z$ across even the CMSSM 68\% CL
Crimea region. \htr{For this reason, much of the 68\% and
95\% CL regions in this panel of Fig.~\ref{fig:Kees} are unshaded, since they lie
more than three current standard deviations away from the prospective measurement}.
The same holds for other electroweak precision observables such
as $M_W$ (\htr{prospective experimental
uncertainty $0.5$~MeV~\cite{TLEP},} lower left panel of Fig.~\ref{fig:PrecisionZ}), $R_\ell$
(\htr{prospective experimental
uncertainty $5 \times 10^{-5}$~\cite{TLEP},} lower right panel of Fig.~\ref{fig:PrecisionZ}) and others not shown.

\begin{figure}[hbtp!]
\centerline{
\includegraphics[height=8cm]{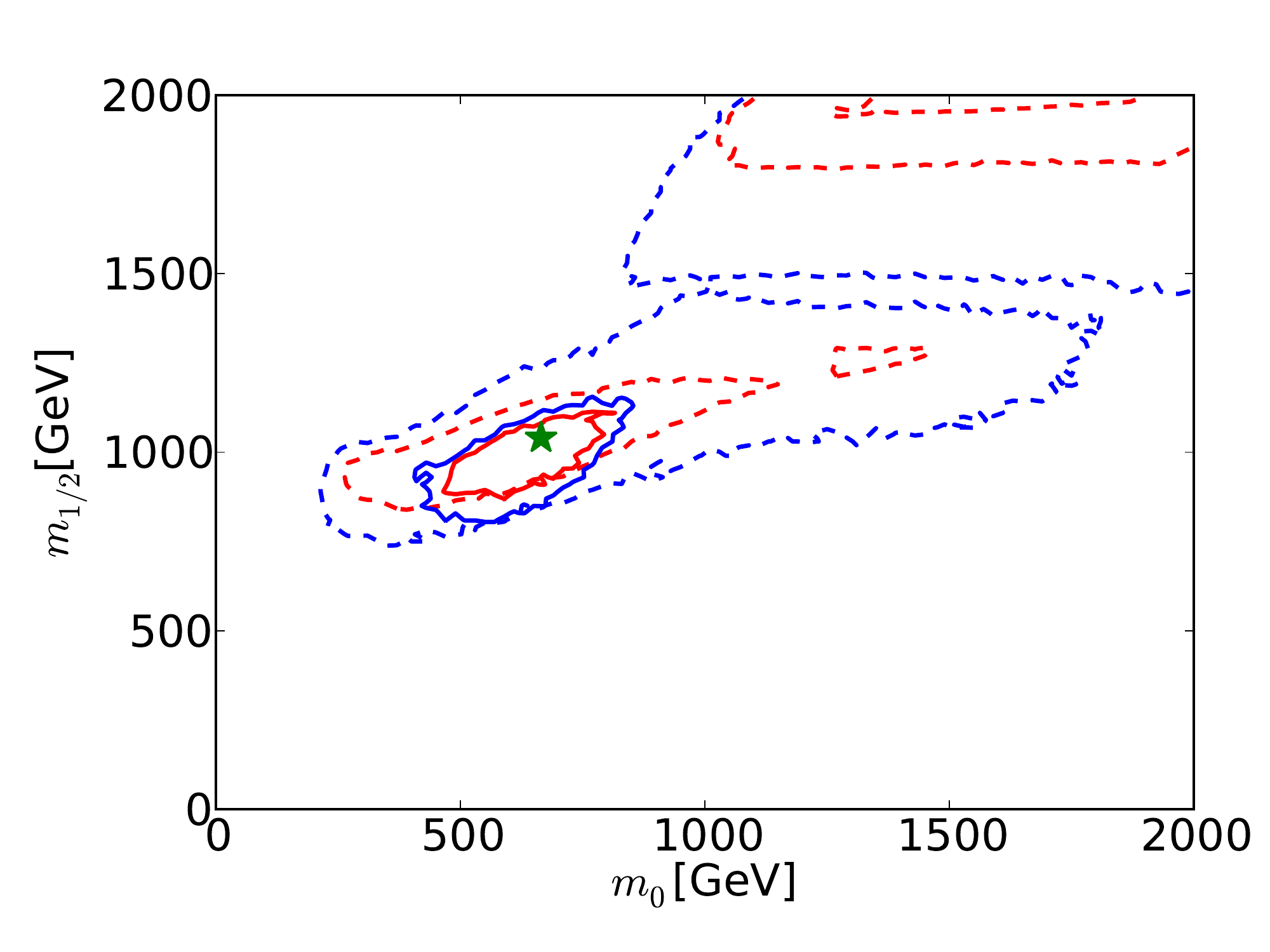}
}
\caption{\label{fig:lowmassprecision}\it
The prospective $\Delta \chi^2 = 2.30$ (68\% CL) and $\Delta \chi^2 = 5.99$ (95\% CL)
contours (solid red and blue lines, respectively) in the $(m_0, m_{1/2})$ plane for
the CMSSM (with the present 68 and 95\% CL contours shown as dashed red and blue lines, respectively), assuming that the electroweak precision
observables are measured at FCC-ee (TLEP) to have the same central values as at the
current low-mass CMSSM best-fit point~\protect\cite{mc9} (shown as the filled green star), and neglecting inevitable improvements in
other constraints on the supersymmetric models.}
\end{figure}

We have made a crude estimate of the impact on the recent global fit to the CMSSM
parameters of these FCC-ee (TLEP) electroweak measurements, neglecting the inevitable
improvements in flavour, dark matter and Higgs observables, and setting aside
the direct measurements of sparticle masses possible at the LHC following
discovery in this optimistic scenario. As we see in 
Fig.~\ref{fig:lowmassprecision}, the electroweak precision measurements would,
by themselves, provide very tight constraints on the CMSSM parameters $m_0$
and $m_{1/2}$.

\begin{figure}[hbt]
\begin{center}
\begin{tabular}{c c}
\includegraphics[height=5.5cm]{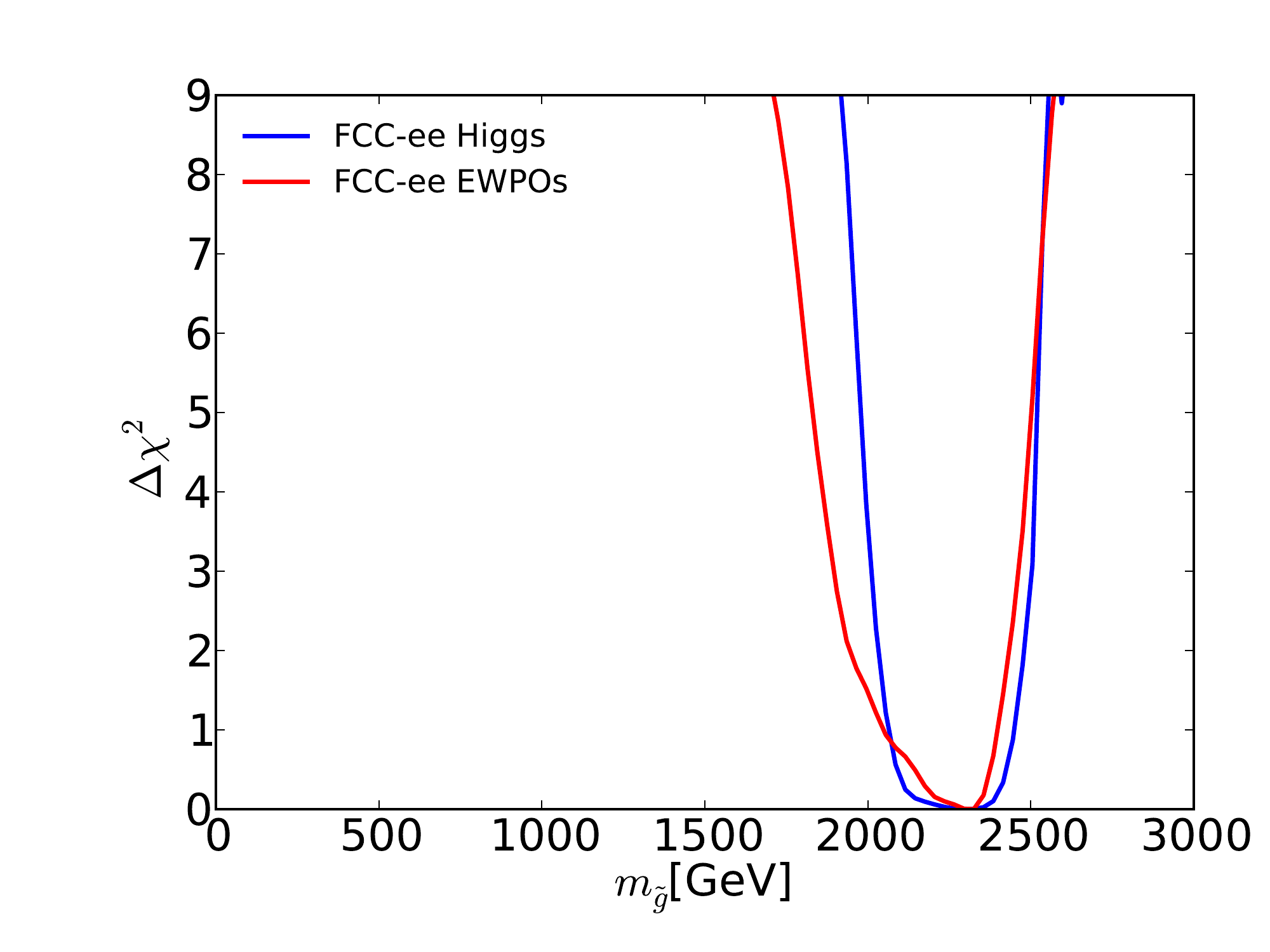} &
\includegraphics[height=5.5cm]{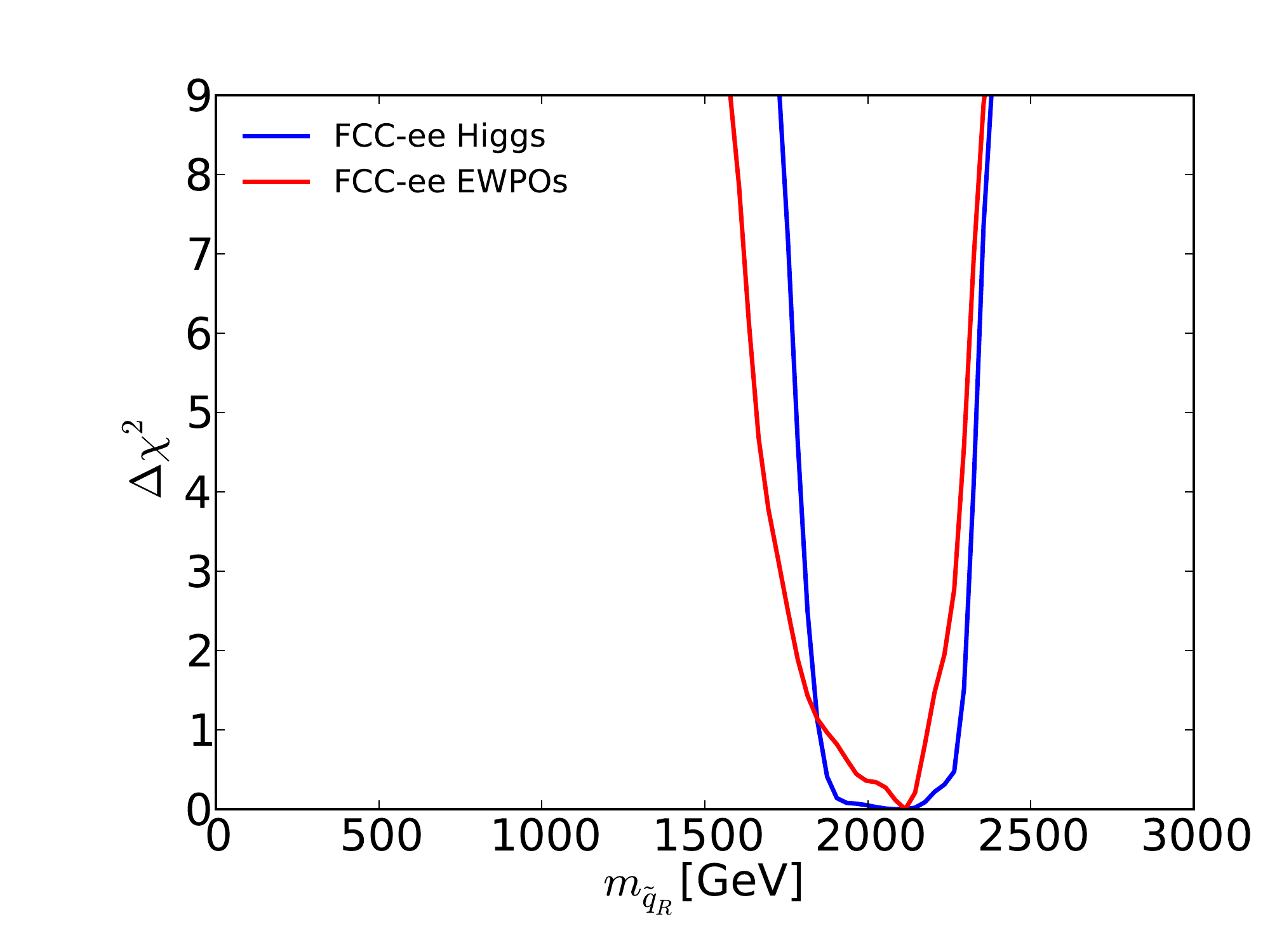} \\
\includegraphics[height=5.5cm]{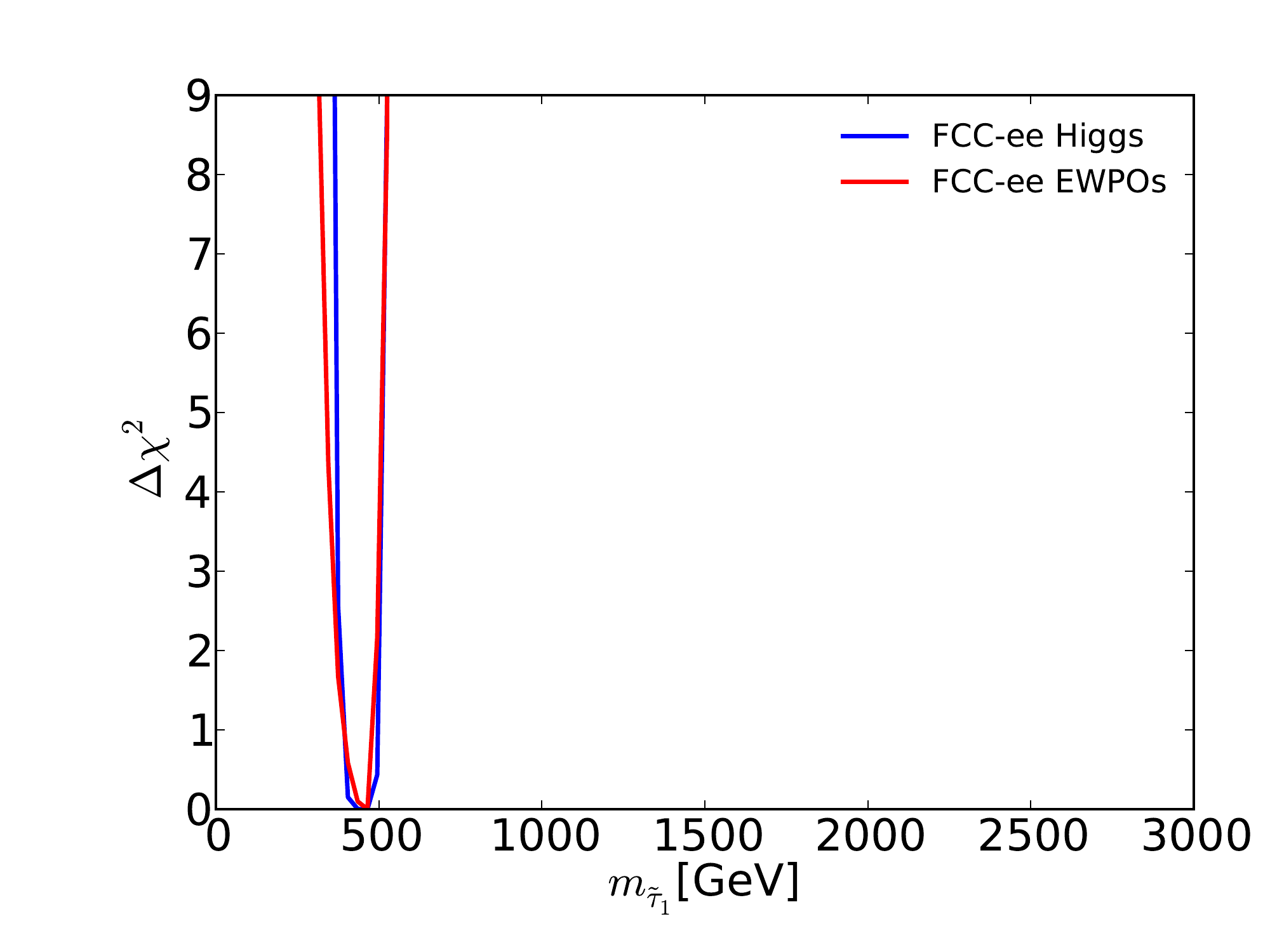} &
\includegraphics[height=5.5cm]{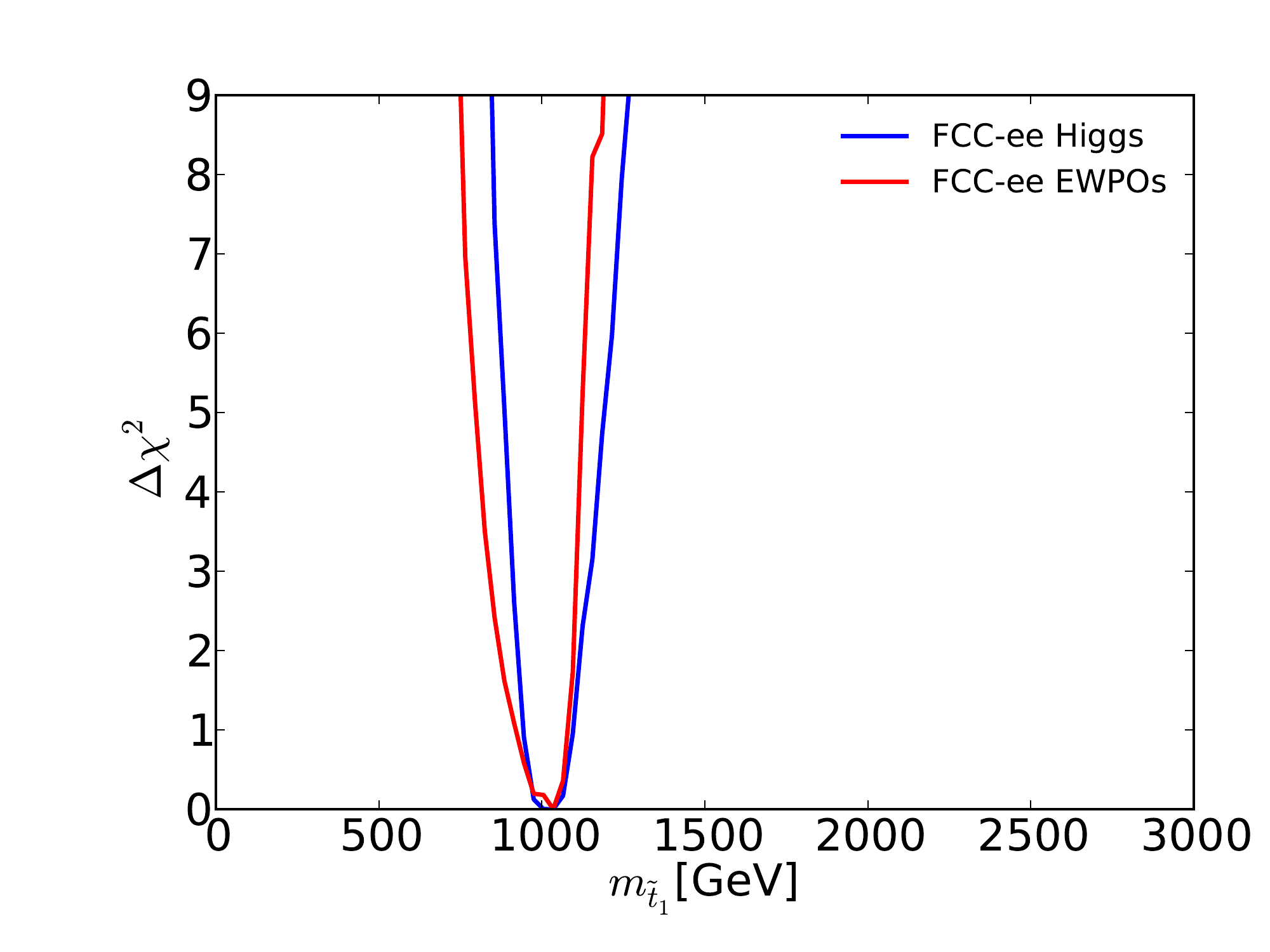} \\
\end{tabular}
\end{center}   
\vspace{-1cm}
\caption{\label{fig:mglmsq}\it
The one-dimensional $\Delta \chi^2$ profile likelihood functions for the gluino mass $m_{\tilde g}$
(upper left panel), the generic first- and second-generation squark mass $m_{\tilde q}$ (upper right panel),
the lighter stop squark mass (lower left panel) and the lighter stau mass (lower right panel),
as obtained using prospective FCC-ee (TLEP)
precision electroweak measurements (solid red lines) and Higgs measurements (solid blue lines)~\protect\cite{TLEP}
with the same central values as the low-mass best-fit CMSSM point~\protect\cite{mc9},
neglecting the inevitable improvements in
other constraints on the supersymmetric models.}
\end{figure}

After inclusion of the FCC-ee (TLEP) measurements~\cite{TLEP}
in this optimistic scenario, only a small part of the low-mass `Crimea' region is 
allowed at the 68 or 95\% CL, as seen in Fig.~\ref{fig:lowmassprecision}. The impact
of the FCC-ee (TLEP) measurements may be translated into the one-dimensional
likelihood functions for various sparticle masses, shown as solid red lines in Fig.~\ref{fig:mglmsq}~\footnote{As
discussed later, the solid blue lines are the corresponding likelihood functions provided by
the prospective Higgs measurements at FCC-ee (TLEP).}.
We see that $m_{\tilde g}$, $m_{\tilde q}$, $m_{\tilde \tau_1}$ and $m_{\tilde t_1}$ could be estimated with interesting accuracy
on the basis of FCC-ee (TLEP):
\begin{eqnarray}
m_{\tilde g}~ & \in & (1680, 2480) \; {\rm GeV} \, , \nonumber \\
m_{\tilde q}~ & \in & (1680, 2280) \; {\rm GeV} \, , \nonumber \\
m_{\tilde \tau_1} & \in & ~~(340, 500) \; {\rm GeV} \, , \nonumber \\
m_{\tilde t_1} & \in & ~(810, 1110) \; {\rm GeV} \, ,
\label{EWPOmasses}
\end{eqnarray}
whereas the nominal values at the best-fit point are 2280, 2080, 450 and 1020~GeV, respectively.
Since, in this optimistic scenario, squarks and gluinos would have been discovered previously
at the LHC, measurements of their masses could be compared with the estimates
based on the FCC-ee (TLEP) measurements. Agreement would constitute a non-trivial
test of the CMSSM at the loop level, analogous to the tests of the Standard Model made
possible by measurements of $m_t$ and $m_H$ and their consistency with predictions
based on LEP and SLC data~\cite{LEPEWWG}. Conversely, any disagreement could be interpreted as
a possible deviation from the CMSSM assumptions of universality for the soft
supersymmetry-breaking parameters.

\subsection{Precision Higgs Observables}

\begin{figure}
\vspace{-2cm}
\begin{center}
\begin{tabular}{c c}
\includegraphics[height=6cm]{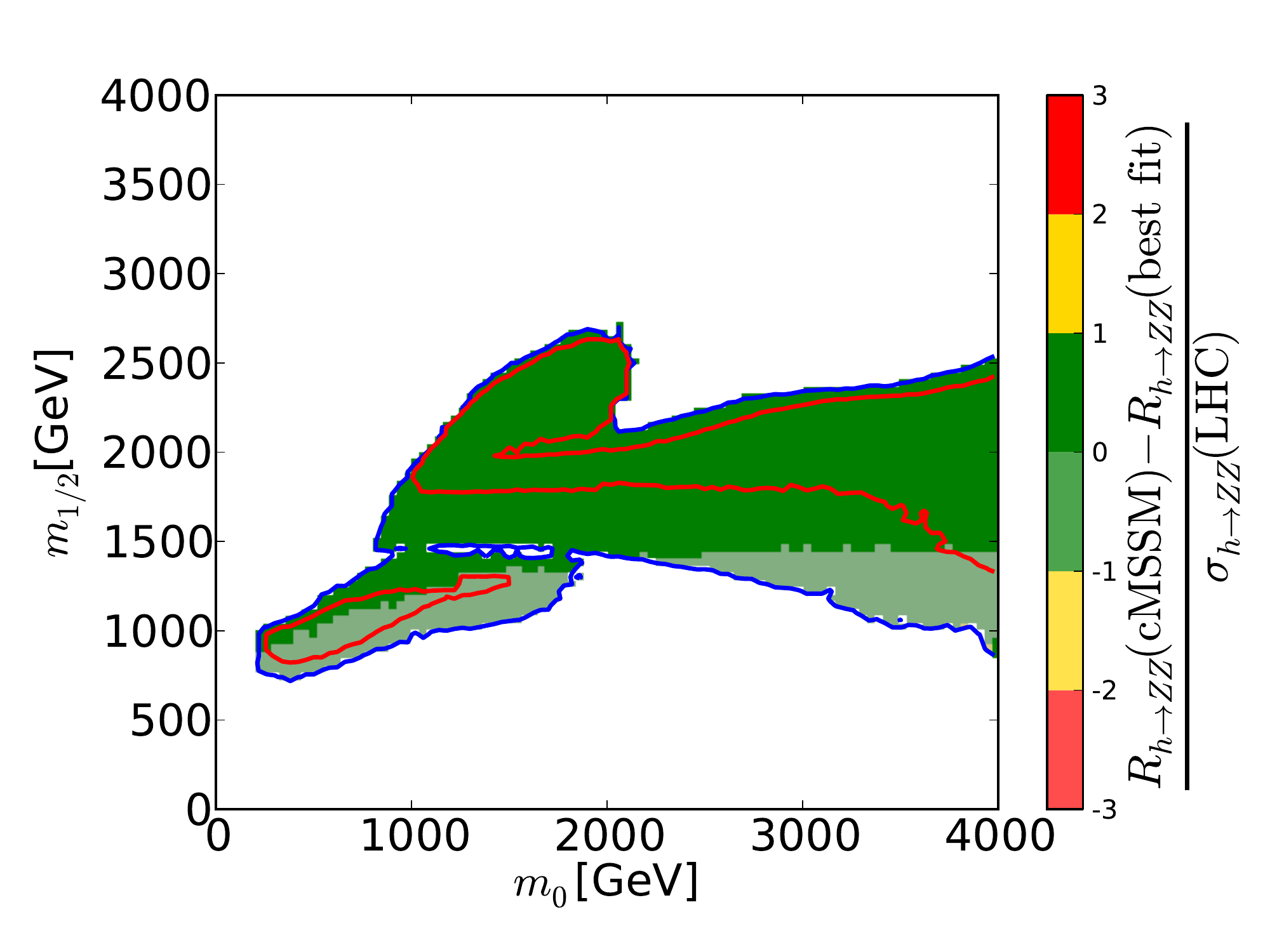} & 
\includegraphics[height=6cm]{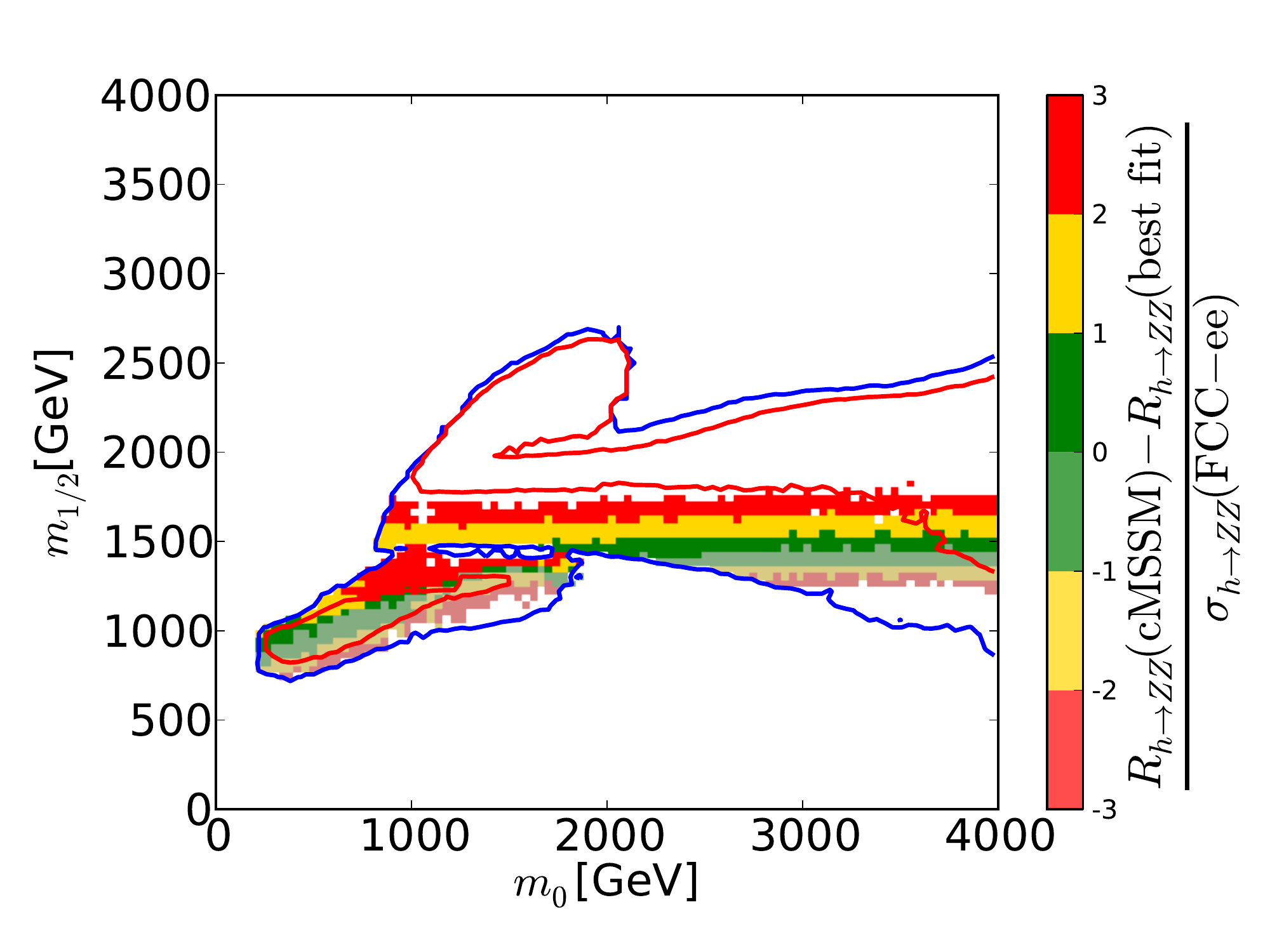} \\
\end{tabular}
\end{center}   
\begin{center}
\begin{tabular}{c c}
\includegraphics[height=6cm]{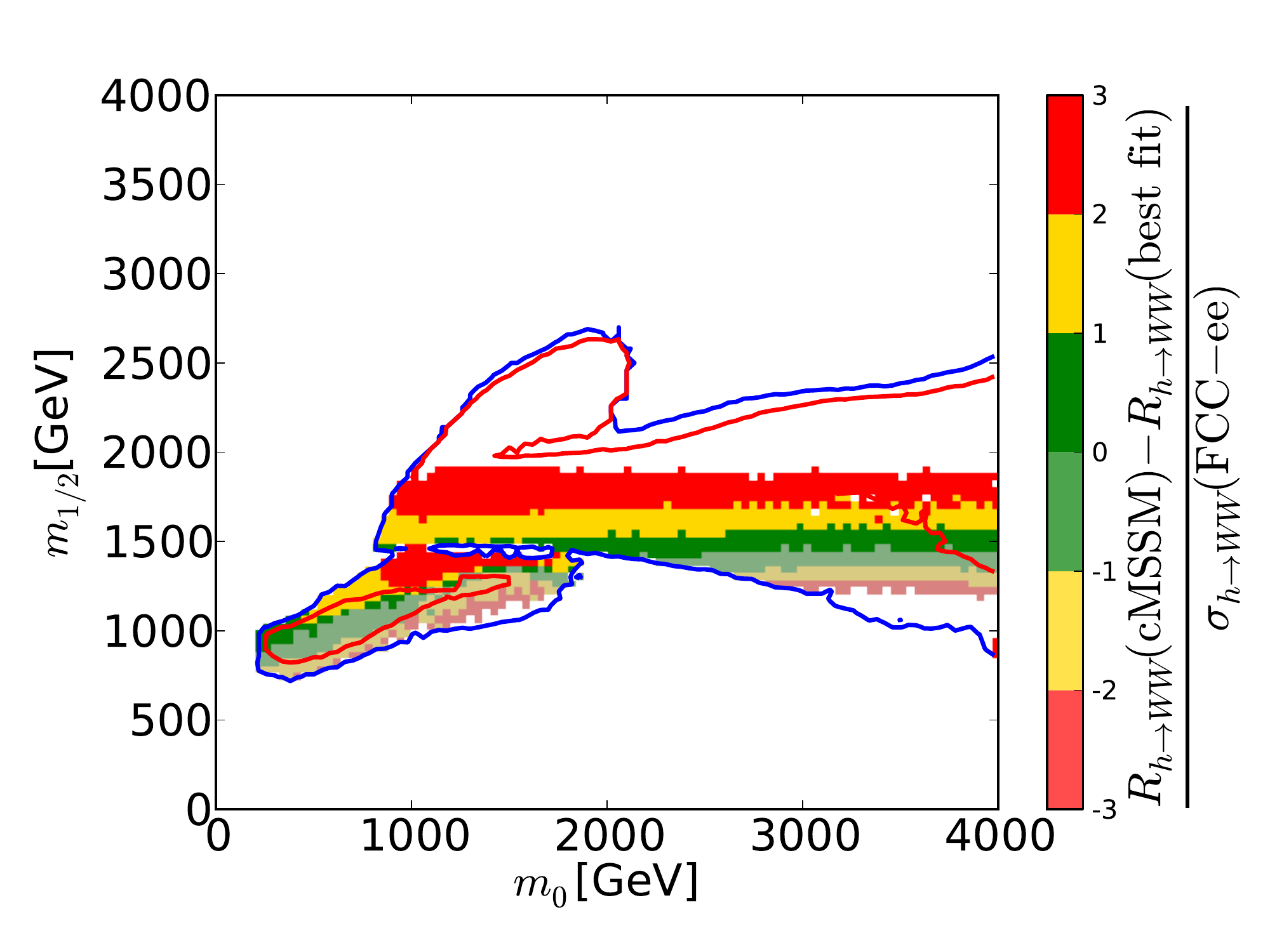} & 
\includegraphics[height=6cm]{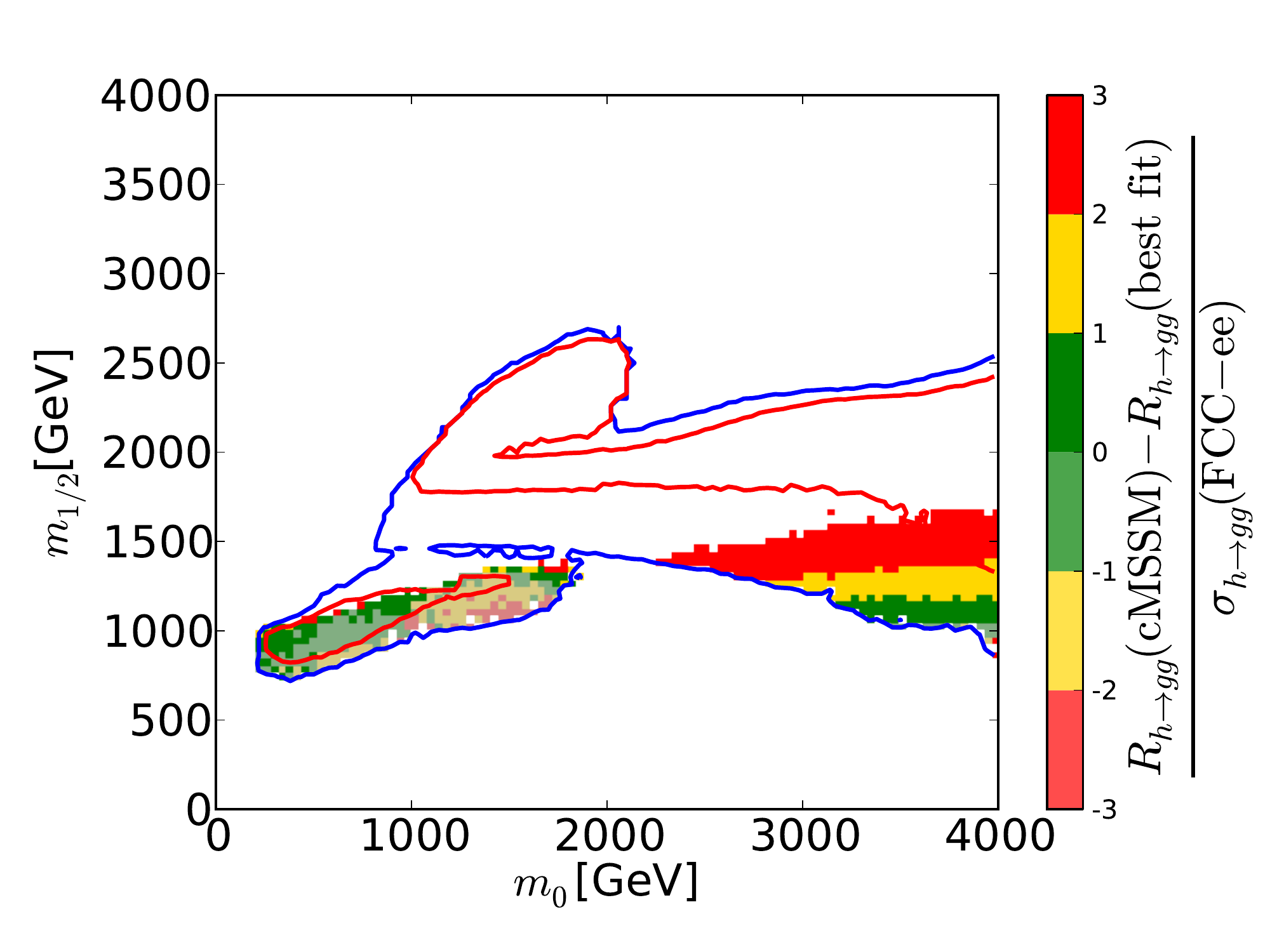} \\
\end{tabular}
\end{center}   
\caption{\label{fig:PrecisionH}\it
The present measurement of BR($H \to ZZ$) (upper left panel), and prospective
FCC-ee (TLEP) measurements~\protect\cite{TLEP} of BR($H \to ZZ$) (upper right), BR($H \to \gamma \gamma$) (lower
left) and BR($H \to gg$) (lower right) are superposed on the $(m_0, m_{1/2})$ plane
in the CMSSM shown previously in Fig.~\protect\ref{fig:Kees}.
The colours represent deviations from the present central value in units
of the present LHC experimental error (upper left panel), and the deviations from the
values at the low-mass best-fit CMSSM point~\protect\cite{mc9} of the values at other points in the $(m_0, m_{1/2})$ plane
in units of the estimated future FCC-ee (TLEP) experimental errors~\protect\cite{TLEP} (other panels).
}
\end{figure}

We have made a similar estimate of the potential impact of the high-precision
Higgs measurements possible with FCC-ee (TLEP)~\cite{TLEP}, as illustrated in the right panel of Fig.~\ref{fig:precision}.
In the upper left panel of
Fig.~\ref{fig:PrecisionH} we display the deviation of the present experimental
value of BR($H \to ZZ$) from the values calculated at points within the 68 and 95\% CL regions in the
$(m_0, m_{1/2})$ plane of the CMSSM, in units of the present experimental error.
In the other panels of Fig.~\ref{fig:PrecisionH}
we show the numbers of FCC-ee (TLEP) $\sigma$'s by which the values of BR($H \to ZZ$) (upper right),
BR($H \to WW$) (lower left) and BR($H \to gg$) (lower right) calculated at other points
in the CMSSM $(m_0, m_{1/2})$ plane differ from the values at the low-mass CMSSM best-fit point.

As in the case of the electroweak precision measurements shown in Fig.~\ref{fig:PrecisionZ}, we see
in the upper left panel of Fig.~\ref{fig:PrecisionH} that the entire
68 and 95\% CL regions of the CMSSM $(m_0, m_{1/2})$ plane lie within a single
current LHC $\sigma$ of the present central value of BR($H \to ZZ$), \htr{and hence are shaded green}, and we
have checked that the same is true for the other Higgs branching ratios measured
currently. For this reason, at the moment the Higgs branching ratios do not make important contributions
to the global likelihood function of the CMSSM. However, we see in the other panels of
Fig.~\ref{fig:PrecisionH} that future measurements of the Higgs branching ratios at FCC-ee (TLEP)
would have the potential to discriminate between different CMSSM parameter sets, \htr{so that
much of the 68\% and 95\% CL regions in these panels are unshaded, since they lie
more than three current standard deviations away from the prospective measurements}.
Specifically, several individual measurements at the central values predicted by the low-mass best-fit
point in the CMSSM would each individually exclude regions at large values of $m_0$ and (particularly)
$m_{1/2}$. As in Fig.~\ref{fig:PrecisionZ} for the electroweak precision observables, we see that prospective measurements 
of the observables studied are compatible with the low-mass best-fit values within
one FCC-ee (TLEP) $\sigma$ only within fractions of the `Crimean' 68 \% CL region.
We also see that only narrow bands of the `Eurasian' 95\% CL regions would
yield values of BR($H \to ZZ, \gamma \gamma$) and BR($H \to gg$) within one $\sigma$
of the low-mass best-fit prediction, and that the band for BR($H \to gg$) does not overlap the others.

\begin{figure}[hbtp!]
\centerline{
\includegraphics[height=8cm]{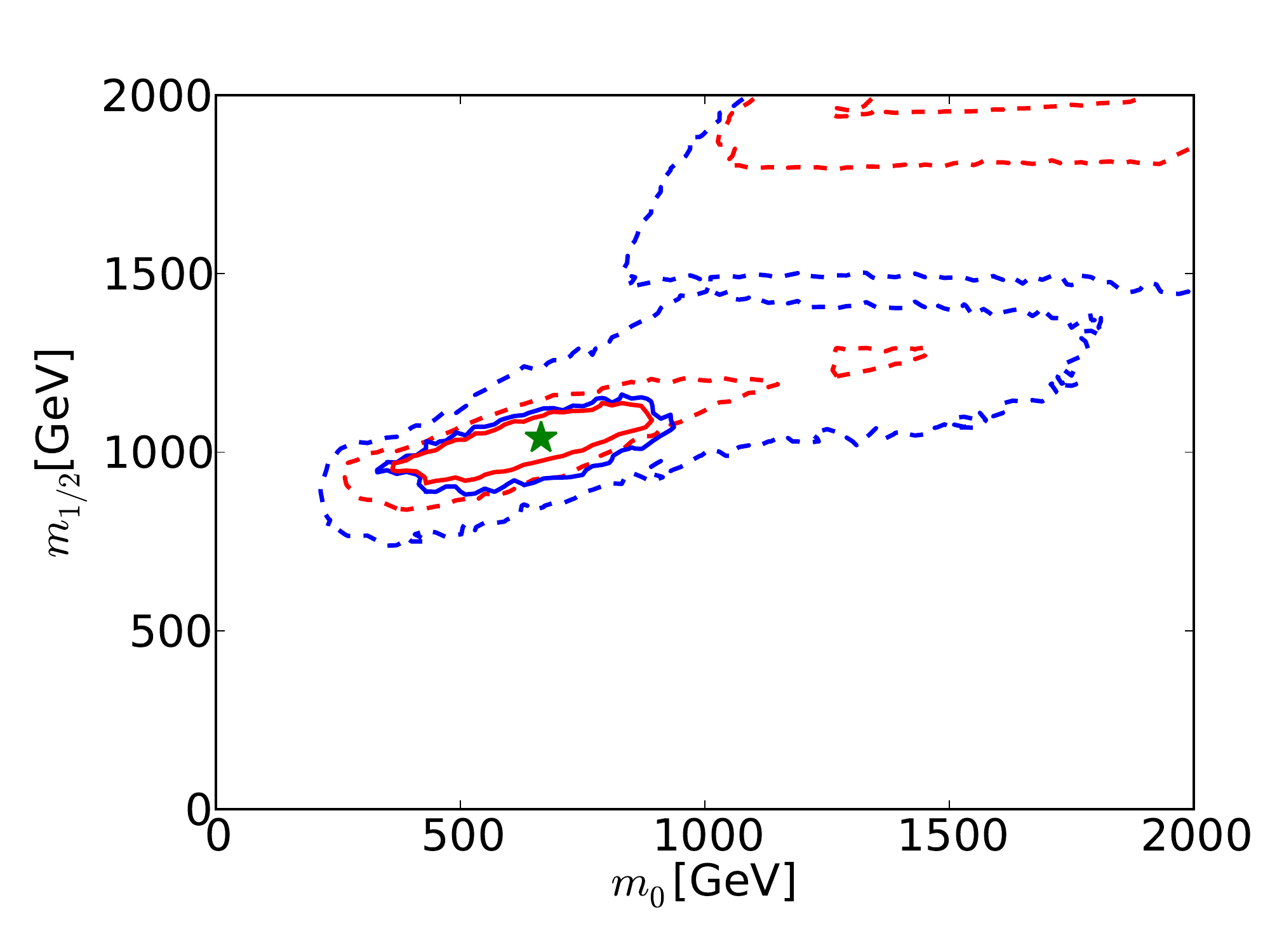}
}
\caption{\label{fig:lowmassprecisionH}\it
The $\Delta \chi^2 = 2.30$ (68\% CL) and $\Delta \chi^2 = 5.99$ (95\% CL)
contours (red and blue, respectively) in the $(m_0, m_{1/2})$ plane for
the CMSSM, assuming that Higgs measurements
at FCC-ee (TLEP)~\protect\cite{TLEP} have the same central values as at the
current low-mass best-fit points, and neglecting inevitable improvements in
other constraints on the supersymmetric models.}
\end{figure}

Also as in the case of the electroweak precision observables discussed above,
we have made a crude estimate of the impact of the prospective FCC-ee (TLEP) Higgs measurements~\cite{TLEP}
on the global $\chi^2$ function for the CMSSM, again neglecting the inevitable
improvements in flavour and dark matter observables, and setting aside the electroweak precision observables
as well as the direct measurements of sparticle masses. As we see in 
Fig.~\ref{fig:lowmassprecisionH}, the high-precision Higgs measurements would,
by themselves, again provide constraints on $m_0$ and $m_{1/2}$, which would
be of comparable importance to those from the electroweak precision observables shown in the corresponding panel of
Fig.~\ref{fig:lowmassprecision}. 

We display as solid blue lines in Fig.~\ref{fig:mglmsq}
the corresponding one-dimensional projections of the contribution of prospective FCC-ee (TLEP) Higgs
measurements to the CMSSM global $\chi^2$ function for $m_{\tilde g}$
(upper left panel), the generic squark mass $m_{\tilde q}$ (upper right panel),
the lighter stop squark mass $m_{\tilde t_1}$ (lower left panel) and the lighter
stau mass $m_{\tilde \tau_1}$ (lower right panel). We see that the Higgs estimates
are comparable with the corresponding
one-dimensional projections of the contribution of prospective FCC-ee (TLEP) electroweak precision
measurements, shown as solid lines in the various panels of Fig.~\ref{fig:mglmsq}. The corresponding 95\% CL
mass ranges are estimated to be
\begin{eqnarray}
m_{\tilde g}~ & \in & (1980, 2500) \; {\rm GeV} \, , \nonumber \\
m_{\tilde q}~ & \in & (1780, 2320) \; {\rm GeV} \, , \nonumber \\
m_{\tilde \tau_1} & \in & ~~(370, 510) \; {\rm GeV} \, , \nonumber \\
m_{\tilde t_1} & \in & ~(890, 1170) \; {\rm GeV} \, ,
\label{Higgsmasses}
\end{eqnarray}
whereas the nominal values at the best-fit point are 2280, 2080, 450 and 1020~GeV, respectively.
The Higgs measurements clearly add another
dimension to the tests of supersymmetric models at the loop level. 

\subsection{Comparison between LHC and $e^+ e^-$ Measurements}

\begin{figure}[ht!]
\centering
\includegraphics[height=5.5cm]{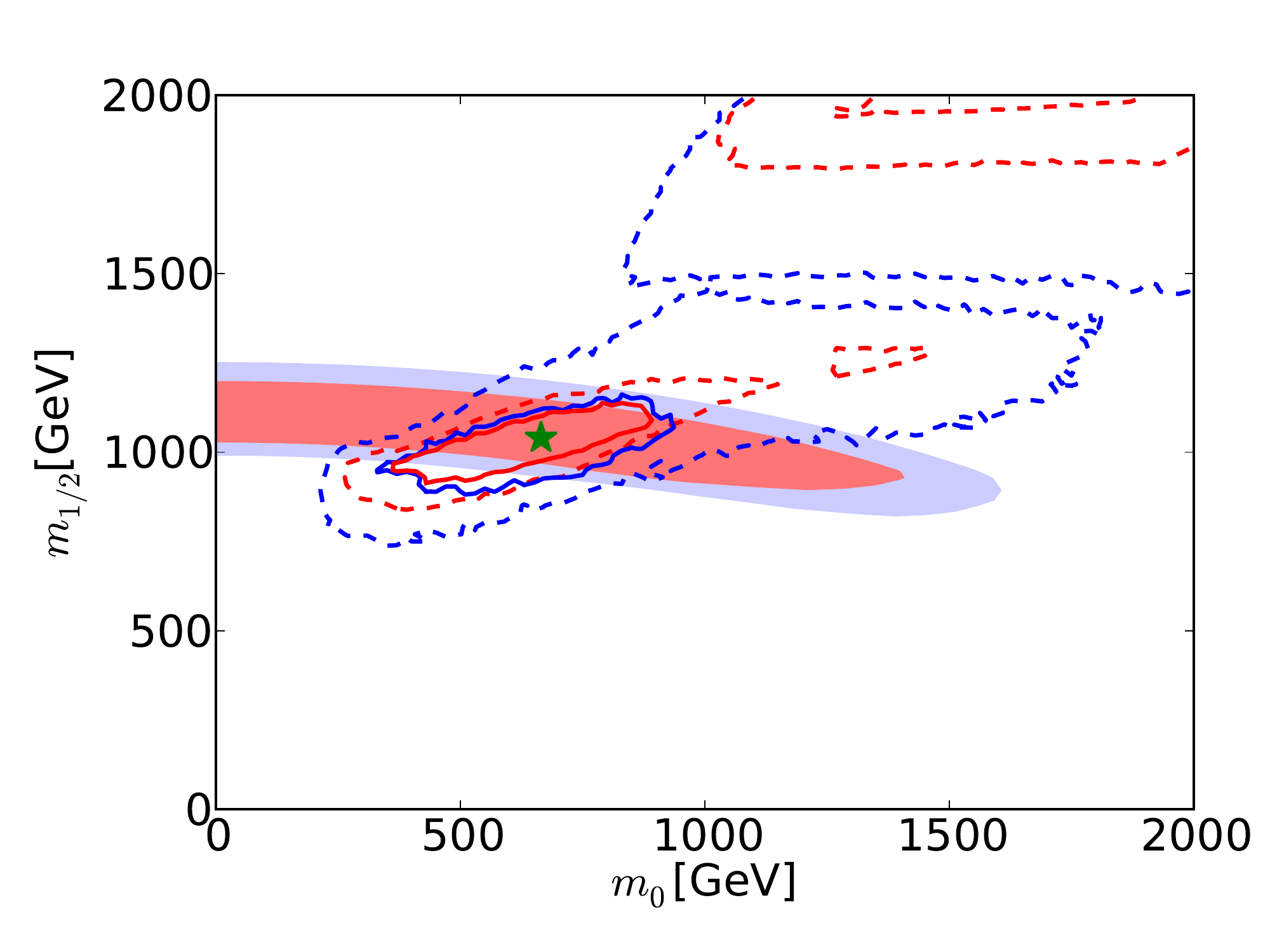}
\includegraphics[height=5.5cm]{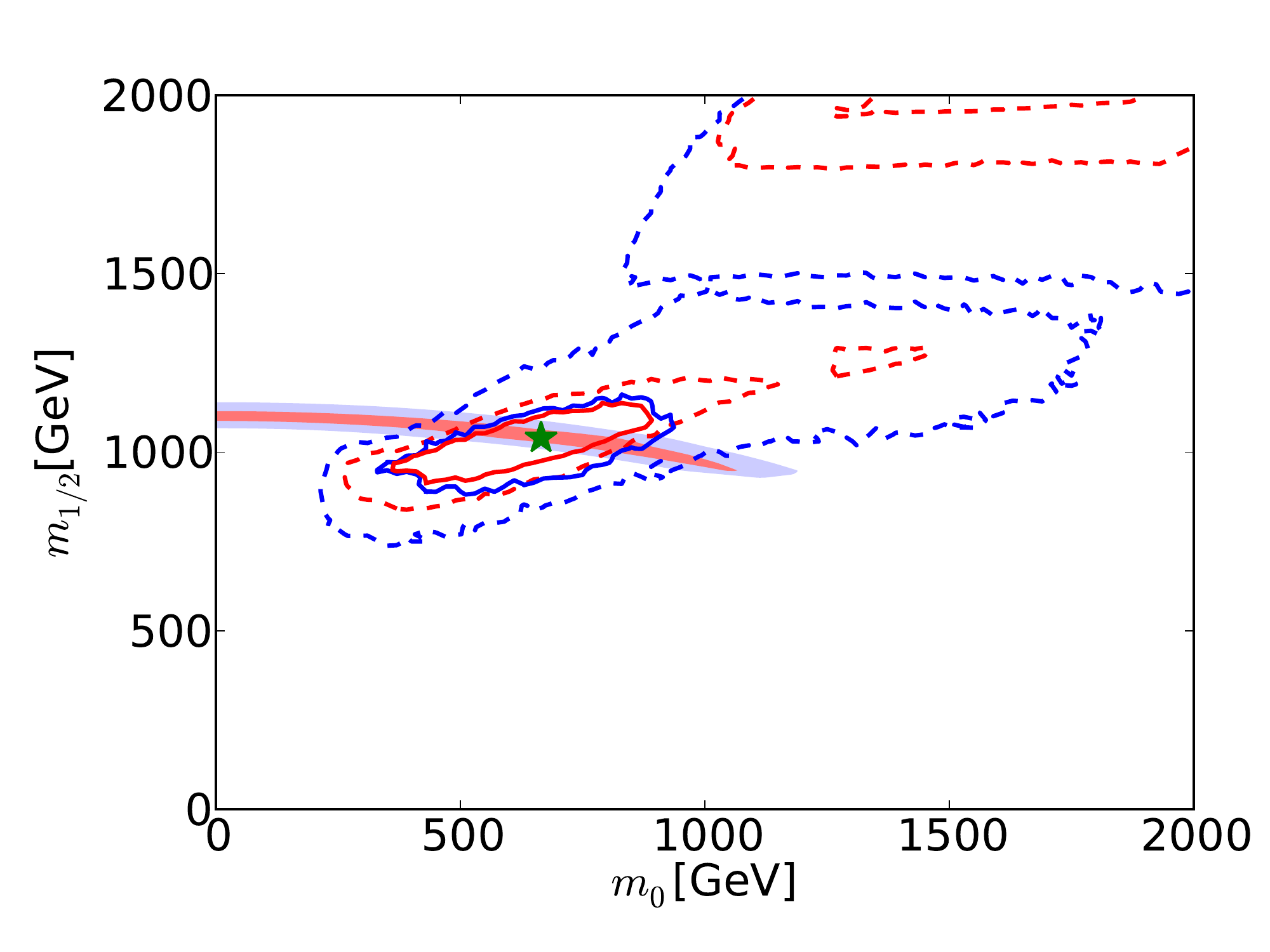} \\
\includegraphics[height=5.5cm]{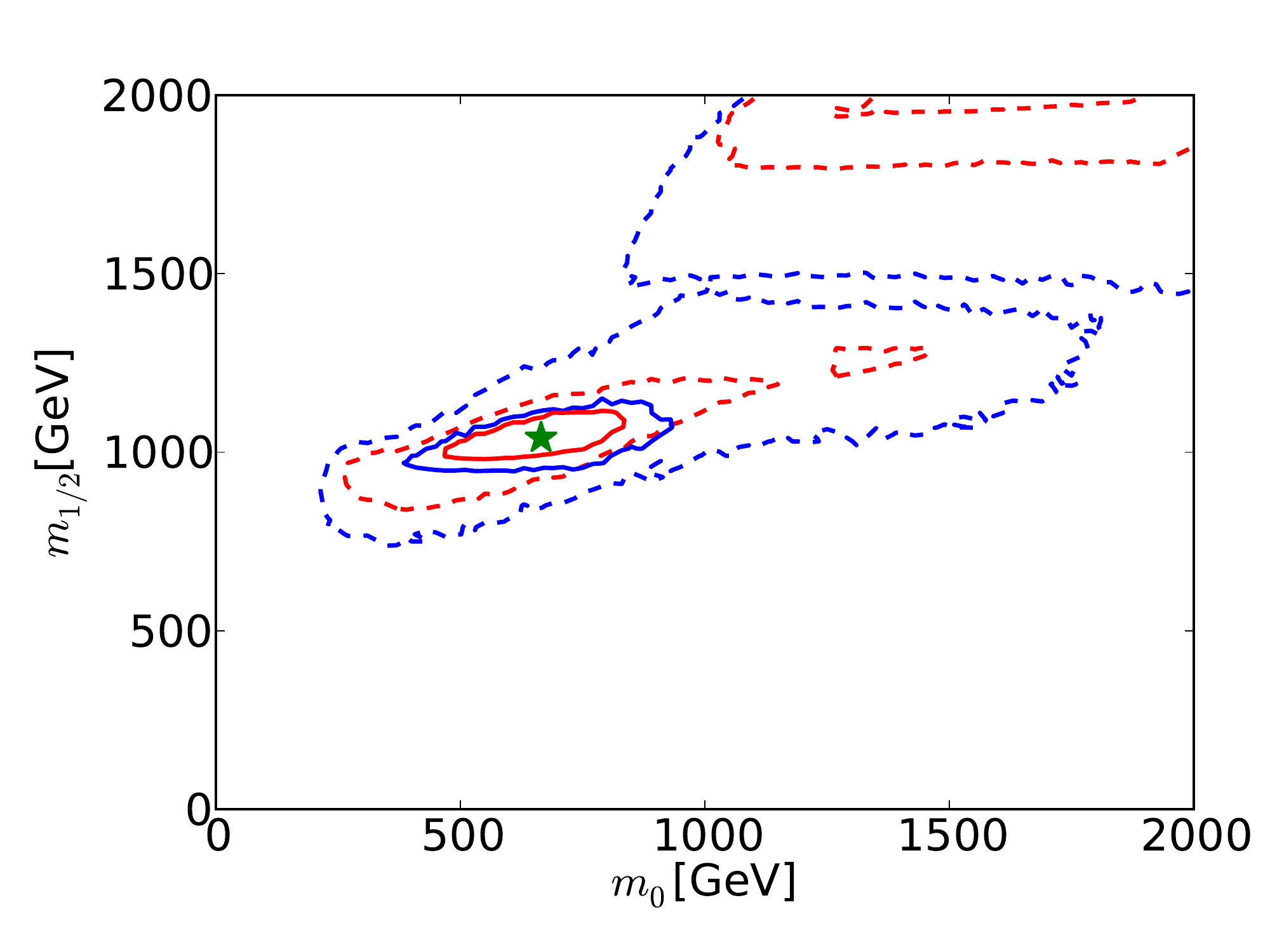}
\includegraphics[height=5.5cm]{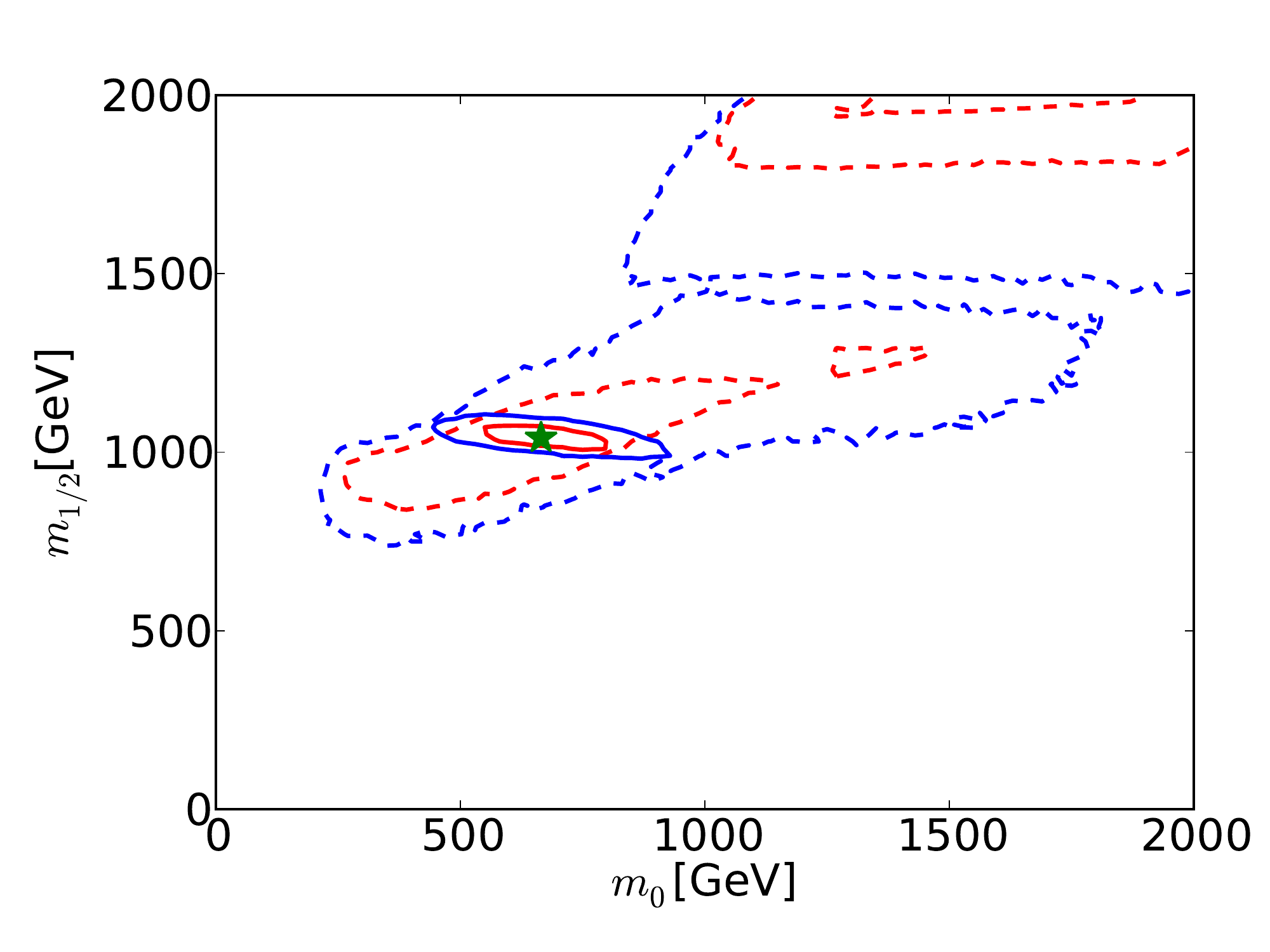} \\
\caption{\label{fig:LHCoverlay}\it
Upper panels: The 68 and 95\% CL regions in the $(m_0, m_{1/2})$ plane of the CMSSM,
overlaying potential direct measurements
at LHC14 with 300/fb (left panel) and 3000/fb (right panel) (pink and blue shading) with indirect
determinations via electroweak precision and Higgs measurements at FCC-ee (TLEP)~\protect\cite{TLEP}
(red and blue solid lines). Lower panels: The 68\% and 95\% CL contours for the combination
of the prospective constraints from LHC14 with 300/fb (left panel) and 3000/fb (right panel) with the indirect
determinations via electroweak precision and Higgs measurements at FCC-ee (TLEP)~\protect\cite{TLEP}
(red and blue solid lines). Also shown are the 68 and 95\% CL regions found in a recent global
fit to the CMSSM (red and blue dashed lines)~\protect\cite{mc9}.}
\end{figure}

The potential comparison between LHC and FCC-ee (TLEP)
measurements in the best-fit low-mass CMSSM scenario can be seen in Fig.~\ref{fig:LHCoverlay},
where we overlay in the CMSSM $(m_0, m_{1/2})$ plane
the potential direct measurements at the LHC \htr{presented earlier (pink and blue shading)}
with indirect determinations at
FCC-ee (TLEP) via EWPOs and Higgs measurements. A triple
coincidence of direct sparticle mass measurements with indirect predictions
from EWPOs and Higgs measurements would be truly impressive, a worthy
successor to the successful predictions of the top and Higgs masses based
on electroweak precision observables at LEP.

\subsection{Probes of Grand Unification}

The precision measurements of electroweak precision observables and the strong coupling $\alpha_s(m_Z)$
at LEP and the SLC opened a new chapter in probes of models of grand unification~\cite{GUTs},
making possible for the first time a clear discrimination between the predictions of
supersymmetric and non-supersymmetric scenarios. As remarked in~\cite{TLEP},
it is clear that FCC-ee (TLEP) measurements could take this confrontation between
experiment and different grand unified theories to a completely new level,
through more accurate determinations of the SU(3), SU(2) and U(1) couplings
by specifying the supersymmetric spectrum and hence TeV-scale threshold
corrections to the running of the couplings. The combination of these measurements
would enable powerful constraints to be placed on the GUT-scale particles in any
specific GUT model.

As an indication of this possibility, we consider the simplest supersymmetric SU(5) GUT,
in which the GUT-scale particles comprise the heavy vector bosons $V$, the 24-plet
Higgs bosons $\Sigma$ and the coloured Higgs triplet bosons $H_c$. By considering
the three one-loop renormalization-group equations (RGEs) for the SU(3), SU(2) and U(1) couplings
in this model, Murayama and Pierce~\cite{MP}~\footnote{For a recent analysis, see~\cite{EW}.}
derived the following one-loop relation between
the low-energy values of the couplings, $\alpha_i(m_Z)$, the supersymmetric threshold
and the mass of the Higgs triplet:
\begin{equation}
 - \frac{2}{\alpha_3(m_Z)} + \frac{3}{\alpha_2(m_Z)} - \frac{1}{\alpha_1(m_Z)} \; = \; 
\frac{1}{2 \pi} \left[ \frac{12}{5} \ln \left( \frac{m_{H_c}}{m_Z}\right) - 2 \ln \left( \frac{m_{SUSY}}{m_Z} \right) \right] \, .
\label{mHc}
\end{equation}
Another combination of the one-loop RGEs gives a similar relation for a combination of
the masses of the heavy vector bosons $V$ and the 24-plet Higgs bosons $\Sigma$:
\begin{equation}
\frac{2}{\alpha_3(m_Z)} + \frac{3}{\alpha_2(m_Z)} - \frac{5}{\alpha_1(m_Z)} \; = \; 
\frac{1}{2 \pi} \left[ - 12 \ln \left( \frac{m^2_{V} m_\Sigma}{m^3_Z}\right) - 8 \ln \left( \frac{m_{SUSY}}{m_Z} \right) \right] \, .
\label{mVSigma}
\end{equation}
These relations are subject to corrections from higher-order terms in the RGEs, etc., but
may be used to estimate the uncertainties in the GUT-scale masses associated with
uncertainties in the low-scale inputs.

For example, inverting (\ref{mHc}) we find
\begin{equation}
\frac{\Delta m_{H_c}}{m_{H_c}} \; \ni \; \frac{5}{6} \frac{\Delta m_{SUSY}}{m_{SUSY}}
-  \frac{5 \pi}{3} \Delta \left( \frac{1}{\alpha_3 (m_Z)} \right)+ \frac{5 \pi}{2} \Delta \left( \frac{1}{\alpha_2 (m_Z)} \right)
-  \frac{5 \pi}{6} \Delta \left( \frac{1}{\alpha_1 (m_Z)} \right) \, ,
\label{DeltamHc}
\end{equation}
and inverting (\ref{mVSigma}) we find
\begin{equation}
2 \frac{\Delta m_V}{m_V} + \frac{\Delta m_\Sigma}{m_\Sigma} \; \ni \; - \frac{2}{3} \frac{\Delta m_{SUSY}}{m_{SUSY}}
-  \frac{\pi}{2} \Delta \left( \frac{1}{\alpha_3 (m_Z)} \right) - \frac{\pi}{3} \Delta \left( \frac{1}{\alpha_2 (m_Z)} \right)
+  \frac{5 \pi}{6} \Delta \left( \frac{1}{\alpha_1(m_Z)} \right) \, .
\label{DeltamVSigma}
\end{equation}
It is estimated that at FCC-ee one could attain uncertainties $\Delta \alpha_3 (m_Z) \sim 10^{-4}$
(corresponding to $\Delta \alpha_3^{-1} (m_Z) \sim 10^{-2}$) 
and $\Delta \sin^2 \theta_W \simeq 10^{-6}$, with an input parametric uncertainty
$\Delta \alpha_{em}^{-1} (m_Z) \simeq 5 \times 10^{-5}$. Using
\begin{equation}
\frac{1}{\alpha_2 (m_Z)} \; = \; \frac{\sin^2 \theta_W}{\alpha_{em}(m_Z)} \, ,
\label{deltaalpha2}
\end{equation}
we find that
\begin{equation}
\Delta \left( \frac{1}{\alpha_2 (m_Z)} \right) \; = \; \Delta \sin^2 \theta_W \times \frac{1}{\alpha_{em}(m_Z)}
+ \sin^2 \theta_W \times \Delta \left( \frac{1}{\alpha_{em} (m_Z)} \right) \, ,
\label{deltaalpha2breakdown}
\end{equation}
and infer that the dominant uncertainty in $\alpha_2^{-1} (m_Z)$ is that due to $\Delta \sin^2 \theta_W$,
giving us the estimate
\begin{equation}
\Delta \left( \frac{1}{\alpha_2 (m_Z)} \right) \; \sim \; 10^{-4} \, .
\label{deltaalpha2number}
\end{equation}
Using then the relationship
\begin{equation}
\frac{1}{\alpha_1 (m_Z)} \; = \; \frac{ 3 \cot^2 \theta_W}{5} \times \frac{1}{\alpha_2 (m_Z)} \, ,
\label{inversealpha1}
\end{equation}
it is evident that the dominant uncertainty in $\alpha_1^{-1} (m_Z)$ is correlated with that
in $\alpha_2^{-1} (m_Z)$:
\begin{equation}
\Delta \left( \frac{1}{\alpha_1 (m_Z)} \right) \; \simeq \; \frac{ 3 \cot^2 \theta_W}{5} \times 
\Delta \left( \frac{1}{\alpha_2 (m_Z)} \right) \; \simeq 2 \times 10^{-4} \, .
\label{Deltainversealpha1}
\end{equation}
It is apparent from the estimates (\ref{deltaalpha2number}) and (\ref{Deltainversealpha1})
that the uncertainties due to the electroweak couplings in the GUT mass estimates
(\ref{DeltamHc}) and (\ref{DeltamVSigma}) are much smaller than the uncertainties due to the
strong coupling. 

Specifically, the precision measurements at FCC-ee (TLEP) should enable
the mass of the colour-triplet to be estimated with an accuracy at the percent level, and similarly
for the combination $m_V^2 m_\Sigma$, assuming that $m_{SUSY}$ can be determined with
similar (or better) precision via direct or indirect measurements. Needless to say, this
possibility of constraining GUT-scale masses
would apply within a specific GUT model, and the implications of the FCC-ee (TLEP)
measurements would depend on the model. However, this analysis makes the point that
high-precision measurements with FCC-ee (TLEP) could impose important constraints
on GUT models, taking to the next level the insights provided previously by LEP
measurements~\cite{GUTs}.

\section{Prospects for the Discovery of Supersymmetry in Pessimistic Scenarios}

We now consider the prospects for discovering supersymmetry in `pessimistic'
high-mass CMSSM scenarios in which the HL-LHC does not discover supersymmetry, but provides only 95\% CL
lower limits on model parameters.

\subsection{Impact of LHC Searches}

\begin{figure}[hbtp!]
\centerline{
\includegraphics[height=8cm]{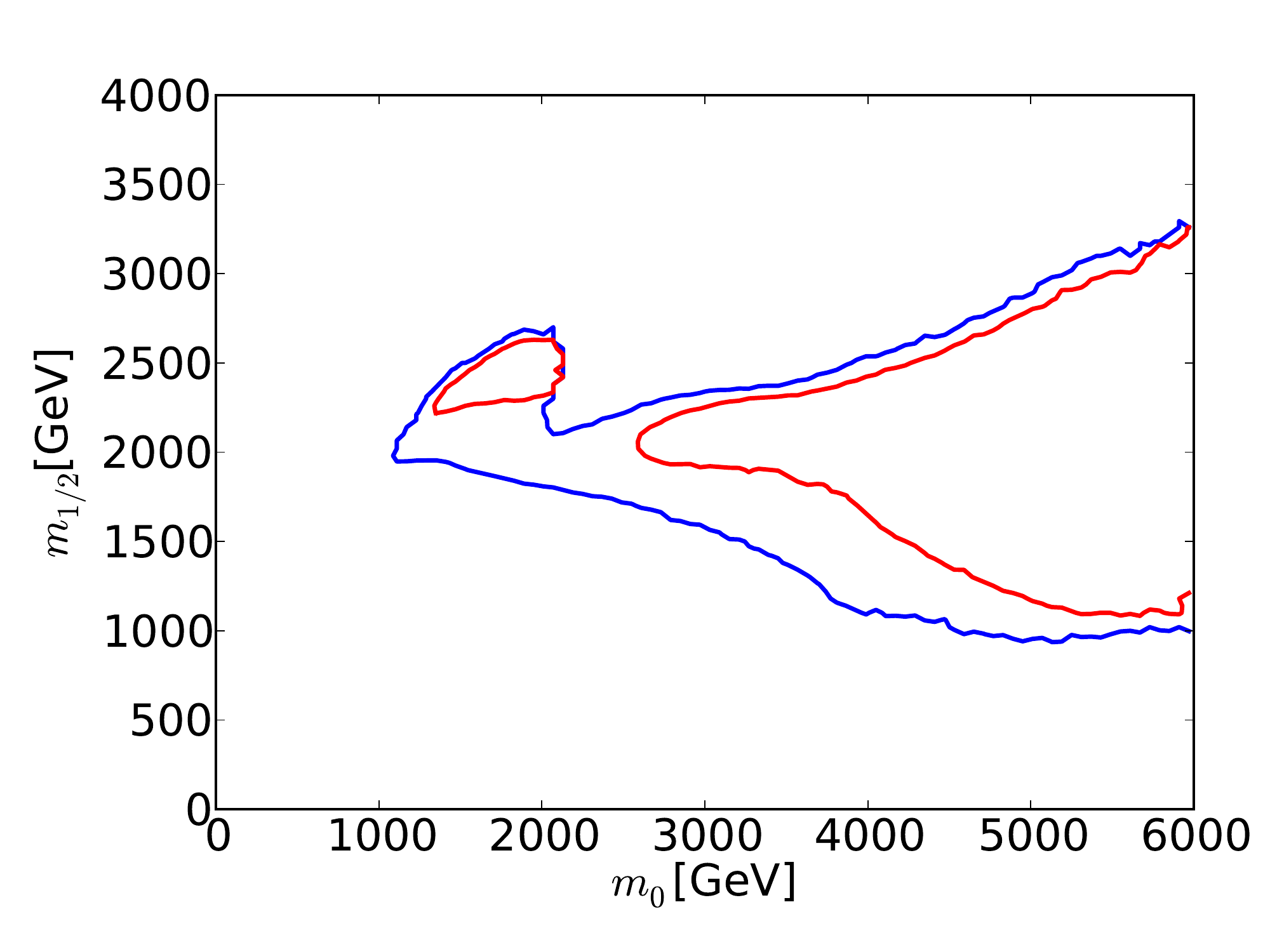}
}
\caption{\label{fig:highmassnoLHC}\it
The $\Delta \chi^2 = 2.30$ (68\% CL) and $\Delta \chi^2 = 5.99$ (95\% CL)
contours (red and blue, respectively) in the $(m_0, m_{1/2})$ plane for
the CMSSM, assuming that supersymmetry has
not been discovered at the LHC with 3000/fb of luminosity, and neglecting 
inevitable improvements in other constraints on the supersymmetric models.}
\end{figure}

To probe this case, we first
make a crude estimate of the impact of such a negative result by including in
the global $\chi^2$ functions for the CMSSM  contributions based on the green lines
in Fig.~\ref{fig:Kees}, which correspond to 95\% CL exclusion by the LHC 
with 3000/fb of data, neglecting again the inevitable improvements in the
measurements of other observables that would provide additional constraints on supersymmetry. The
resulting 68\% and 95\% CL contours (red and blue, respectively) are
shown in Fig.~\ref{fig:highmassnoLHC}. The `Crimea' region has now disappeared
completely, and only the `Eurasia' region remains. However, in the CMSSM, although this
region is unified at the 95\% CL, it is divided at the 68\% CL into regions at lower and
higher values of $m_0$.

\begin{figure}[hbt]
\begin{center}
\begin{tabular}{c c}
\includegraphics[height=6cm]{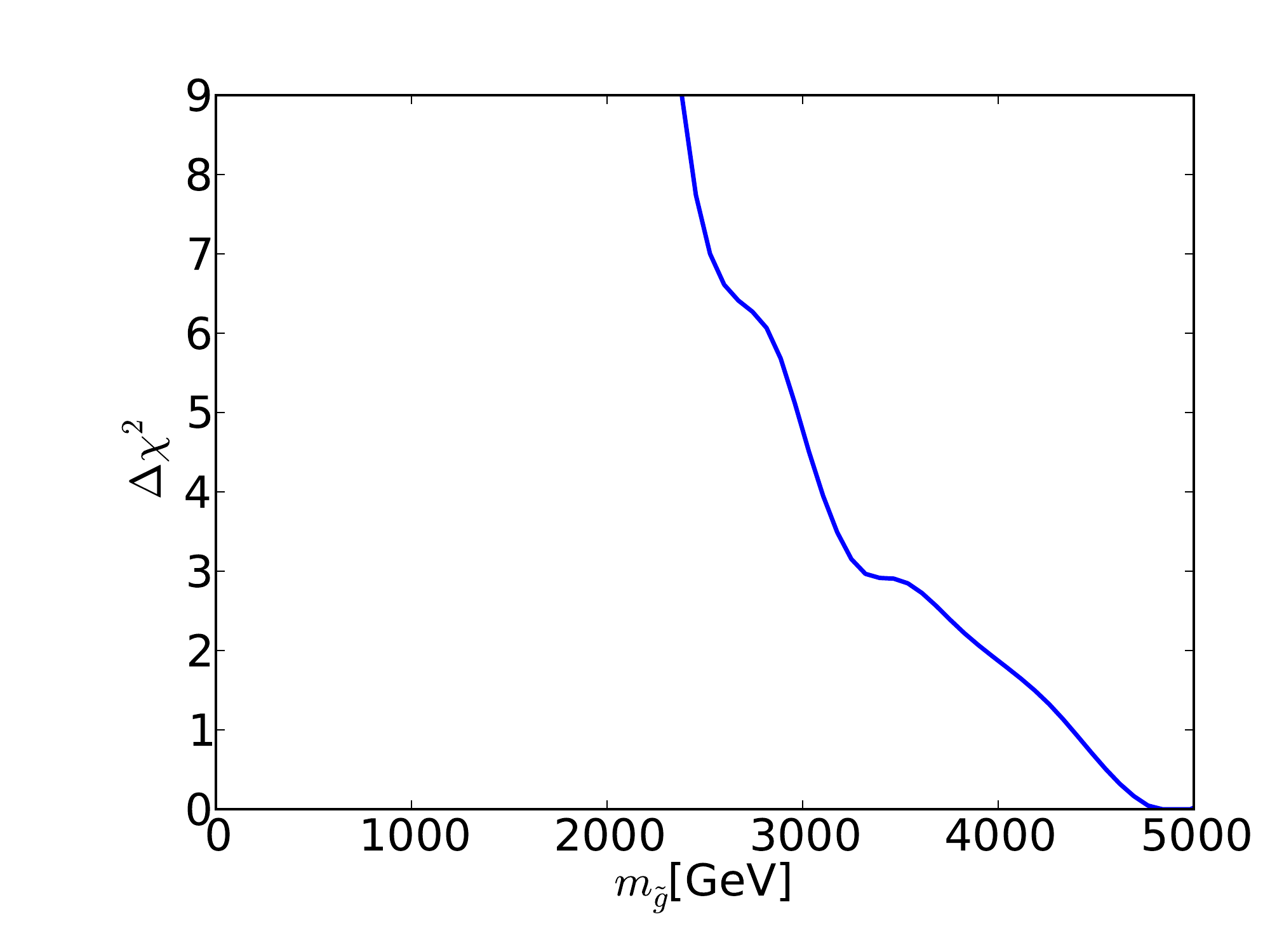} &
\includegraphics[height=6cm]{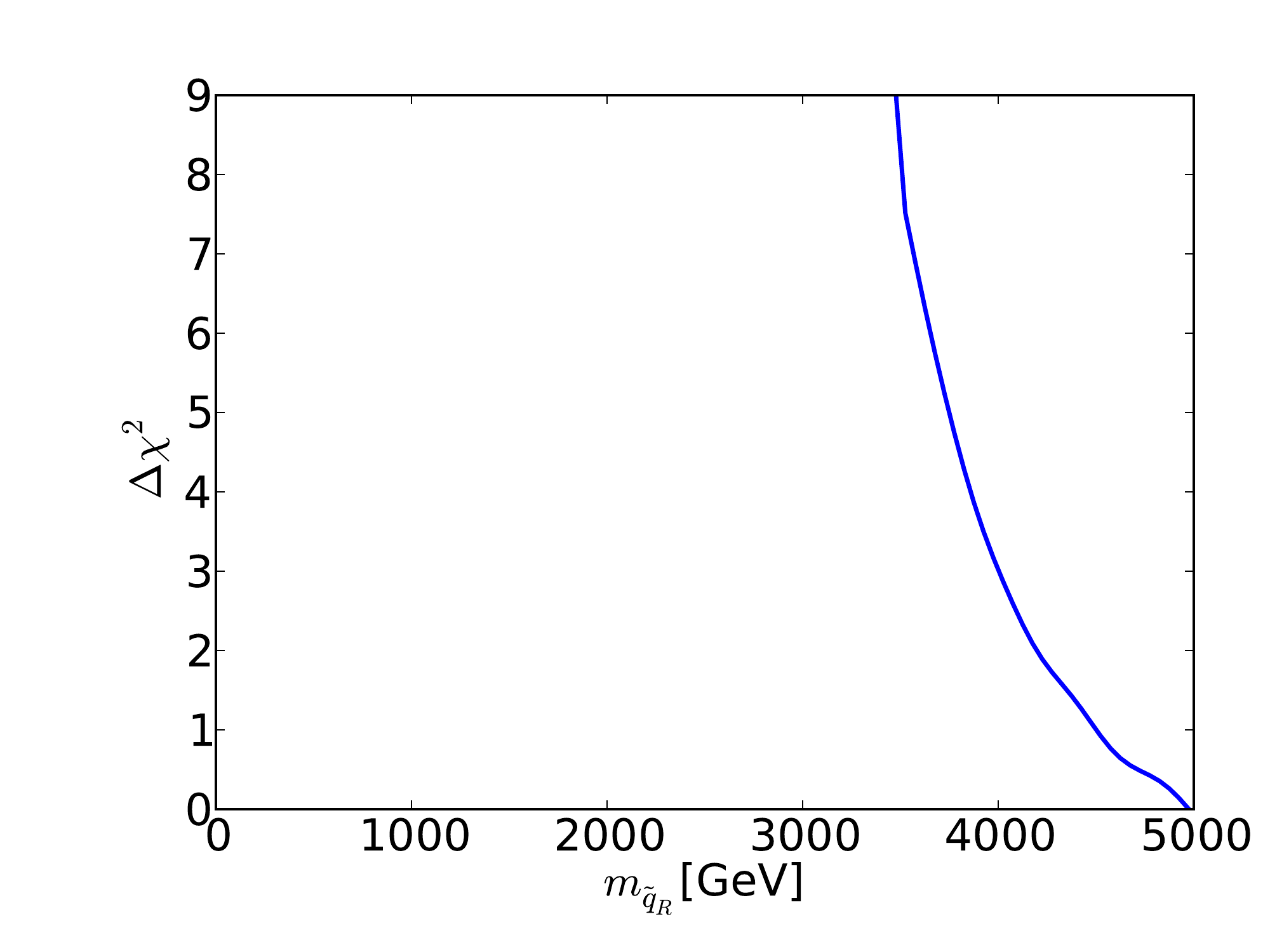} \\
\includegraphics[height=6cm]{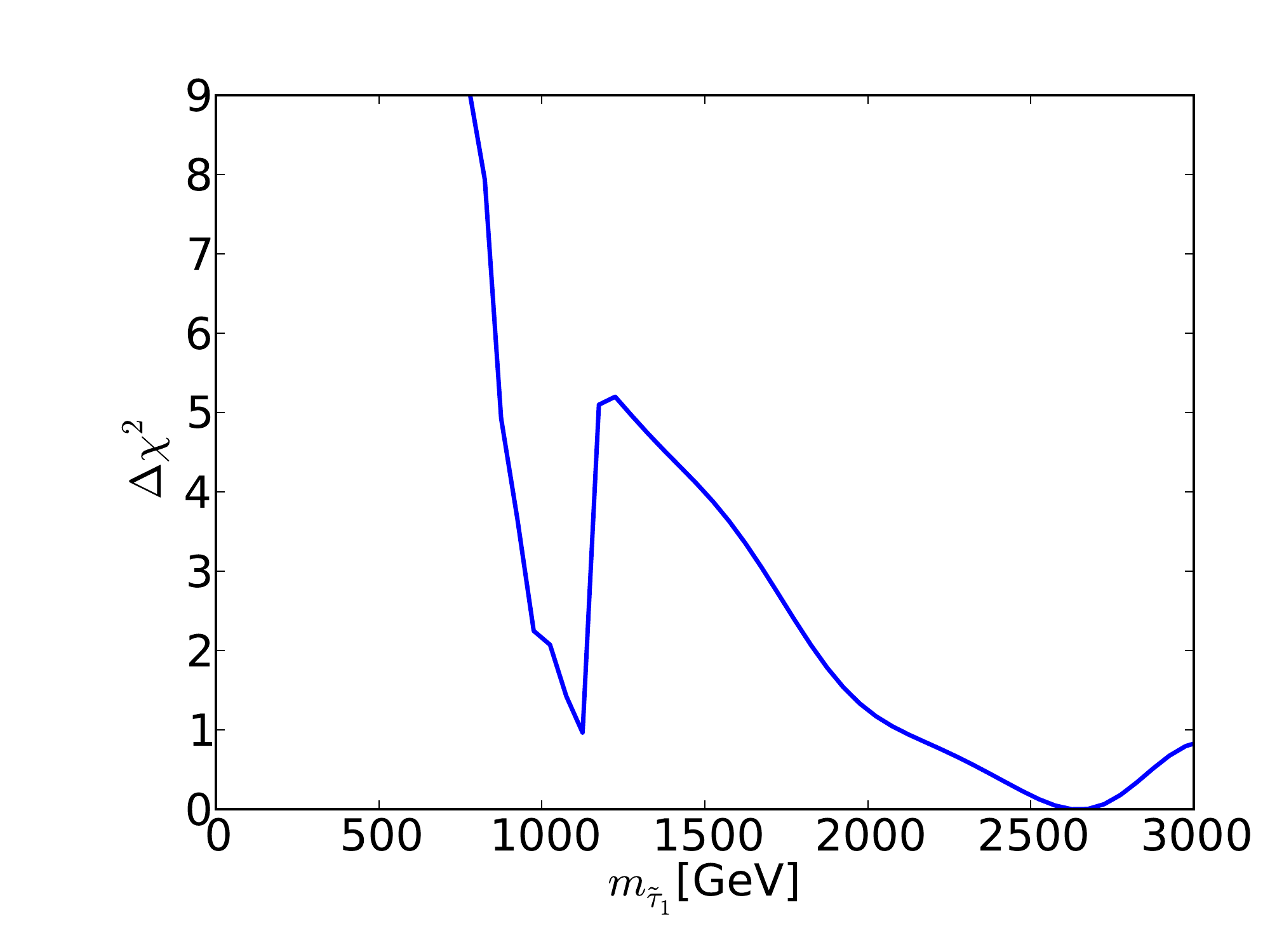} &
\includegraphics[height=6cm]{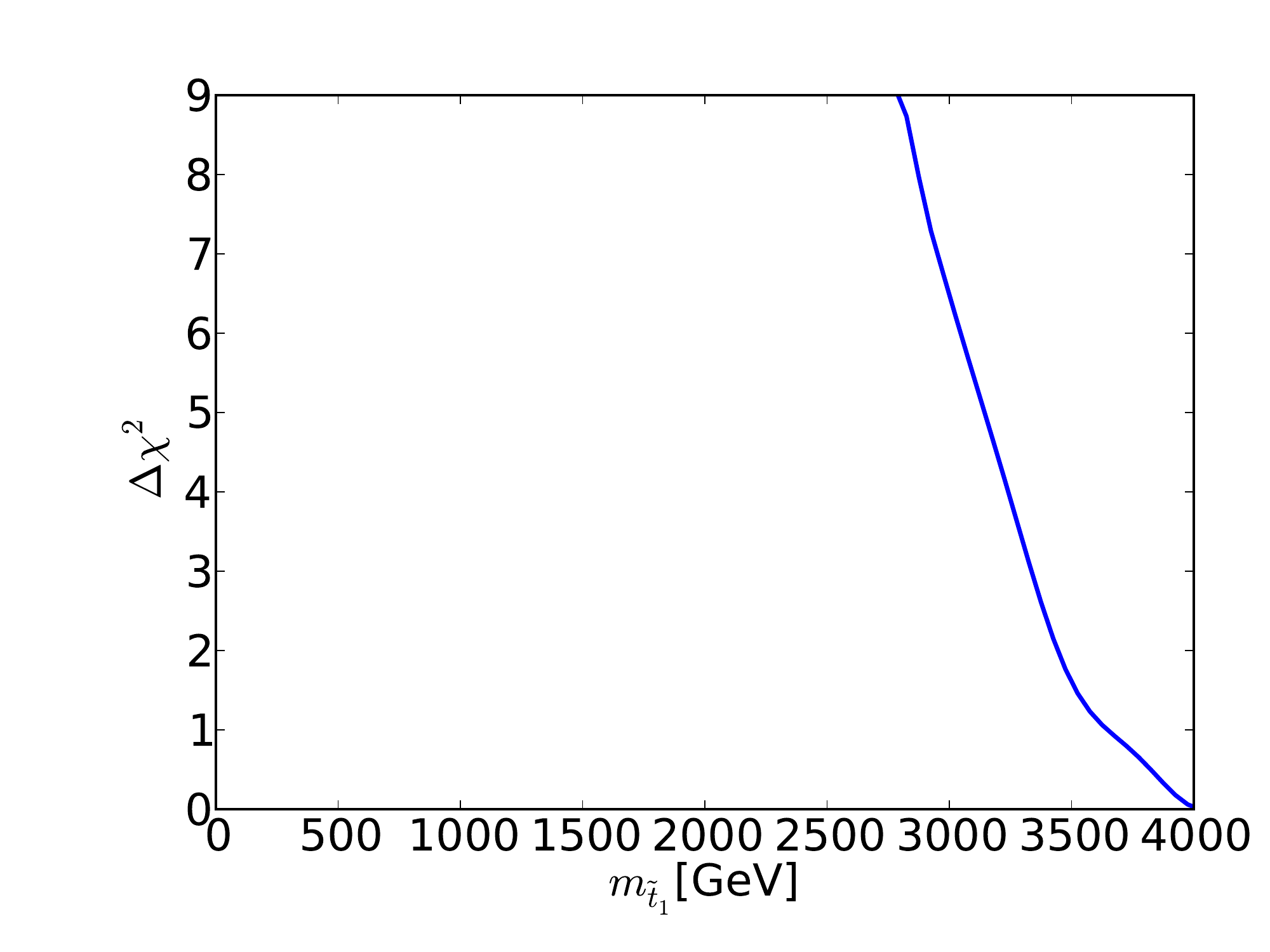} \\
\end{tabular}
\end{center}   
\caption{\label{fig:heavymasses}\it
The one-dimensional profile likelihood functions for $m_{\tilde g}$ (upper left panel),
$m_{\tilde q}$ (upper right panel), $m_{\tilde \tau_1}$ (lower left panel) and $m_{\tilde t_1}$ (lower right panel)
in the CMSSM, assuming that supersymmetry has not been discovered at the LHC with 3000/fb of luminosity, and neglecting 
inevitable improvements in other constraints on the supersymmetric models.}
\end{figure}

Fig.~\ref{fig:heavymasses} shows the corresponding one-dimensional profile
likelihood functions for $m_{\tilde g}$ (upper left panel) and $m_{\tilde q}$ (upper right panel).
In the case of the gluino, we find a prospective 95\% CL lower limit $m_{\tilde g} \gtrsim 3$~TeV,
and a lower limit $m_{\tilde q} \gtrsim 4$~TeV for the squark mass. The limitations of
our CMSSM sample~\cite{mc9} are such that we do not have any useful information about
the likelihood functions for large masses, where they are expected to be quite flat.
The lower left panel of Fig.~\ref{fig:heavymasses} shows the corresponding one-dimensional profile
likelihood function for $m_{\tilde \tau_1}$: the dip at $m_{\tilde \tau_1} \sim 1000$~GeV
corresponds to the `cockscomb' feature visible as an isolated 68\% CL region
with $(m_0, m_{1/2}) \sim (1800, 2400)$~GeV in Fig.~\ref{fig:highmassnoLHC}, with the local peak at
$m_{\tilde \tau_1} \sim 1300$~GeV corresponding to the gap between the `cockscomb' and the Eurasia region.
As is also apparent in Fig.~\ref{fig:stau}, within the CMSSM
there is significant likelihood that $m_{\tilde \tau_1} < 1500$~GeV, so that ${\tilde \tau_1}$
pair-production would be possible at CLIC, even if the LHC fails to discover supersymmetry.

\subsection{Direct Sparticle Searches at a Higher-Energy Proton-Proton Collider}

We now turn to the potential of a future higher-energy hadron collider for discovering
supersymmetry within the CMSSM framework. To this end, we first analyze the
nature of the CMSSM parameter space for large values of $m_0$ and $m_{1/2}$,
taking into account the cold dark matter density constraint and the measurement
of $m_h$, which are the only constraints capable of imposing upper limits on $m_0$ and $m_{1/2}$.
Generally speaking, bringing the relic density down into the astrophysical range
when these mass parameters are large requires some specific features in the
sparticle spectrum such as near-degeneracy between the LSP, the NLSP and perhaps
other supersymmetric particles, so as to suppress the relic LSP density via
coannihilation, or the existence of a massive Higgs boson that acts as an s-channel
resonance and thereby suppresses the LSP density by enhancing LSP annihilation.

One such possibility is the stau coannihilation strip~\cite{stau-co}, which appears at low values of
the ratio $m_0/m_{1/2}$, adjacent to
the stau LSP region. Its length depends on $\tan \beta$ and $A_0$, extending
as far as $m_{1/2} \sim 1400$~GeV for $\tan \beta = 40$ and $A_0 = 2.5 m_0$~\cite{eo6}.
A recent study has shown that all this strip may be explored by Run~2 of the
LHC at 14~TeV~\cite{ehow+}, as also discussed above.

\begin{figure}[ht!]
\begin{center}
\begin{tabular}{c c}
\includegraphics[height=7.5cm]{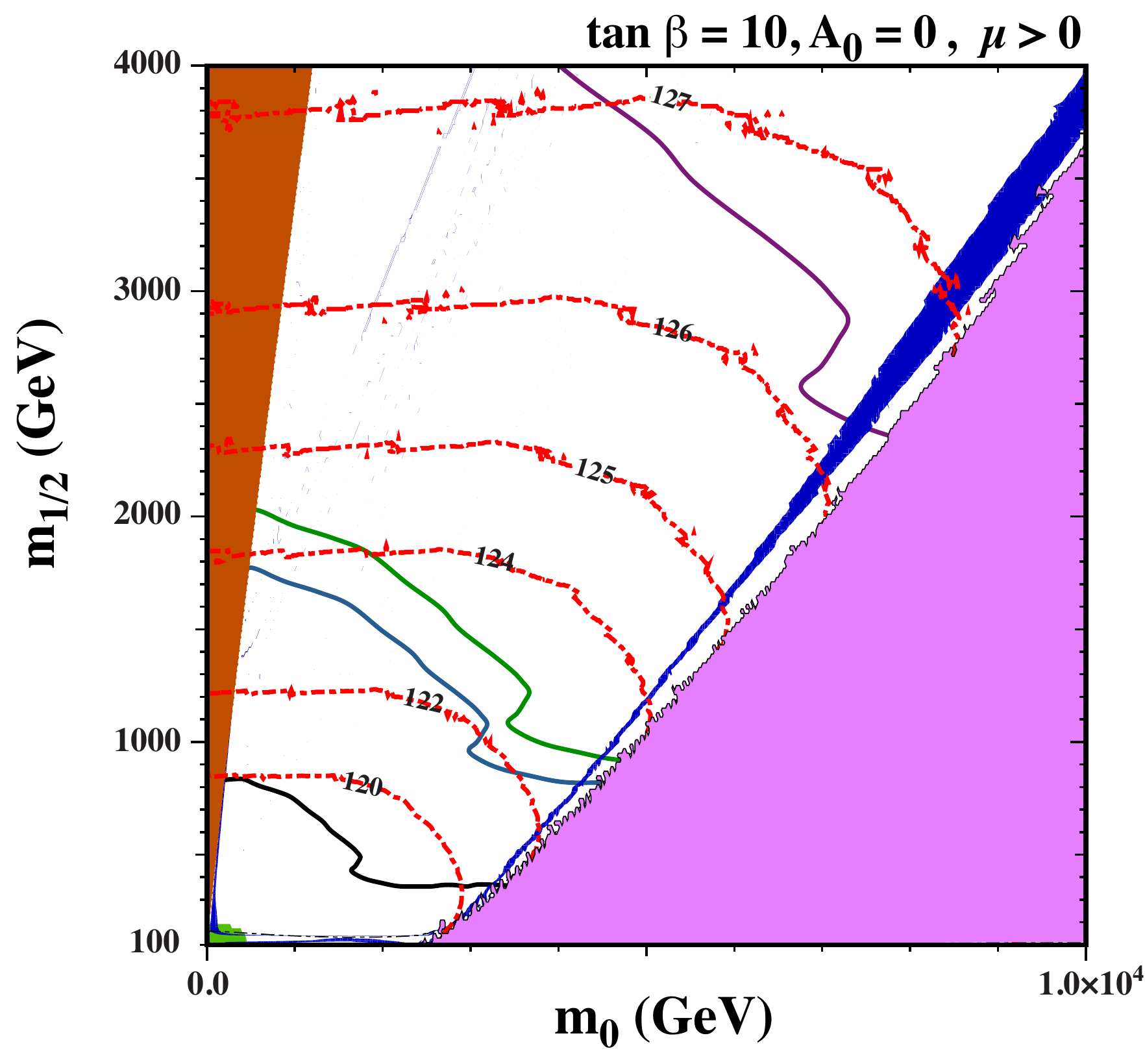} &
\includegraphics[height=7.5cm]{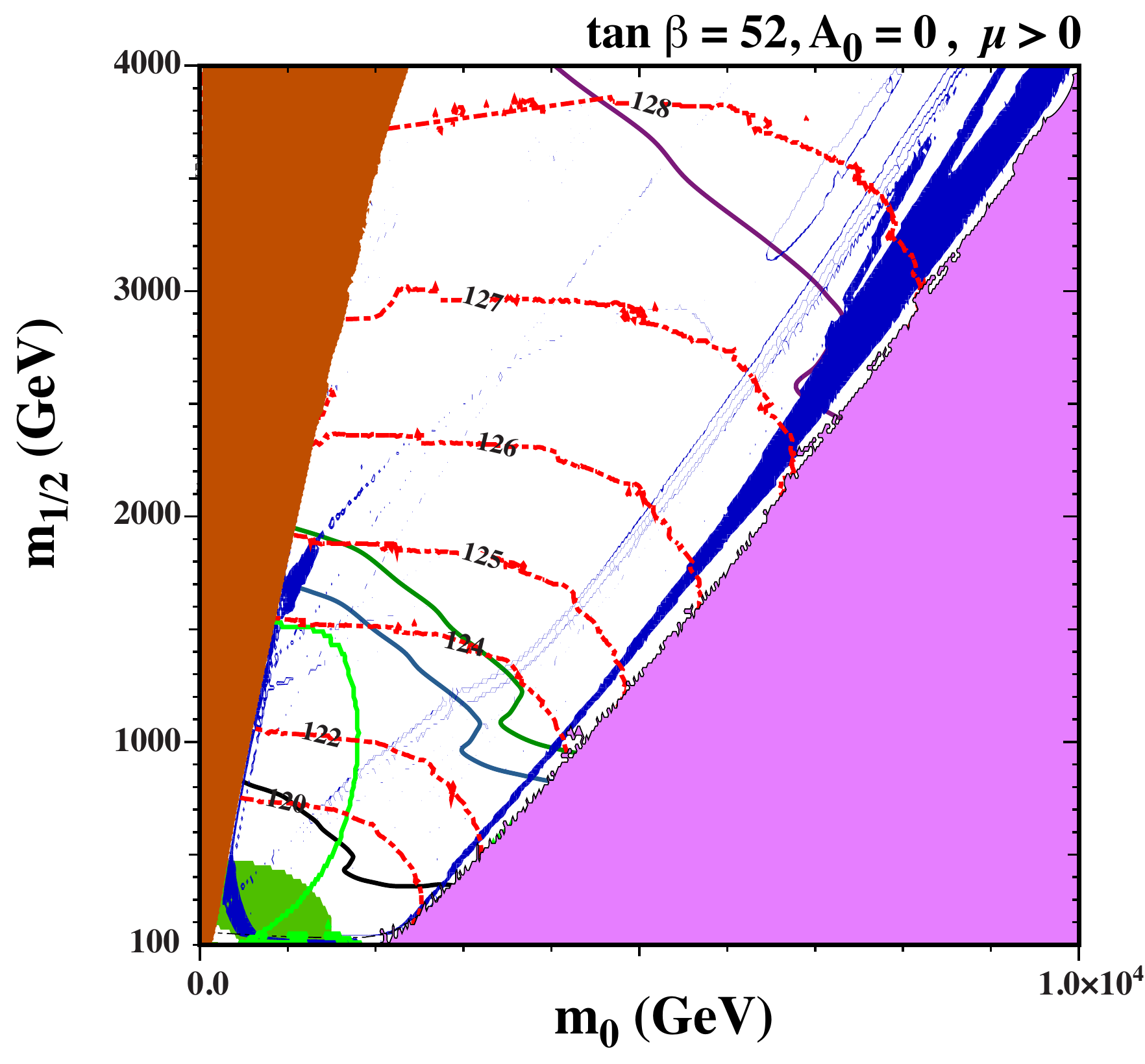} \\
\includegraphics[height=7.5cm]{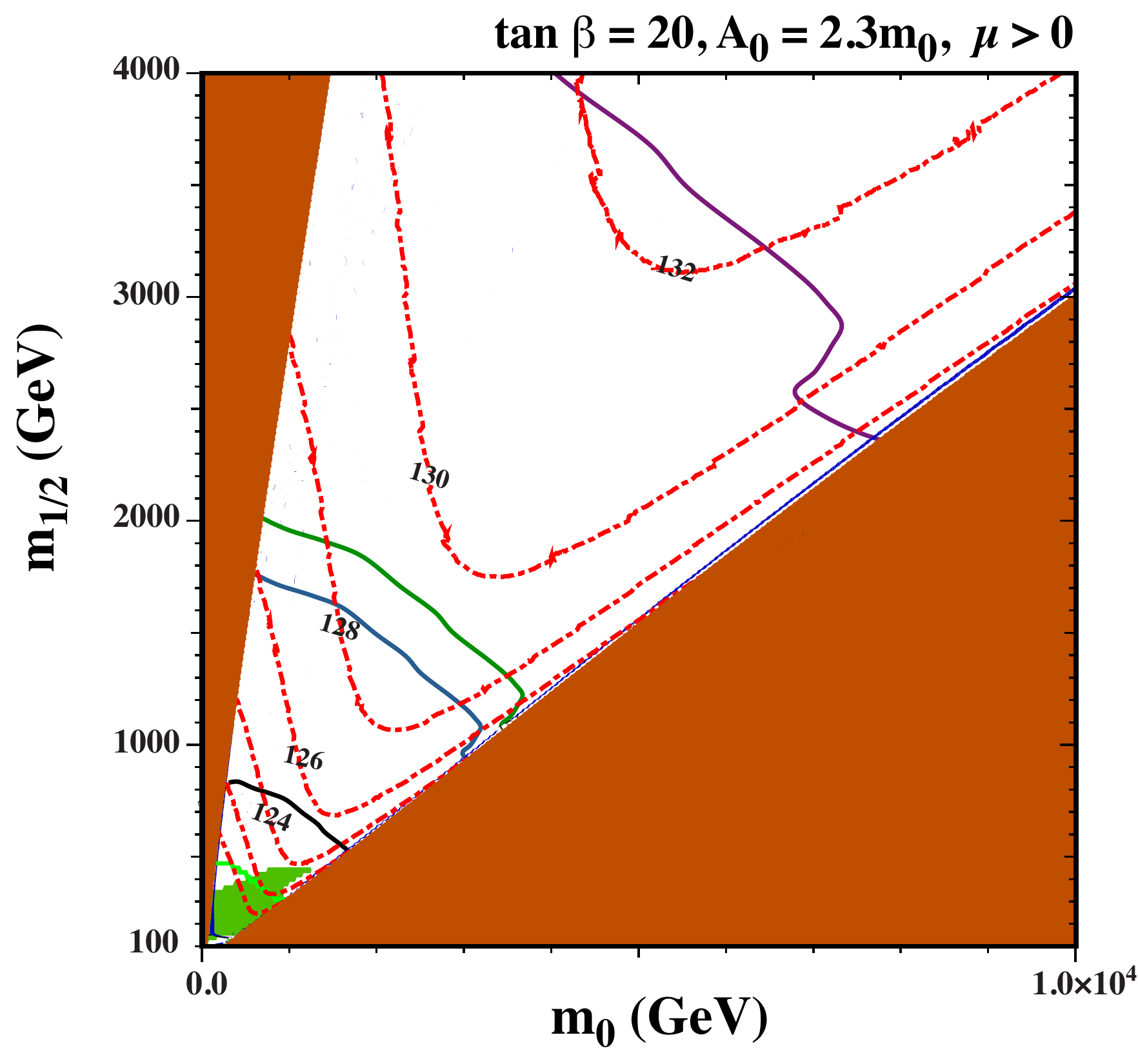} &
\includegraphics[height=7.5cm]{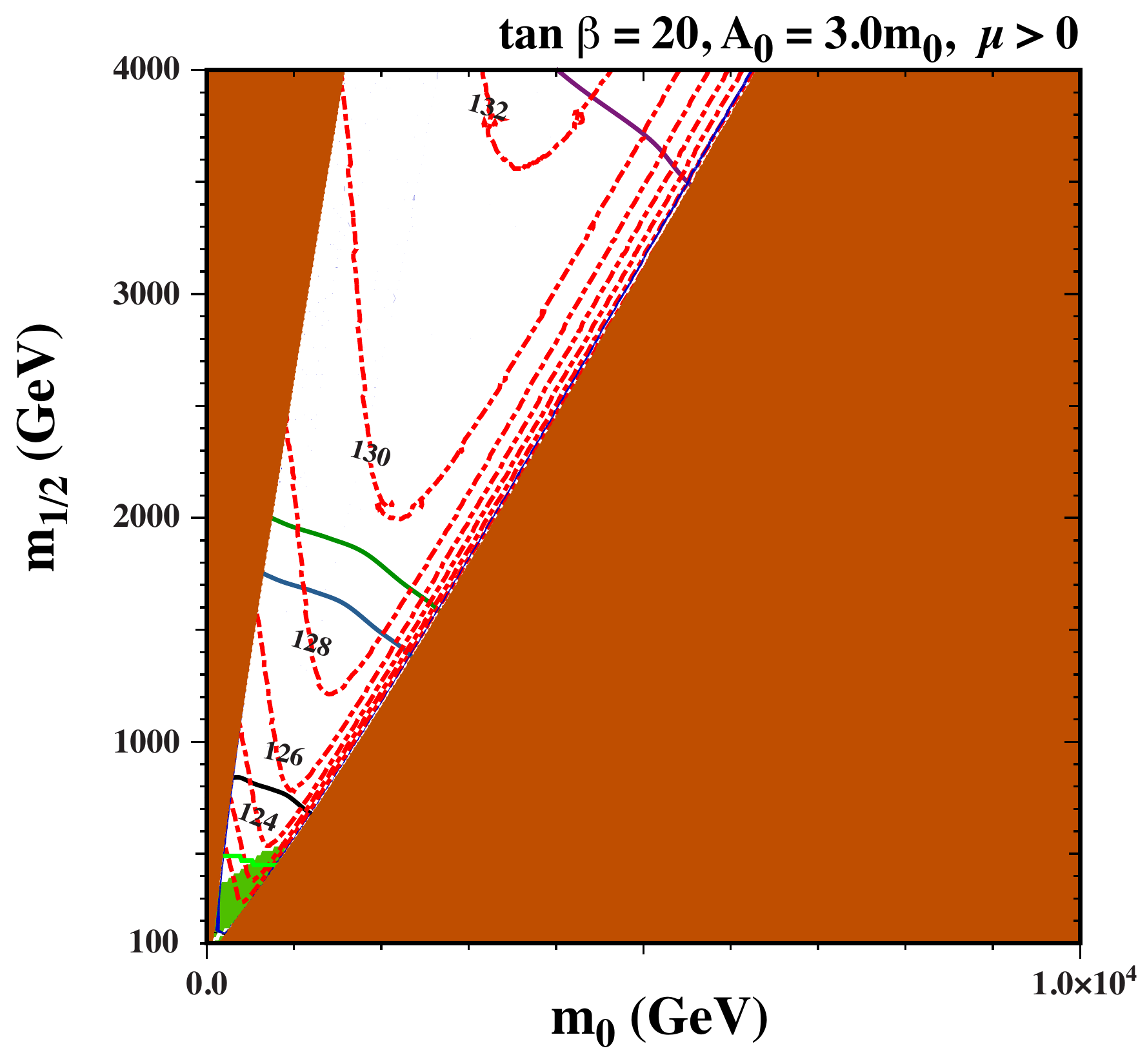} \\
\end{tabular}
\end{center}   
\caption{\label{fig:focuspt}\it
Upper panels: the $(m_0, m_{1/2})$ planes for $A_0 = 0$ and $\tan \beta = 10$ (left panel)
and $40$ (right panel), displaying focus-point strips.
Lower panels: the $(m_0, m_{1/2})$ planes for $\tan \beta = 20$ and $A_0/m_0 = 2.3$  (left panel),
and $A_0/m_0 = 3.0$ (right panel).
In each panel, the ochre regions are excluded because of a charged LSP and/or a tachyon, and the
green regions are excluded by $b \to s \gamma$ decay. There is no consistent electroweak vacuum
in the purple regions in the upper panels. In the dark blue strips the relic LSP density lies within
the range allowed by cosmology, and the dashed red lines are contours of $m_h$ as calculated using
{\tt FeynHiggs~2.10.0}. The solid black, blue, green and purple lines in each panel are
particle exclusion reaches for $\ETslash$ searches with LHC at 8~TeV,
300 and 3000/fb with LHC at 14~TeV, and 3000/fb with HE-LHC at 33~TeV, respectively.}
\end{figure}

Another possibility is the focus-point strip \cite{oldfp,fp}, which appears at higher values of
the ratio $m_0/m_{1/2}$, adjacent to the boundary of the region where one can find a
consistent electroweak vacuum, \htr{and generally lies beyond the reach of the LHC searches
discussed earlier}. Along this strip, the Higgsino component of
the neutralino LSP is enhanced, and its annihilations and coannihilations
with heavier neutralinos and charginos are enhanced. Various studies have
shown that the focus-point strip may extend to very large values of $m_0$
and $m_{1/2}$, with $m_0/m_{1/2} \sim 3$ and $A_0 \lsim m_0$. The upper panels of
Fig.~\ref{fig:focuspt} display a pair of focus-point strips for $A_0 = 0$ and
$\tan \beta = 10$ (left panel) and $52$ (right panel), assuming 
$m_t = 173.2$~GeV~\footnote{Larger (smaller) values
of  $m_0/m_{1/2}$ are found for larger (smaller) values of $m_t$.}.
The regions of these planes where there is no consistent electroweak vacuum
are coloured purple, the ochre regions at lower $m_0/m_{1/2}$
are excluded because of a charged LSP and/or a tachyon, and the green shaded
regions are excluded by $b \to s \gamma$ decay~\footnote{The lighter green line
in the upper right panel of Fig.~\ref{fig:focuspt} delineates the region allowed
by $B_s \to \mu^+ \mu^-$.}.

In both the upper panels of Fig.~\ref{fig:focuspt}, we see a (dark blue)
focus-point strip hugging the boundary of the region at $m_0/m_{1/2} \gsim 3$
where electroweak symmetry-breaking is not possible. In the right panel, we
also see a rapid-annihilation funnel~\cite{funnel} projecting out of the stau coannihilation
strip at low $m_0/m_{1/2}$ and extending to $(m_0, m_{1/2}) \sim (2500, 1800)$~GeV.
We also see in both panels that the contours of $m_h$ calculated using {\tt FeynHiggs~2.10.0}~\cite{newFH}
(shown as red dashed lines) are almost
orthogonal to the focus-point strips. We note that the LHC measurement of $m_h$, even allowing for a 3~GeV
uncertainty in the {\tt FeynHiggs~2.10.0} calculation, excludes $m_{1/2} \gsim 4000$~GeV.
The solid black, blue, green and purple lines in each panel are
particle exclusion reaches for $\ETslash$ searches with LHC at 8~TeV,
300 and 3000/fb with LHC at 14~TeV, and 3000/fb with HE-LHC at 33~TeV, respectively.
The focus-point strip extends beyond the reach of the LHC, even with 3000/fb at 14~TeV in the centre of mass,
and even beyond the reach of the HE-LHC with 3000/fb at 33~TeV. However, the portion allowed by the Higgs mass 
constraint lies comfortably within the reach of the FCC-hh with 3000/fb at 100~TeV,
as discussed below.

As shown by two examples in the lower panels of Fig.~\ref{fig:focuspt},
at larger values of $A_0/m_0 \gsim 2.2$ there are wedges at
larger $m_0/m_{1/2}$ that are excluded because the lighter stop squark is the LSP.
Hugging the boundaries of these wedges there are narrow stop coannihilation
strips~\cite{stop,eoz} where $\delta m_{\tilde t_1} - m_\chi$ is small.
The opening angles of the stop LSP wedges have little dependence on $\tan \beta$,
and both the planes we show have $\tan \beta = 20$.
On the other hand, the opening angles of the stop LSP wedges increase with $A_0/m_0$,
with the result that the wedge at intermediate $m_0/m_{1/2}$
where the LSP is the lightest neutralino is closed off for $A_0/m_0 \gsim 5.5$.

We display in the lower panels of Fig.~\ref{fig:focuspt} the cases $\tan \beta = 2.3$ (left panel) and 3 (right panel).
As we discuss later, the stop coannihilation strips extend far beyond the ranges of
$m_0$ and $m_{1/2}$ that we display in these two panels of Fig.~\ref{fig:focuspt}.
These panels also display contours of $m_h$ calculated using {\tt FeynHiggs~2.10.0}.
We note that, in contrast to the focus-point cases displayed in Fig.~\ref{fig:focuspt},
the $m_h$ contours cross the dark matter strip at a much smaller angle. As a consequence,
the allowed ranges of $m_0$ and $m_{1/2}$ are much larger than in the focus-point case,
after allowing for the 3~GeV uncertainty estimated within {\tt FeynHiggs~2.10.0} for a given input SLHA file~\cite{SLHA}.
Additionally, we find that the value $m_h$ calculated along the stop coannihilation strip
is very sensitive to the codes used to evolve the CMSSM input parameters down to low energies and
calculate the spectra used as SLHA inputs in the {\tt FeynHiggs~2.10.0} calculation of $m_h$.
This introduces an additional uncertainty of several GeV in the value of $m_h$ corresponding to
any given set of CMSSM input parameters. For this reason, no
portions of these stop coannihilation strips can be excluded
on the basis of the LHC measurement of $m_h$~\footnote{We
assume in calculating these parameter planes that $\alpha_s (m_Z) = 0.1184$ and $\sin^2 \theta_W
= 0.2325$. With these choices, the SU(3), SU(2) and U(1) couplings are in general not quite equal at the GUT
scale, but the deviations are generally compatible with zero when the uncertainties in the inputs are taken
into account, and any apparent non-universality could be cancelled by GUT threshold corrections
and/or produced by higher-dimensional operator contributions. The comparable planes shown in~\cite{eoz}
assumed $\sin^2 \theta_W = 0.2325$ and strict coupling-constant unification, and consequently had
a range of values of $\alpha_s(m_Z)$. This is the reason for the differences between these planes and comparable
planes in~\cite{eoz}.}.

As in the upper panels of Fig.~\ref{fig:focuspt}, the solid black, blue, green and purple lines in the lower panels are
particle exclusion reaches for $\ETslash$ searches with LHC at 8~TeV,
300 and 3000/fb with LHC at 14~TeV, and 3000/fb with HE-LHC at 33~TeV, respectively,
now for CMSSM scenarios with stop coannihilation strips.
We see that the LHC sensitivity contours in the lower panels of Fig.~\ref{fig:focuspt} include only portions
of these stop coannihilation strips extending to $m_{1/2} \sim 500 \, (700)$~GeV for $\tb = 20$ and $A_0 = 2.3 \, (3) \, m_0$. The HE-LHC
sensitivity contour with 3000/fb at 33~TeV extends to $m_{1/2} \sim 2400 \, (3500)$~GeV along
the stop coannihilation strips for $\tb = 20$ and $A_0 = 2.3 \, (3) \, m_0$, but large fractions of these strips lie beyond its reach.

\begin{figure}[ht!]
\vspace{1cm}
\begin{center}
\begin{tabular}{c c}
\includegraphics[height=6.2cm]{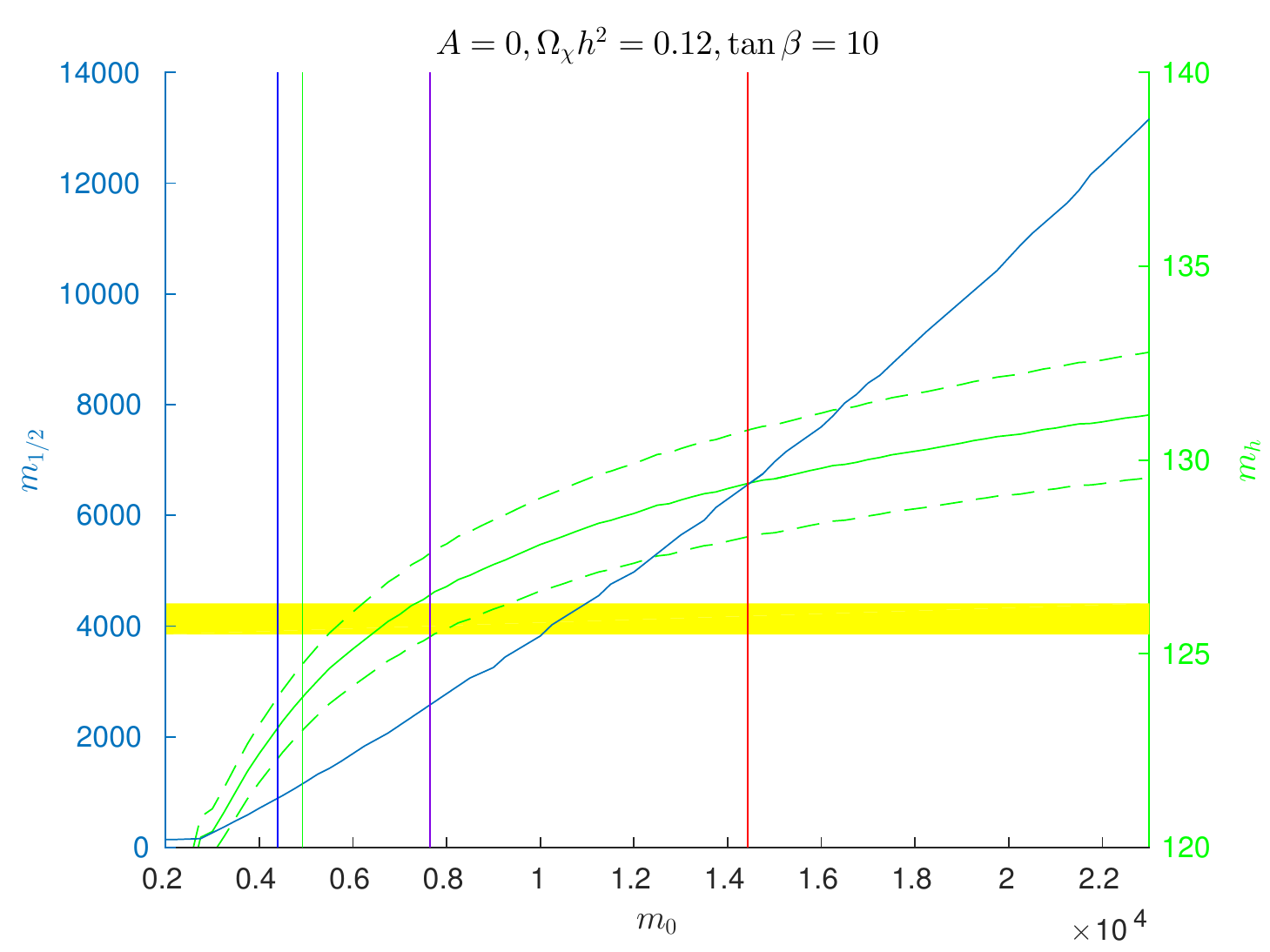} &
\includegraphics[height=6.2cm]{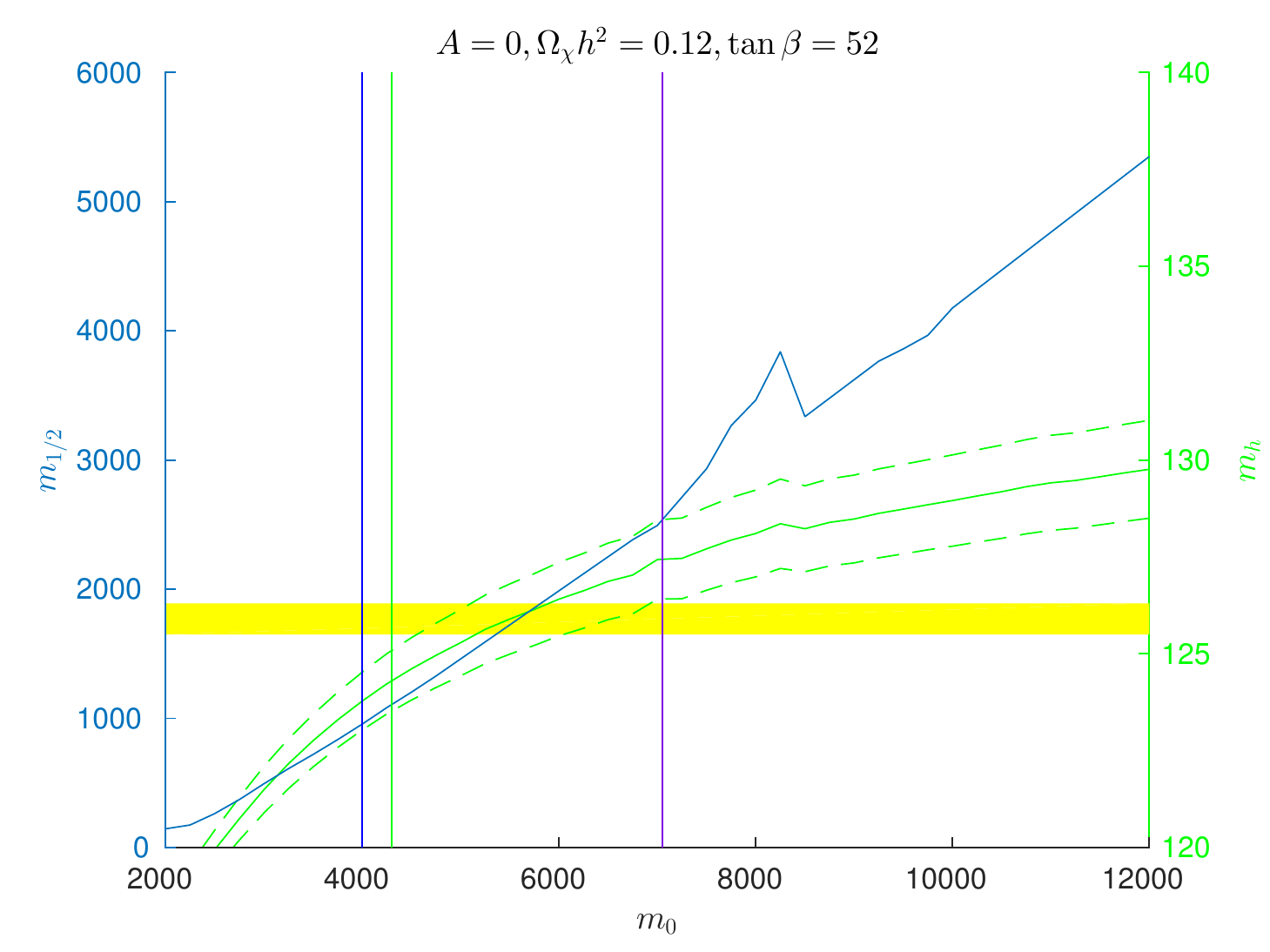} \\
\includegraphics[height=6.2cm]{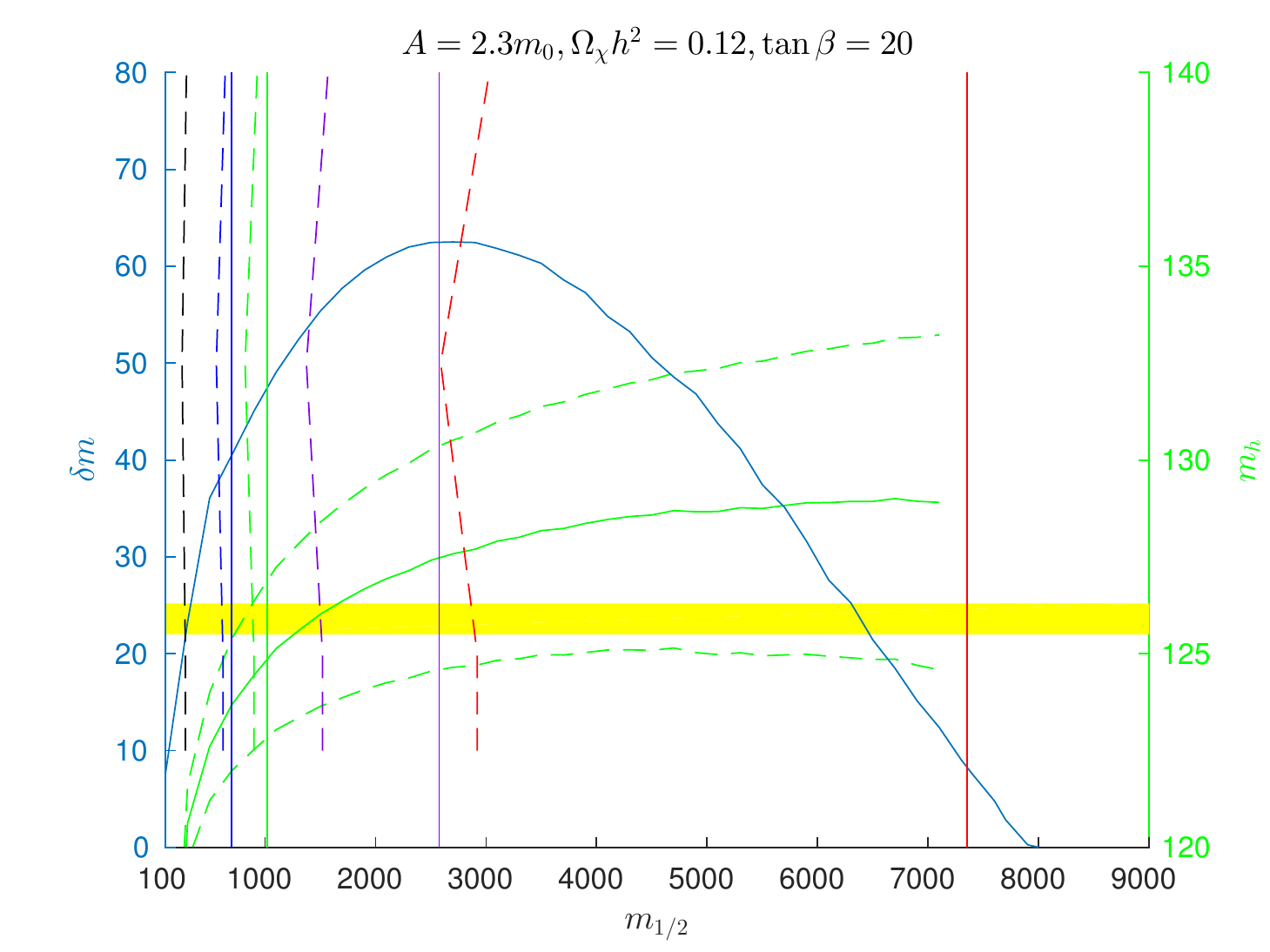} &
\includegraphics[height=6.2cm]{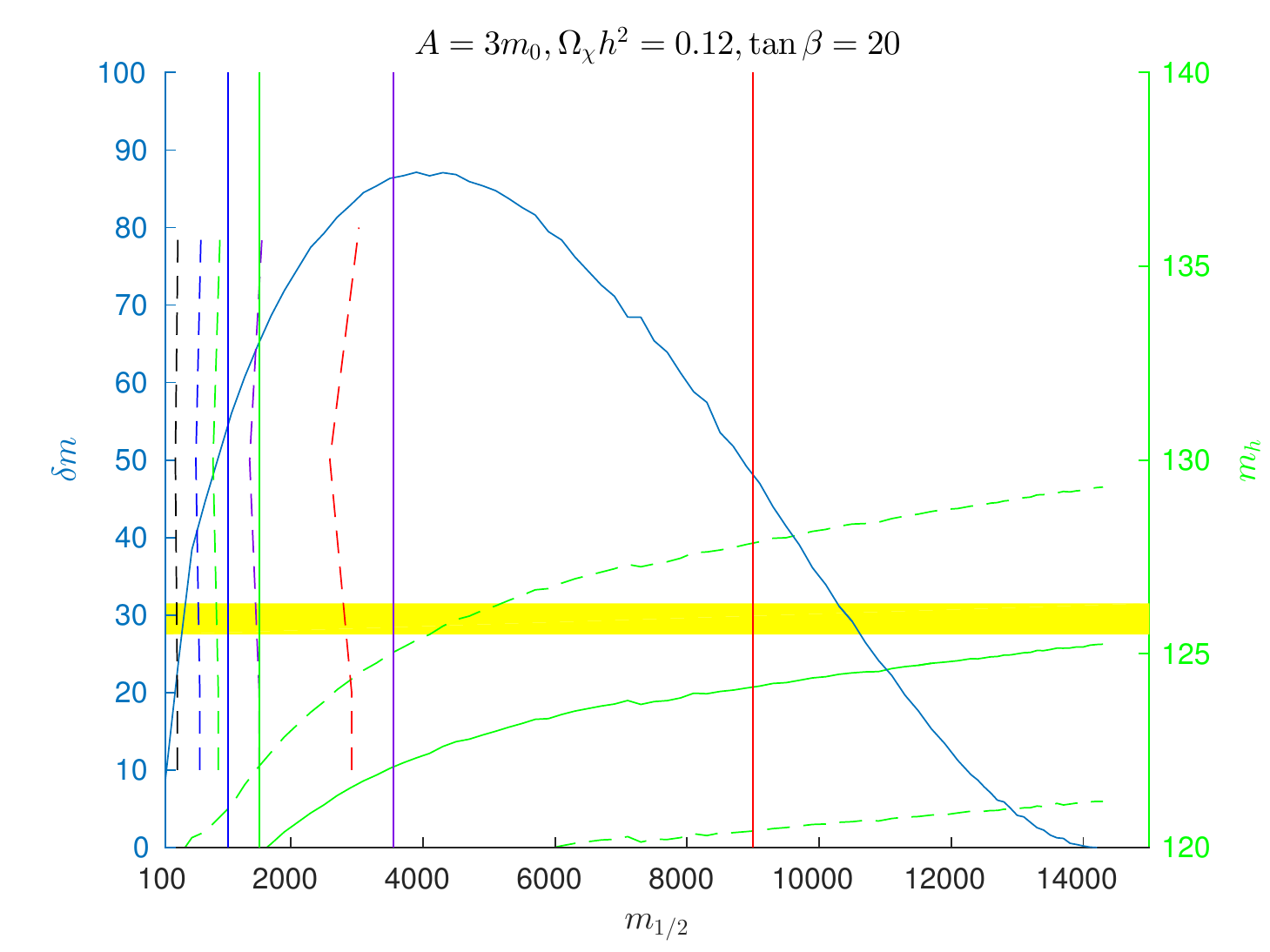} \\
\end{tabular}
\end{center}   
\vspace{1cm}
\caption{\label{fig:profiles}\it
Upper panels: the solid blue lines are the profiles in the $(m_0, m_{1/2})$ plane
of the focus-point strips for $A_0 = 0$ and $\tan \beta = 10$ (left panel),
and $A_0 = 0$ and $\tan \beta = 52$ (right panel). Lower panels: 
the solid blue lines are the profiles in the $(m_{1/2}, \delta m \equiv m_{\tilde t_1} - m_\chi)$ plane
of the stop coannihilation strips for $A_0/m_0 = 2.3$ and $\tan \beta = 20$ (left panel),
and $A_0/m_0 = 3.0$ and $\tan \beta = 20$
(right panel). The \htr{near-vertical} black, blue, green, purple and red lines in each panel are
particle exclusion reaches for particle searches with LHC at 8~TeV,
300 and 3000/fb with LHC at 14~TeV, 3000/fb with HE-LHC at 33~TeV
and 3000/fb with FCC-hh at 100~TeV, respectively.
The solid lines are for generic $\ETslash$ searches, and (in the lower panels) the dashed lines are for dedicated
stop searches. The solid (dashed) \htr{near-horizontal} green lines are central values (probable ranges)
of $m_h$ calculated using {\tt FeynHiggs~2.10.0}, and the yellow band represents the experimental value of $m_h$.}
\end{figure}

\begin{figure}[hbt]
\begin{center}
\vspace{-3cm}
\includegraphics[height=15cm]{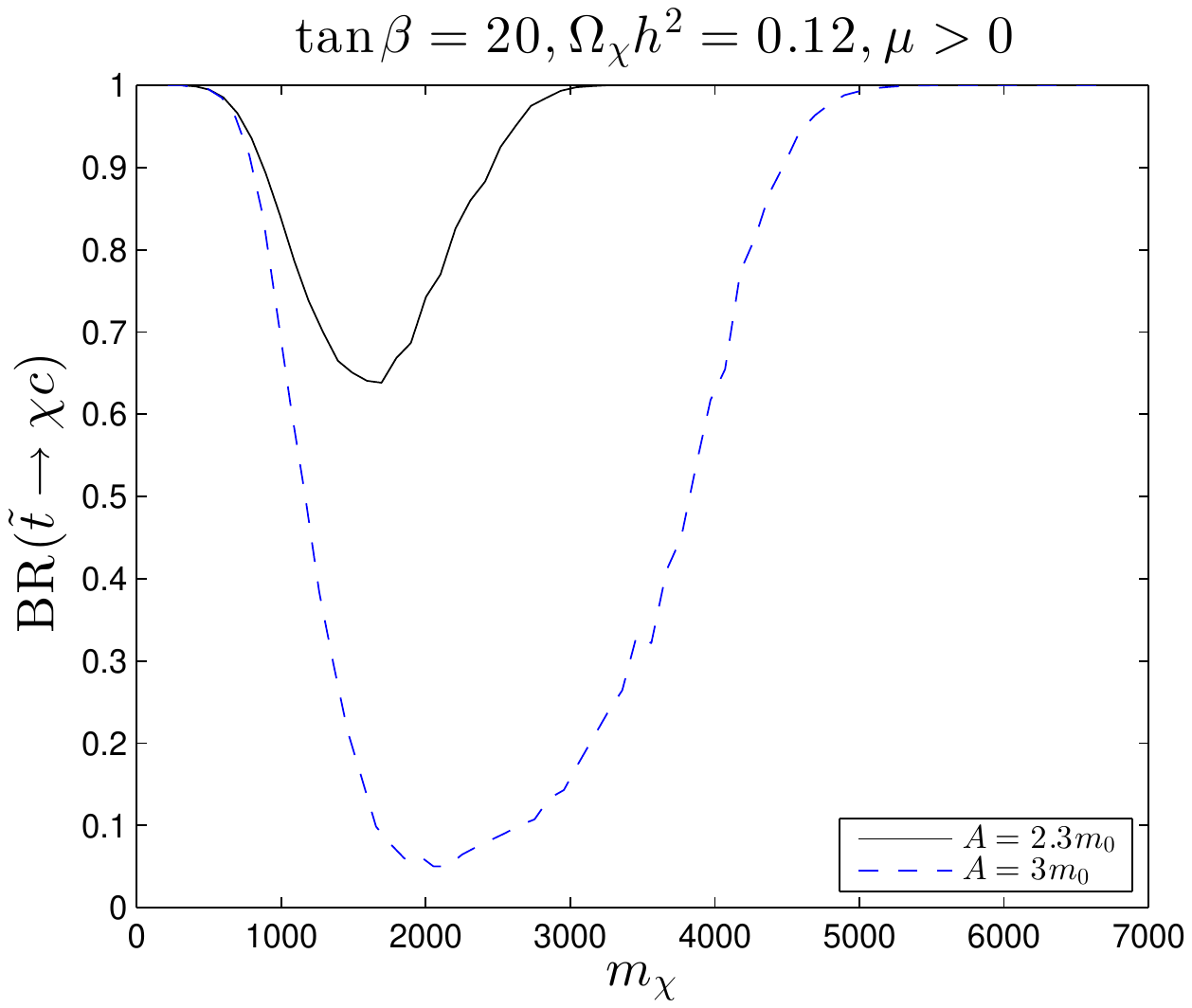}
\end{center}   
\vspace{-4cm}
\caption{\label{fig:BR}\it
The branching ratio for ${\tilde t_1} \to \chi + c$ along the stop coannihilation strips
for $\tan \beta = 20$ and $A_0/m_0 = 2.3$ (solid black line) and $A_0/m_0 = 3.0$ (dashed blue line).
In the latter case the branching ratio drops to a minimum $< 0.1$ when
$m_{\tilde t_1} - m_\chi > m_W + m_b$, as seen in the lower right panel of
Fig.~\protect\ref{fig:profiles}.}
\end{figure}

Fig.~\ref{fig:profiles}~\footnote{The upper panels of Fig.~\ref{fig:profiles} in the published version are corrected in an erratum} displays the profiles of the focus-point strips in
Fig.~\ref{fig:focuspt} (upper panels) and of the stop coannihilation strips
in Fig.~\ref{fig:focuspt} (lower panels), along their full lengths. 
Both pairs of profiles exhibit the values of $m_h$ calculated using SLHA files
obtained using {\tt SSARD} as inputs to {\tt FeynHiggs~2.10.0} (\htr{near-horizontal} solid green lines),
including uncertainty estimates of $\pm 3$~GeV (\htr{near-horizontal dashed green lines}). As already noted, only portions of the
focus-point strips are compatible with the LHC measurement of $m_h$ (yellow bands)
within these uncertainties, whereas in the cases of the stop
coannihilation strips there are significant additional uncertainties associated with the
RGE running, and all portions of the strips are compatible with $m_h$.
In the cases of the stop coannihilation strips in the lower panels of Fig.~\ref{fig:profiles}, we also display as blue lines the
mass difference $\delta m \equiv m_{\tilde t_1} - m_\chi$ along the strips~\footnote{This
is generally larger than in the cases with strict gauge coupling unification assumed in~\cite{eoz},
with the corollary that the strips extend to larger $m_{1/2}$ than shown there.}. In the examples shown,
this mass difference is generally $< m_W + m_b$, so that the branching ratio for two-body
${\tilde t_1} \to \chi + c$ decay usually dominates over that for four-body ${\tilde t_1} \to \chi + W + b + \nu$
decay. However, this is not always the case, as illustrated by examples in~\cite{eoz}
and by Fig.~\ref{fig:BR} for the stop coannihilation strip with $A_0/m_0 = 3.0$ and $\tan \beta = 20$.
The branching ratio for ${\tilde t_1} \to \chi + W + b + \nu$ decay may dominate when
$m_{\tilde t_1} - m_\chi > m_W + m_b$, as seen in the lower right panel of
Fig.~\ref{fig:profiles}. Thus, a complete search for supersymmetry
at FCC-hh should include searches for both the ${\tilde t_1} \to \chi + c$ and 
${\tilde t_1} \to \chi + W + b + \nu$ decay signatures.

The (near-)vertical lines in Fig.~\ref{fig:profiles}
mark our estimates of the sensitivities of the LHC (black - 8~TeV, blue - 300/fb at 14~TeV,
green - 3000/fb at 14~TeV), 3000/fb at HE-LHC (purple) and 3000/fb at
FCC-hh (red) along the stop coannihilation strips. The solid lines represent
the extrapolated reaches of the generic jets + $\ETslash$ searches, and the dashed lines in the lower panels represent
the extrapolated reaches of dedicated searches for ${\tilde t_1} \to c + \chi$ decays, which
lose some sensitivity as $\delta m$ increases because of the increase in the ${\tilde t_1} \to \chi + W + b + \nu$ decay
branching ratio. We see that the FCC-hh would be sensitive to the
full extents of the focus-point strips (upper panels) and of the stop coannihilation strip for $A_0 = 2.3 \, m_0$ (lower left panel),
but not all the stop coannihilation strip for $A_0 = 3.0 \, m_0$ (lower right panel): this is true in general for
$A_0/m_0 \gsim 2.5$. 

\subsection{Impacts of Electroweak and Higgs Precision Observables}

We have also studied the possible impact of EWPOs
and Higgs precision measurements along the focus-point and stop
coannihilation strips discussed in the previous Subsection. We find that the
contributions to the global $\chi^2$ function of the present EWPOs and Higgs measurements
do not vary strongly along the strips, so do not discuss them further. Instead, we
focus on the potential impacts of FCC-ee measurements, choosing benchmark points on these strips.
These benchmark points are chosen to have values of $m_H$,
as calculated with {\tt FeynHiggs~2.10.0}, that are highly
compatible with the central experimental value $m_h \simeq 125$~GeV~\footnote{We
recall that calculations of $m_h$ with {\tt FeynHiggs~2.10.0} have uncertainties $\sim \pm 3$~GeV so that,
as discussed in~\cite{eoz}, substantial fractions of the focus-point and stop coannihilation strips are compatible
with the $m_h$ constraint.}.

\begin{figure}[hbt]
\begin{center}
\begin{tabular}{c c}
\includegraphics[height=5cm]{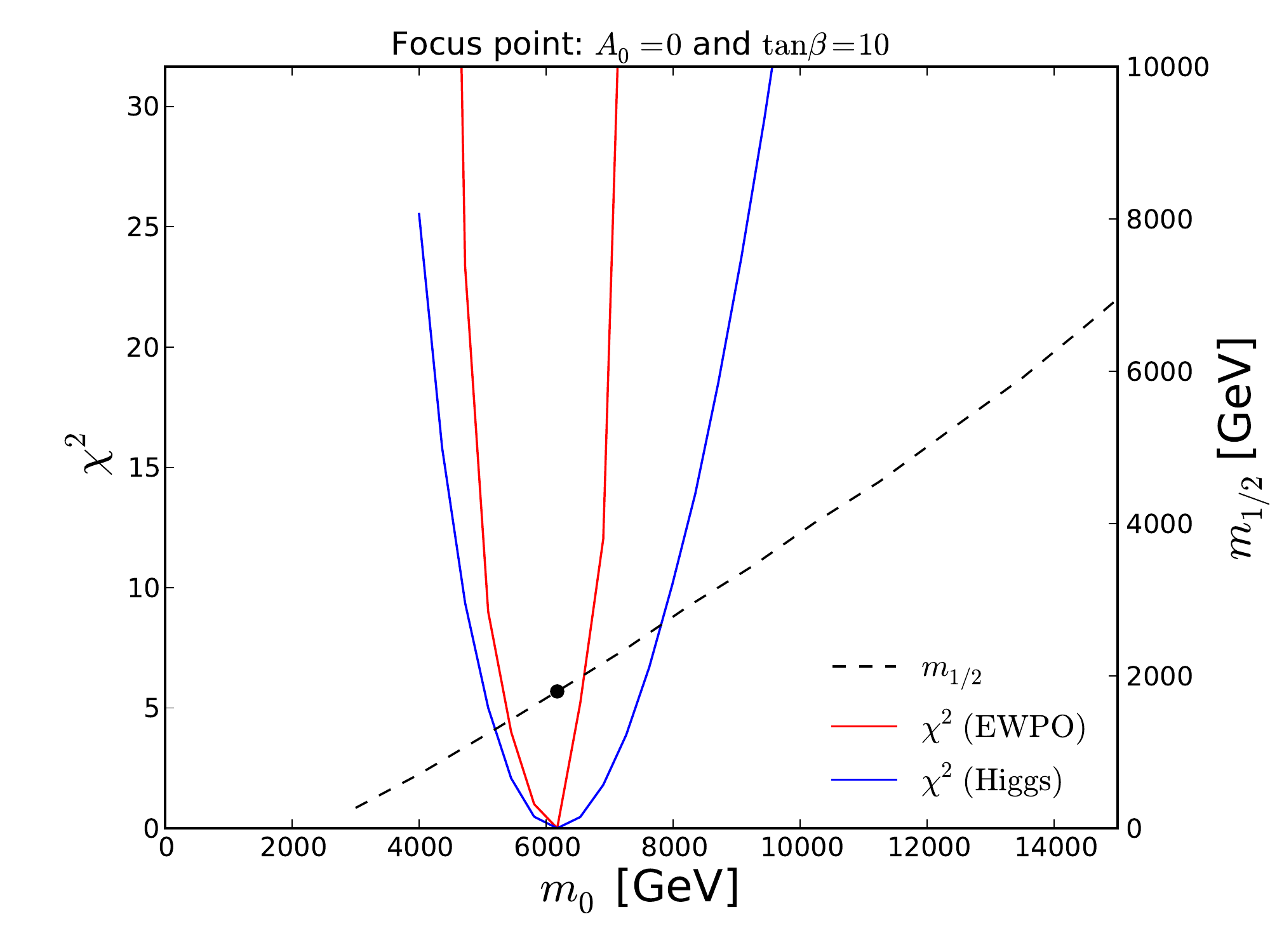} &
\includegraphics[height=5cm]{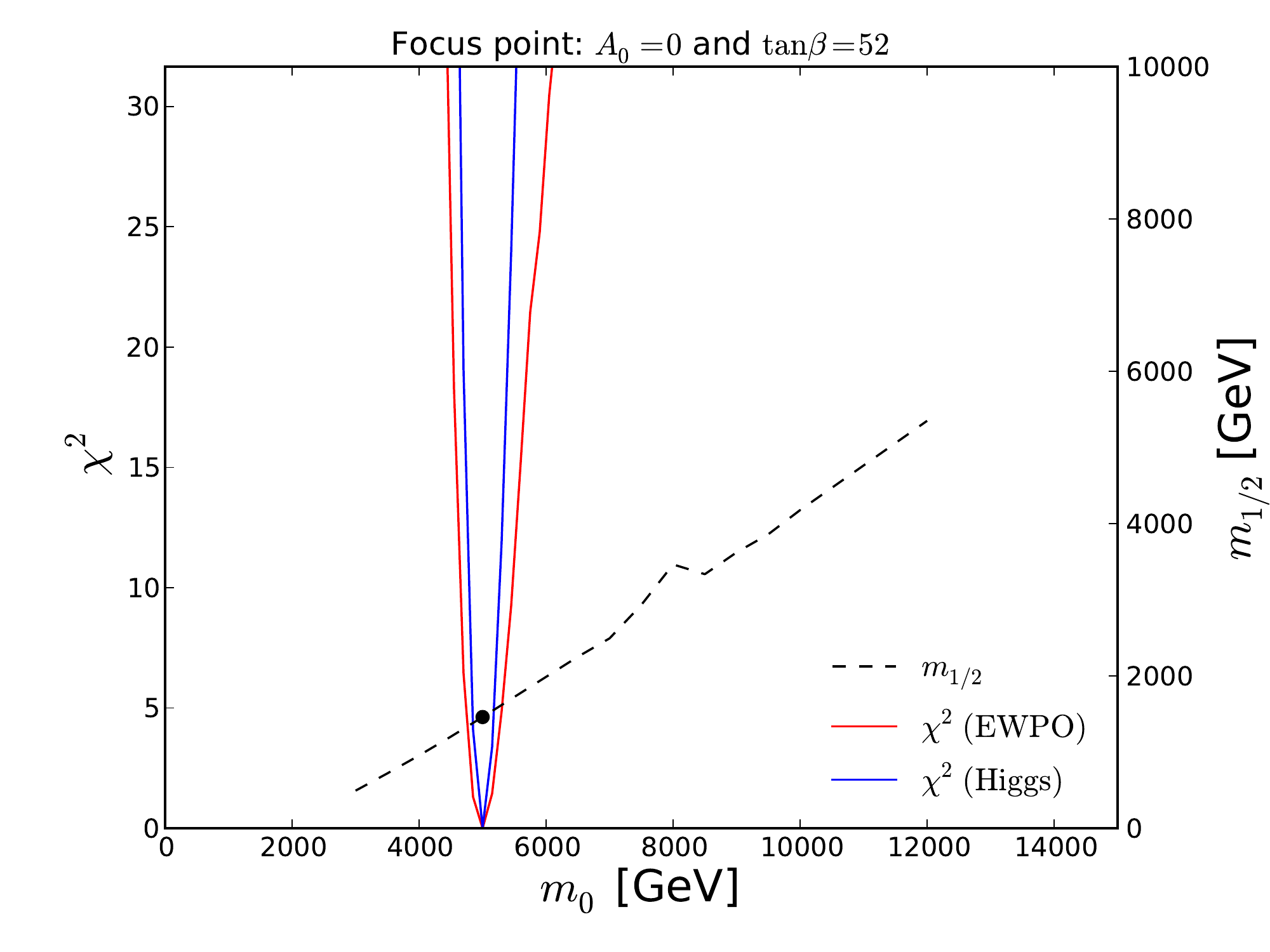} \\
\includegraphics[height=5cm]{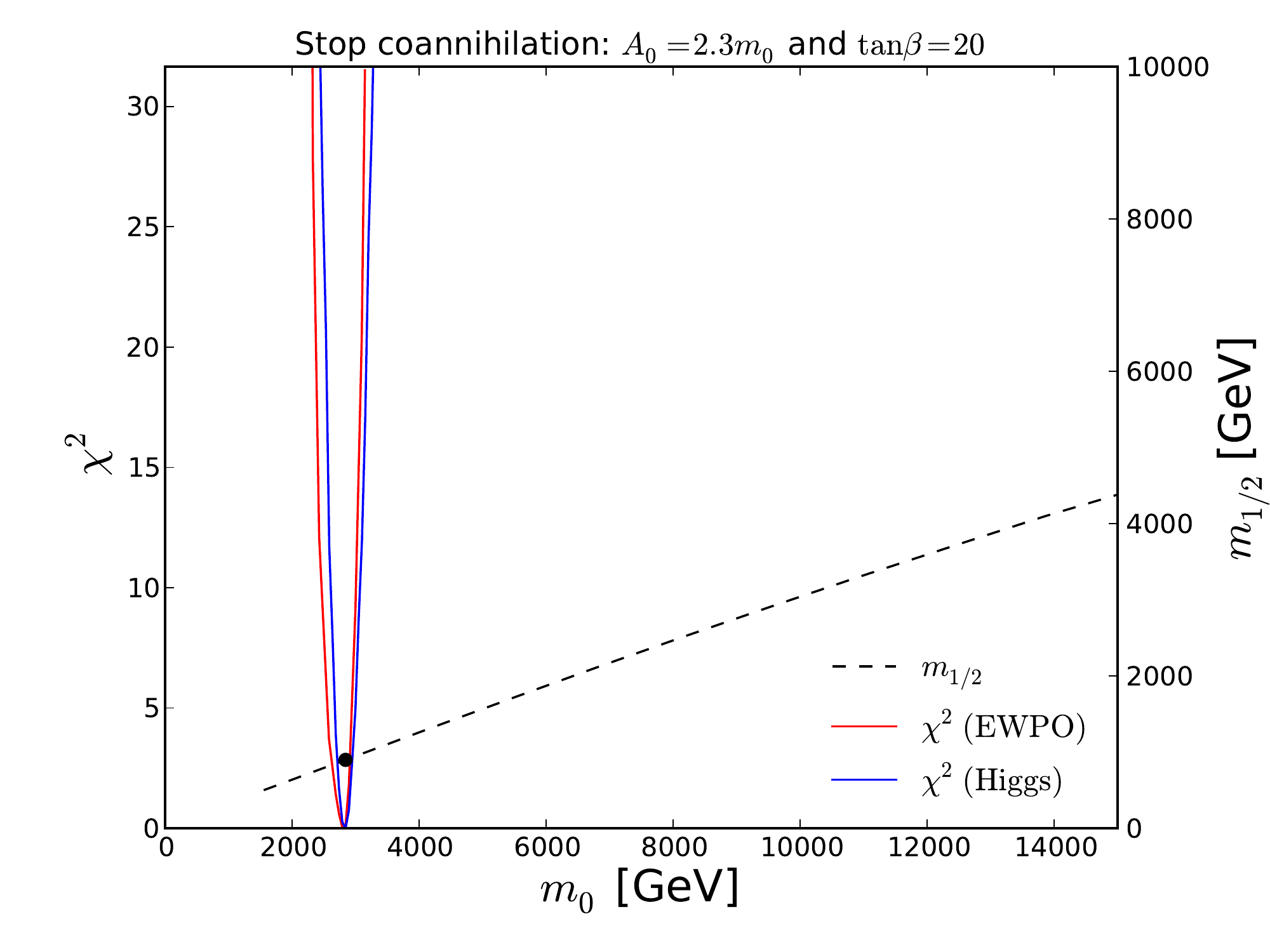} &
\includegraphics[height=5cm]{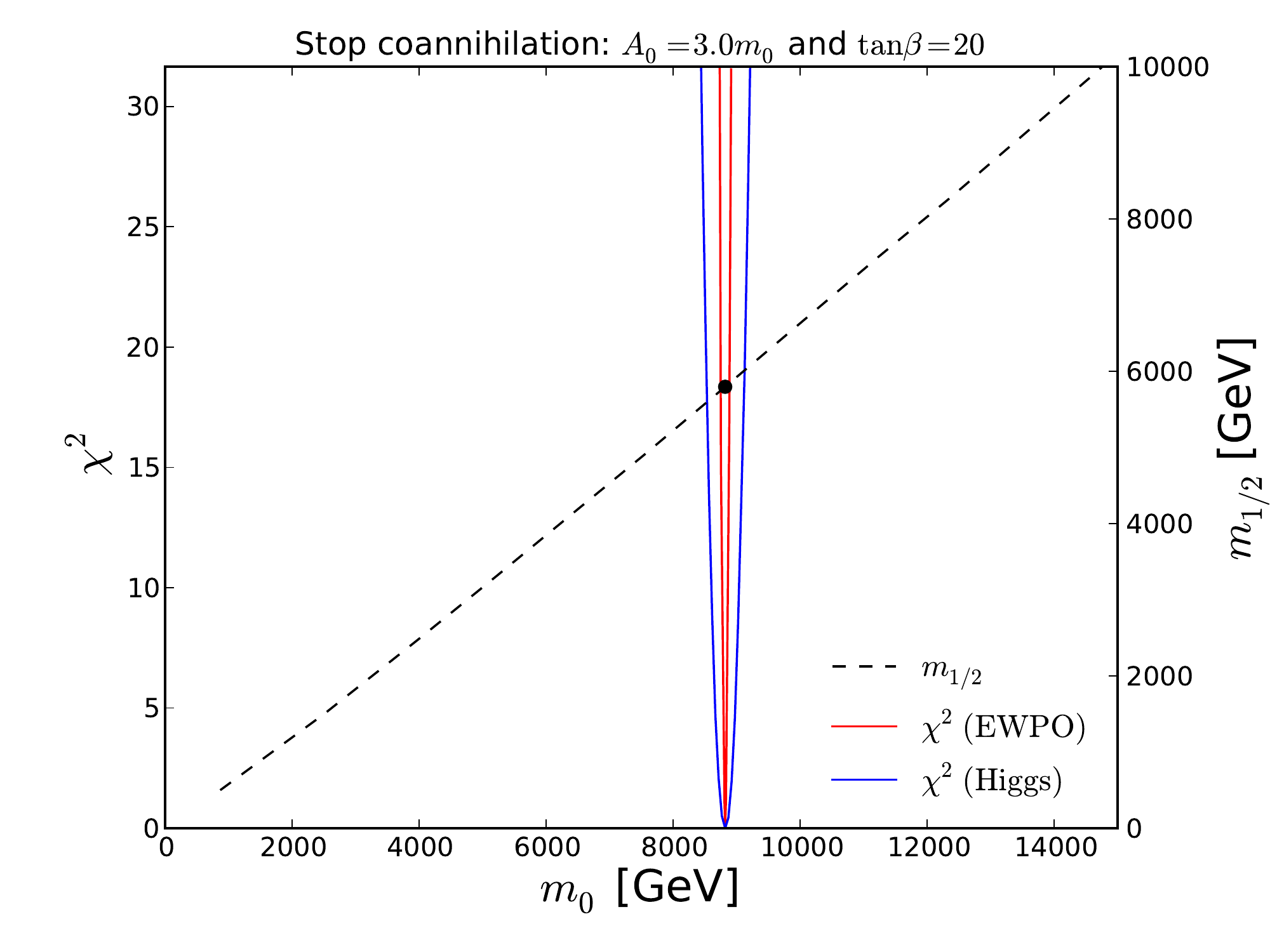} \\
\end{tabular}
\end{center}   
\caption{\label{fig:EWPOHTLEP}\it
Upper panels: the estimated contributions of the EWPOs and Higgs observables
measured at FCC-ee (TLEP) to the global $\chi^2$ function (solid red and blue lines, respectively)
assuming the parameters of benchmark points along the focus-point strips for $A_0 = 0$ and $\tan \beta = 10$ (left panel),
and $A_0 = 0$ and $\tan \beta = 52$ (right panel). Lower panels: the same for the
stop coannihilation strips for $A_0/m_0 = 2.3$ and $\tan \beta = 20$ (left panel),
and $A_0/m_0 = 3.0$ and $\tan \beta = 20$ (right panel). The diagonal black dashed lines show
the values of $m_{1/2}$ along the corresponding strips (right-hand vertical axes), and the black spots show the parameters
of the corresponding benchmark points.}
\end{figure}

Fig.~\ref{fig:EWPOHTLEP} shows the estimated contributions of the EWPOs and Higgs observables
measured at FCC-ee (TLEP) (red and blue lines, respectively) to the global $\chi^2$ functions along the focus-point and
stop coannihilation strips, which are plotted using $m_0$ as the horizontal axis.
The diagonal dashed black lines show the corresponding values of $m_{1/2}$ using the scale shown
on the right-hand vertical axis.
In each case, we have assumed measurements with FCC-ee (TLEP)
uncertainties and central values coinciding with those
calculated using CMSSM model parameters at the benchmark points shown as the black spots.
In each case, we see that the FCC-ee (TLEP) measurements would be capable of
specifying with an accuracy that is greatest for the stop coannihilation strip with
$A_0/m_0 = 3.0$ and $\tan \beta = 20$ (lower right panel) and least for the focus-point strip
with $A_0 = 0$ and $\tan \beta = 10$ (upper left panel).

All of these benchmark points lie within reach of generic $\ETslash$ searches at FCC-hh,
as seen in Fig.~\ref{fig:profiles}. (Indeed, the fixed-point benchmark points lie with the reach of
HE-LHC.) Therefore, for each of these benchmark points it would be possible to make a
comparison between the direct determination of the CMSSM parameters with those inferred
indirectly from FCC-ee measurements, much as we discussed earlier for the case of the
low-mass CMSSM benchmark point, the LHC and FCC-ee. In all cases, FCC-ee would make
possible tests of supersymmetry at the loop level, even in pessimistic scenarios where the LHC
does not discover supersymmetry.

\section{Summary}

We have explored in this paper the interplay between direct and indirect searches for
supersymmetry in future runs of the LHC and at proposed future colliders. 
This is clearly a very broad topic, so we have restricted our attention to the CMSSM.
A recent global fit to the CMSSM has found two favoured regions of its parameter
space: a low-mass `Crimea' region and a high-mass `Eurasia' region. In the
optimistic low-mass case, extrapolating the sensitivities of supersymmetry searches with LHC Run~1,
we have found that future runs of the LHC with 300 or 3000/fb of data at 14~TeV
should be able to discover gluinos and squarks if Nature is described by the CMSSM
in the Crimea region. Moreover, the LHC experiments should be able to measure the
gluino and squark masses with high accuracy and hence also the soft supersymmetry-breaking
parameters of the CMSSM.

In this optimistic scenario, electroweakly-interacting sparticles could be discovered at the
CLIC $e^+ e^-$ collider and their masses measured very accurately, providing an important
cross-check on the consistency of the CMSSM with its universal input soft supersymmetry-breaking
parameters. On the other hand, an $e^+ e^-$ collider with centre-of-mass energy limited to 0.5~TeV
would not produce supersymmetric particles even in this optimistic scenario, though a collider with
a centre-of-mas energy of 1~TeV could detect the lighter stau slepton. 

However, an
$e^+ e^-$ collider capable of high-luminosity running on the $Z$ peak as well as producing
large numbers of Higgs bosons, such as FCC-ee (TLEP), would provide two sets of high-precision measurements that
could be used to constrain supersymmetric loop corrections and hence, in conjunction with the
LHC measurements, check supersymmetric predictions at the quantum level in two
independent ways. The combination of direct and indirect measurements possible in this
optimistic scenario would test the CMSSM in a way reminiscent of the use of precision
measurements from LEP and elsewhere to predict successfully the masses of the top
quark and the Higgs boson, albeit in time-reversed order.

In the pessimistic scenario in which sparticles are too heavy to be produced
at the LHC, the CLIC $e^+ e^-$ collider might still be able to discover the lighter stau slepton.
We have also assessed the capability of a higher-energy proton-proton collider such as FCC-hh
to discover supersymmetry, and the ability of high-precision FCC-ee (TLEP) measurements to
provide any indirect evidence. For this part of our analysis, we concentrate on the narrow
strips in the CMSSM parameter space that extend to large sparticle masses, namely the
stop-coannihilation strip and the focus-point strip. We find that a 33-TeV collider such as
HE-LHC would be able to discover supersymmetry via $\ETslash$ searches along some fractions of
these strips, and that a 100-TeV collider such as FCC-hh would be able to discover supersymmetry
along most of the extents of the strips examined. By studying specific benchmark points along
these strips, we have shown that FCC-ee measurements could in principle determine
indirectly CMSSM model parameters also in the pessimistic scenario, and the combination
with FCC-hh measurements could test supersymmetry at the loop level.

In both the optimistic and pessimistic scenarios, we find that high-precision measurements
with a high-luminosity $e^+ e^-$ collider can play r{\^o}les that are complementary to direct
particle searches with proton-proton colliders, and could enable supersymmetry to be tested at
the quantum (loop) level). Run~2 of the LHC will provide us with some
valuable pointers indicating which of these scenarios may be realised in Nature.


\section*{Acknowledgements}

The work of K.J.dV., O.B. and J.E. is supported in part by
the London Centre for Terauniverse Studies (LCTS), using funding from
the European Research Council 
via the Advanced Investigator Grant 267352.
The work of J.E. is also supported in part by STFC
(UK) via the research grants ST/J002798/1 and ST/L000326/1. 
The work of K.A.O. is supported in part by DOE grant
DE-SC0011842 at the University of Minnesota.


\begin{thebibliography}{99}



\bibitem{lhch}
G.~Aad {\it et al.}  [ATLAS Collaboration],
  Phys.\ Lett.\ B {\bf 716}, 1 (2012)
  [arXiv:1207.7214 [hep-ex]];
   S.~Chatrchyan {\it et al.}  [CMS Collaboration],
  Phys.\ Lett.\ B {\bf 716}, 30 (2012)
  [arXiv:1207.7235 [hep-ex]].
  
 \bibitem{ATLAS20}
G.~Aad {\it et al.}  [ATLAS Collaboration],
  JHEP {\bf 1409} (2014) 176
  [arXiv:1405.7875 [hep-ex]];
{full ATLAS Run~1 results can be found at} \\ 
{\tt https://twiki.cern.ch/twiki/bin/view/AtlasPublic/SupersymmetryPublicResults}.   

\bibitem{CMS20}
S.~Chatrchyan {\it et al.}  [CMS Collaboration],
  JHEP {\bf 1406} (2014) 055
  [arXiv:1402.4770 [hep-ex]];
{full CMS Run~1 results can be found at}
 {\tt https://twiki.cern.ch/twiki/bin/view/CMSPublic/PhysicsResultsSUS}.   
  
  \bibitem{funnel}  
M.~Drees and M.~M.~Nojiri,
Phys.\ Rev.\ D {\bf 47} (1993) 376 [arXiv:hep-ph/9207234];
  H.~Baer and M.~Brhlik,
Phys.\ Rev.\ D {\bf 53} (1996) 597 [arXiv:hep-ph/9508321];
  Phys.\ Rev.\  D {\bf 57} (1998) 567
  [arXiv:hep-ph/9706509];
   H.~Baer, M.~Brhlik, M.~A.~Diaz, J.~Ferrandis, P.~Mercadante, P.~Quintana and X.~Tata,
    Phys.\ Rev.\  D {\bf 63} (2001) 015007
  [arXiv:hep-ph/0005027];
  J.~R.~Ellis, T.~Falk, G.~Ganis, K.~A.~Olive and M.~Srednicki,
  Phys.\ Lett.\ B {\bf 510}, 236 (2001)
  [hep-ph/0102098].


\bibitem{cmssm}
 G.~L.~Kane, C.~F.~Kolda, L.~Roszkowski and J.~D.~Wells,
  Phys.\ Rev.\  D {\bf 49} (1994) 6173
  [arXiv:hep-ph/9312272];
  J.~R.~Ellis, T.~Falk, K.~A.~Olive and M.~Schmitt,
Phys.\ Lett.\ B {\bf 388} (1996) 97
[arXiv:hep-ph/9607292];
Phys.\ Lett.\ B {\bf 413} (1997) 355
[arXiv:hep-ph/9705444];
J.~R.~Ellis, T.~Falk, G.~Ganis, K.~A.~Olive and M.~Schmitt,
Phys.\ Rev.\ D {\bf 58} (1998) 095002
[arXiv:hep-ph/9801445];
V.~D.~Barger and C.~Kao,
Phys.\ Rev.\ D {\bf 57} (1998) 3131
[arXiv:hep-ph/9704403];
J.~R.~Ellis, T.~Falk, G.~Ganis and K.~A.~Olive,
Phys.\ Rev.\ D {\bf 62} (2000) 075010
[arXiv:hep-ph/0004169];
L.~Roszkowski, R.~Ruiz de Austri and T.~Nihei,
JHEP {\bf 0108} (2001) 024
[arXiv:hep-ph/0106334];
  A.~Djouadi, M.~Drees and J.~L.~Kneur,
JHEP {\bf 0108} (2001) 055
[arXiv:hep-ph/0107316];
U.~Chattopadhyay, A.~Corsetti and P.~Nath,
Phys.\ Rev.\ D {\bf 66} (2002) 035003
[arXiv:hep-ph/0201001];
J.~R.~Ellis, K.~A.~Olive and Y.~Santoso,
New Jour.\ Phys.\  {\bf 4} (2002) 32
[arXiv:hep-ph/0202110].

  \bibitem{ehnos} H.~Goldberg,
                Phys.\ Rev.\ Lett.\ {\bf 50} (1983) 1419;
                J.~Ellis, J.~Hagelin, D.~Nanopoulos, K.~Olive and M.~Srednicki,
                Nucl.\ Phys.\ B {\bf 238} (1984) 453.
                
                
\bibitem{cmssmwmap}
J.~R.~Ellis, K.~A.~Olive, Y.~Santoso and V.~C.~Spanos,
Phys.\ Lett.\ B {\bf 565} (2003) 176
[arXiv:hep-ph/0303043];
H.~Baer and C.~Balazs,
  JCAP {\bf 0305}, 006 (2003)
  [arXiv:hep-ph/0303114];
A.~B.~Lahanas and D.~V.~Nanopoulos,
  Phys.\ Lett.\  B {\bf 568}, 55 (2003)
  [arXiv:hep-ph/0303130];
U.~Chattopadhyay, A.~Corsetti and P.~Nath,
  Phys.\ Rev.\  D {\bf 68}, 035005 (2003)
  [arXiv:hep-ph/0303201];
   C.~Munoz,
  Int.\ J.\ Mod.\ Phys.\  A {\bf 19}, 3093 (2004)
  [arXiv:hep-ph/0309346];
    R.~Arnowitt, B.~Dutta and B.~Hu,
arXiv:hep-ph/0310103;
   J.~Ellis and K.~A.~Olive,
  in {\it Particle dark matter}, ed. G.~Bertone (Cambridge University Press, 2010) pp142-163
  [arXiv:1001.3651 [astro-ph.CO]].

 \bibitem{eo6}
  J.~Ellis and K.~A.~Olive,
  Eur.\ Phys.\ J.\ C {\bf 72}, 2005 (2012)
  [arXiv:1202.3262 [hep-ph]].


\bibitem{ehow+}
O.~Buchmueller, M.~J.~Dolan, J.~Ellis, T.~Hahn, S.~Heinemeyer, W.~Hollik, J.~Marrouche and K.~A.~Olive {\it et al.},
  Eur.\ Phys.\ J.\ C {\bf 74} (2014) 3,  2809
  [arXiv:1312.5233 [hep-ph]].
    
  \bibitem{mc75} 
O.~Buchmueller, 
{\it et al.},
Eur.\ Phys.\ J.\ C {\bf 72} (2012) 2020
 [arXiv:1112.3564 [hep-ph]].
  
\bibitem{mc8}
O.~Buchmueller, 
{\it et al.},
Eur.\ Phys.\ J.\ C {\bf 72} (2012) 2243
  [arXiv:1207.7315 [hep-ph]].

\bibitem{mc9}
O.~Buchmueller, R.~Cavanaugh, A.~De Roeck, M.~J.~Dolan, J.~R.~Ellis, H.~Flacher, S.~Heinemeyer and G.~Isidori {\it et al.},
  Eur.\ Phys.\ J.\ C {\bf 74} (2014) 6,  2922
  [arXiv:1312.5250 [hep-ph]].

  \bibitem{post-mh}
   H.~Baer, V.~Barger and A.~Mustafayev,
  Phys.\ Rev.\ D {\bf 85}, 075010 (2012)
  [arXiv:1112.3017 [hep-ph]];
    T.~Li, J.~A.~Maxin, D.~V.~Nanopoulos and J.~W.~Walker,
  Phys.\ Lett.\ B {\bf 710} (2012) 207
  [arXiv:1112.3024 [hep-ph]];
      S.~Heinemeyer, O.~Stal and G.~Weiglein,
  Phys.\ Lett.\ B {\bf 710}, 201 (2012)
  [arXiv:1112.3026 [hep-ph]];
A.~Arbey, M.~Battaglia, A.~Djouadi, F.~Mahmoudi and J.~Quevillon,
  Phys.\ Lett.\ B {\bf 708} (2012) 162
  [arXiv:1112.3028 [hep-ph]];
   P.~Draper, P.~Meade, M.~Reece and D.~Shih,
  Phys.\ Rev.\ D {\bf 85}, 095007 (2012)
  [arXiv:1112.3068 [hep-ph]];
  S.~Akula, B.~Altunkaynak, D.~Feldman, P.~Nath and G.~Peim,
  Phys.\ Rev.\ D {\bf 85} (2012) 075001
  [arXiv:1112.3645 [hep-ph]];
M.~Kadastik, K.~Kannike, A.~Racioppi and M.~Raidal,
  JHEP {\bf 1205} (2012) 061
  [arXiv:1112.3647 [hep-ph]];
  C.~Strege, G.~Bertone, D.~G.~Cerdeno, M.~Fornasa, R.~Ruiz de Austri and R.~Trotta,
  JCAP {\bf 1203}, 030 (2012)
  [arXiv:1112.4192 [hep-ph]];
  J.~Cao, Z.~Heng, D.~Li and J.~M.~Yang,
  Phys.\ Lett.\ B {\bf 710} (2012) 665
  [arXiv:1112.4391 [hep-ph]];
   L.~Aparicio, D.~G.~Cerdeno and L.~E.~Ibanez,
  JHEP {\bf 1204}, 126 (2012)
  [arXiv:1202.0822 [hep-ph]]; 
  H.~Baer, V.~Barger and A.~Mustafayev,
  JHEP {\bf 1205} (2012) 091
  [arXiv:1202.4038 [hep-ph]];
   P.~Bechtle, T.~Bringmann, K.~Desch, H.~Dreiner, M.~Hamer, C.~Hensel, M.~Kramer and N.~Nguyen {\it et al.},
  JHEP {\bf 1206}, 098 (2012)
  [arXiv:1204.4199 [hep-ph]];
C.~Balazs, A.~Buckley, D.~Carter, B.~Farmer and M.~White,
  arXiv:1205.1568 [hep-ph];
D.~Ghosh, M.~Guchait, S.~Raychaudhuri and D.~Sengupta,
  Phys.\ Rev.\ D {\bf 86}, 055007 (2012)
  [arXiv:1205.2283 [hep-ph]];
   A.~Fowlie, M.~Kazana, K.~Kowalska, S.~Munir, L.~Roszkowski, E.~M.~Sessolo, S.~Trojanowski and Y.~-L.~S.~Tsai,
  Phys.\ Rev.\ D {\bf 86}, 075010 (2012)
  [arXiv:1206.0264 [hep-ph]];
   K.~Kowalska {\it et al.}  [BayesFITS Group Collaboration],
  Phys.\ Rev.\ D {\bf 87}, no. 11, 115010 (2013)
  [arXiv:1211.1693 [hep-ph]];
   C.~Strege, G.~Bertone, F.~Feroz, M.~Fornasa, R.~Ruiz de Austri and R.~Trotta,
  JCAP {\bf 1304}, 013 (2013)
  [arXiv:1212.2636 [hep-ph]];
  M.~E.~Cabrera, J.~A.~Casas and R.~R.~de Austri,
  JHEP {\bf 1307} (2013) 182
  [arXiv:1212.4821 [hep-ph]];
  T.~Cohen and J.~G.~Wacker,
  JHEP {\bf 1309} (2013) 061
  [arXiv:1305.2914 [hep-ph]];
S.~Henrot-Versill�, R�m.~Lafaye, T.~Plehn, M.~Rauch, D.~Zerwas, 
S.~�p.~Plaszczynski, B.~Rouill� d'Orfeuil and M.~Spinelli,
  Phys.\ Rev.\ D {\bf 89}, 055017 (2014)
  [arXiv:1309.6958 [hep-ph]];
  P.~Bechtle, K.~Desch, H.~K.~Dreiner, M.~Hamer, M.~Kr確er, B.~O'Leary, W.~Porod and X.~Prudent {\it et al.},
  arXiv:1310.3045 [hep-ph];
  L.~Roszkowski, E.~M.~Sessolo and A.~J.~Williams,
  arXiv:1405.4289 [hep-ph].
  
  
  \bibitem{fp}
   J.~L.~Feng, K.~T.~Matchev and D.~Sanford,
  Phys.\ Rev.\ D {\bf 85}, 075007 (2012)
  [arXiv:1112.3021 [hep-ph]];
  P.~Draper, J.~Feng, P.~Kant, S.~Profumo and D.~Sanford,
  Phys.\ Rev.\ D {\bf 88}, 015025 (2013)
  [arXiv:1304.1159 [hep-ph]].
  
  \bibitem{eoz}
  J.~Ellis, K.~A.~Olive and J.~Zheng,
  Eur.\ Phys.\ J.\ C {\bf 74} (2014) 2947
  [arXiv:1404.5571 [hep-ph]].
    
  \bibitem{1304.2825}
  M.~Battaglia, J.~J.~Blaising, J.~S.~Marshall, S.~Poss, A.~Sailer, M.~Thomson and E.~van der Kraaij,
  JHEP {\bf 1309} (2013) 001
  [arXiv:1304.2825 [hep-ex]].
 
  \bibitem{CLIC}
 P.~Lebrun, L.~Linssen, A.~Lucaci-Timoce, D.~Schulte, F.~Simon, S.~Stapnes, N.~Toge and H.~Weerts {\it et al.},
 {\it The CLIC Programme: Towards a Staged $e^+ e^-$ Linear Collider Exploring the Terascale : CLIC Conceptual Design Report},
  arXiv:1209.2543 [physics.ins-det].
  
  \bibitem{LEPEWWG}
  S.~Schael {\it et al.}  [ALEPH, DELPHI, L3, OPAL and SLD Collaborations, LEP Electroweak Working Group and SLD Electroweak and Heavy Flavour Groups],
  Phys.\ Rept.\  {\bf 427} (2006) 257
  [hep-ex/0509008].

  \bibitem{ATLASmu}
ATLAS Collaboration, {\tt https://twiki.cern.ch/twiki/bin/} \\
{\tt view/AtlasPublic/HiggsPublicResults}.

\bibitem{CMSmu}
S.~Chatrchyan {\it et al.}  [CMS Collaboration],
  JHEP {\bf 1306} (2013) 081
  [arXiv:1303.4571 [hep-ex]].

\bibitem{Tevatronmu}
CDF and D0 Collaborations, {\tt http://tevnphwg.fnal.gov}.

\bibitem{TLEP}
M.~Bicer {\it et al.}  [TLEP Design Study Working Group Collaboration],
  JHEP {\bf 1401} (2014) 164
  [arXiv:1308.6176 [hep-ex]].
  
  \bibitem{FCC-hh}
{\tt https://espace2013.cern.ch/fcc/Pages/default.aspx}.

\bibitem{LW}
For some studies with similar motivations, see
 A.~Fowlie and M.~Raidal,
  Eur.\ Phys.\ J.\ C {\bf 74} (2014) 2948
  [arXiv:1402.5419 [hep-ph]];
  M.~Low and L.~T.~Wang,
  JHEP {\bf 1408} (2014) 161
  [arXiv:1404.0682 [hep-ph]];
 B.~S.~Acharya, K.~Bo{\.z}ek, C.~Pongkitivanichkul and K.~Sakurai,
  JHEP {\bf 1502} (2015) 181
  [arXiv:1410.1532 [hep-ph]];
  S.~Gori, S.~Jung, L.~T.~Wang and J.~D.~Wells,
  JHEP {\bf 1412} (2014) 108
  [arXiv:1410.6287 [hep-ph]];
  J.~Bramante, P.~J.~Fox, A.~Martin, B.~Ostdiek, T.~Plehn, T.~Schell and M.~Takeuchi,
  Phys.\ Rev.\ D {\bf 91} (2015) 5,  054015
  [arXiv:1412.4789 [hep-ph]].
  For some other studies of supersymmetry at a 100-TeV $pp$ collider, see
  T.~Cohen, T.~Golling, M.~Hance, A.~Henrichs, K.~Howe, J.~Loyal, S.~Padhi and J.~G.~Wacker,
  JHEP {\bf 1404} (2014) 117
  [arXiv:1311.6480 [hep-ph]];
   T.~Cohen, R.~T.~D'Agnolo, M.~Hance, H.~K.~Lou and J.~G.~Wacker,
  JHEP {\bf 1411} (2014) 021
  [arXiv:1406.4512 [hep-ph]];
  C.~Borschensky, M.~Kr�mer, A.~Kulesza, M.~Mangano, S.~Padhi, T.~Plehn and X.~Portell,
  Eur.\ Phys.\ J.\ C {\bf 74} (2014) 12,  3174
  [arXiv:1407.5066 [hep-ph]];
  S.~A.~R.~Ellis, G.~L.~Kane and B.~Zheng,
  arXiv:1408.1961 [hep-ph];
 H.~Beauchesne, K.~Earl and T.~Gregoire,
  arXiv:1503.03099 [hep-ph].

\bibitem{nuhm1}
H.~Baer, A.~Mustafayev, S.~Profumo, A.~Belyaev and X.~Tata,
  Phys.\ Rev.\  D {\bf 71}, 095008 (2005)
  [arXiv:hep-ph/0412059] and
               JHEP {\bf 0507} (2005) 065, 
               hep-ph/0504001.
               
 \bibitem{eos}              
 J.~R.~Ellis, K.~A.~Olive and P.~Sandick,
  Phys.\ Rev.\  D {\bf 78}, 075012 (2008)
  [arXiv:0805.2343 [hep-ph]];
 J.~Ellis, F.~Luo, K.~A.~Olive and P.~Sandick,
  Eur.\ Phys.\ J.\ C {\bf 73}, 2403 (2013)
  [arXiv:1212.4476 [hep-ph]].


 \bibitem{nuhm2}
  J.~Ellis, K.~Olive and Y.~Santoso,
Phys.\ Lett.\  B~{\bf 539}, 107 (2002)
[arXiv:hep-ph/0204192];
J.~R.~Ellis, T.~Falk, K.~A.~Olive and Y.~Santoso,
Nucl.\ Phys.\ B {\bf 652}, 259 (2003)
[arXiv:hep-ph/0210205].

\bibitem{mc10}
O.~Buchmueller, R.~Cavanaugh, M.~Citron, A.~De Roeck, M.~J.~Dolan, J.~R.~Ellis, H.~Flaecher and S.~Heinemeyer {\it et al.},
  Eur.\ Phys.\ J.\ C {\bf 74} (2014) 3212
  [arXiv:1408.4060 [hep-ph]].
  
\bibitem{mc11}
K.~J.~de Vries, E.~A.~Bagnaschi, O.~Buchmueller, R.~Cavanaugh, M.~Citron, A.~De Roeck, M.~J.~Dolan and J.~R.~Ellis {\it et al.},
  arXiv:1504.03260 [hep-ph].
  
 \bibitem{LHCmH}
 G.~Aad {\it et al.}  [CMS Collaboration],
  arXiv:1503.07589 [hep-ex].

  \bibitem{FeynHiggs}
 G.~Degrassi, S.~Heinemeyer, W.~Hollik, P.~Slavich and G.~Weiglein,
  Eur.\ Phys.\ J.\ C {\bf 28} (2003) 133
  [arXiv:hep-ph/0212020];
   S.~Heinemeyer, W.~Hollik and G.~Weiglein,
  Eur.\ Phys.\ J.\ C {\bf 9} (1999) 343
  [arXiv:hep-ph/9812472];
  S.~Heinemeyer, W.~Hollik and G.~Weiglein,
  Comput.\ Phys.\ Commun.\  {\bf 124} (2000) 76
  [arXiv:hep-ph/9812320];
   M.~Frank {\it et al.}, 
  JHEP {\bf 0702} (2007) 047
  [arXiv:hep-ph/0611326];
  T.~Hahn, S.~Heinemeyer, W.~Hollik, H.~Rzehak and G.~Weiglein,
  Comput.\ Phys.\ Commun.\  {\bf 180} (2009) 1426.
  see {\tt http://www.feynhiggs.de}~.

  
   \bibitem{newFH} T.~Hahn, S.~Heinemeyer, W.~Hollik, H.~Rzehak and
                G.~Weiglein,
                [arXiv:1312.4937 [hep-ph]].
                
\bibitem{newBNL} [The Muon g-2 Collaboration],
                 {\it Phys. Rev. Lett.} {\bf 92} (2004) 161802, 
                 hep-ex/0401008;
                 G.~Bennett et al.\ [The Muon g-2 Collaboration],
                  Phys.\ Rev. D {\bf 73} (2006) 072003
                  [arXiv:hep-ex/0602035].
                  
  
   \bibitem{bsgex}
S.~Chen {\it et al.}  [CLEO Collaboration],
Phys.\ Rev.\ Lett.\  {\bf 87} (2001) 251807
[arXiv:hep-ex/0108032];
P.~Koppenburg {\it et al.}  [Belle Collaboration],
Phys.\ Rev.\ Lett.\  {\bf 93} (2004) 061803
[arXiv:hep-ex/0403004];
B.~Aubert {\it et al.}  [BaBar Collaboration],
arXiv:hep-ex/0207076;
E.~Barberio {\it et al.}  [Heavy Flavor Averaging Group (HFAG)],
  arXiv:hep-ex/0603003.
                  

\bibitem{bmm}
V.~M.~Abazov {\it et al.}  [D0 Collaboration],
 Phys.\ Lett.\ B {\bf 693} (2010) 539
 [arXiv:1006.3469 [hep-ex]];
T.~Aaltonen {\it et al.}  [CDF Collaboration],
 Phys.\ Rev.\ Lett.\  {\bf 107} (2011) 191801
  [Publisher-note {\bf 107} (2011) 239903]
 [arXiv:1107.2304 [hep-ex]];
   G.~Aad {\it et al.}  [ATLAS Collaboration],
 Phys.\ Lett.\ B {\bf 713} (2012) 387
 [arXiv:1204.0735 [hep-ex]].
 
 
 \bibitem{LHCbBsmm}
 R.Aaij {\it et al.}  [LHCb Collaboration],
 Phys.\ Rev.\ Lett.\  {\bf 111} (2013) 101805
 [arXiv:1307.5024 [hep-ex]].
 
 \bibitem{CMSBsmm}
  S.~Chatrchyan {\it et al.}  [CMS Collaboration],
 Phys.\ Rev.\ Lett.\  {\bf 111} (2013) 101804
 [arXiv:1307.5025 [hep-ex]].
 
 \bibitem{BsmmComb}
V.~Khachatryan {\it et al.}  [CMS and LHCb Collaborations], DOI: 10.1038/nature14474,
  arXiv:1411.4413 [hep-ex].
  
  \bibitem{LUX}
D.~S.~Akerib {\it et al.}  [LUX Collaboration],
  Phys.\ Rev.\ Lett.\  {\bf 112}, 091303 (2014)
  [arXiv:1310.8214 [astro-ph.CO]].

  
  \bibitem{Planck}
  P.~A.~R.~Ade {\it et al.}  [Planck Collaboration],
  Astron.\ Astrophys.\  {\bf 571} (2014) A16
  [arXiv:1303.5076 [astro-ph.CO]].
Using the Planck 2015 results: P.~A.~R.~Ade {\it et al.}  [Planck Collaboration],
  arXiv:1502.01589 [astro-ph.CO]
 would have no visible effect on our analysis.
 
\bibitem{Snowmass}
Y.~Gershtein, M.~Luty, M.~Narain, L.-T.~Wang, D.~Whiteson, K.~Agashe, L.~Apanasevich and G.~Artoni {\it et al.},
{\it Working Group Report: New Particles, Forces, and Dimensions},
  arXiv:1311.0299 [hep-ex].
  
 \bibitem{CMS:2013xfa}
 CMS Collaboration, arXiv:1307.7135 [hep-ex].

 \bibitem{Malik:2014ggr}
 S.~Malik, C.~McCabe, H.~Araujo, A.~Belyaev, C.~Boehm, J.~Brooke, O.~Buchmueller and G.~Davies {\it et al.},
  arXiv:1409.4075 [hep-ex].

 \bibitem{Buchmueller:2014yoa}
O.~Buchmueller, M.~J.~Dolan, S.~A.~Malik and C.~McCabe,
  JHEP {\bf 1501} (2015) 037
  [arXiv:1407.8257 [hep-ph]];
 
 \bibitem{colliderreach}
 G.~Salam and A.~Weiler, 
{\tt http://collider-reach.web.cern.ch/collider-reach/}.

 \bibitem{ATLASHL}
ATLAS Collaboration,  {\tt https://cdsweb.cern.ch/record/1472518/files/} \\ {\tt ATL-PHYS-PUB-2012-001.pdf}.
    
  \bibitem{PYTHIA}
T.~Sjostrand, S.~Mrenna and P.~Z.~Skands,
  Comput.\ Phys.\ Commun.\  {\bf 178} (2008) 852
  [arXiv:0710.3820 [hep-ph]].

\bibitem{py8susy}
N.~Desai and P.~Z.~Skands,
  Eur.\ Phys.\ J.\ C {\bf 72}, 2238 (2012)
  [arXiv:1109.5852 [hep-ph]].

\bibitem{MSTW}
A.~D.~Martin, W.~J.~Stirling, R.~S.~Thorne and G.~Watt,
  Eur.\ Phys.\ J.\ C {\bf 63} (2009) 189
  [arXiv:0901.0002 [hep-ph]].
  
\bibitem{NNPDF}
R.~D.~Ball, V.~Bertone, S.~Carrazza, C.~S.~Deans, L.~Del Debbio, S.~Forte, A.~Guffanti and N.~P.~Hartland {\it et al.},
  Nucl.\ Phys.\ B {\bf 867} (2013) 244
  [arXiv:1207.1303 [hep-ph]], {\tt https://nnpdf.hepforge.org}.
  
\bibitem{MT2}
C.~G.~Lester and D.~J.~Summers,
  Phys.\ Lett.\ B {\bf 463} (1999) 99
  [hep-ph/9906349];
  A.~Barr, C.~Lester and P.~Stephens,
  J.\ Phys.\ G {\bf 29} (2003) 2343
  [hep-ph/0304226].
  
\bibitem{GFitter}
M.~Baak {\it et al.}  [Gfitter Group],
  Eur.\ Phys.\ J.\ C {\bf 74} (2014) 3046
  [arXiv:1407.3792 [hep-ph]].
  
\bibitem{ICFA}
A.~Blondel, A.~Chao, W.~Chou, J.~Gao, D.~Schulte and K.~Yokoya,
{\it Report of the ICFA Beam Dynamics Workshop 'Accelerators for a Higgs Factory: Linear vs. Circular' (HF2012)},
  arXiv:1302.3318 [physics.acc-ph].

\bibitem{GUTs}  
J.~Ellis, S.~Kelley and D.V. Nanopoulos,
Phys.\ Lett.\ B {\bf 249} (1990) 441;
Phys.\ Lett.\ B {\bf 260} (1991) 131;
U.~Amaldi, W.~de~Boer and H.~Furstenau, 
Phys.\ Lett.\ B {\bf 260} (1991) 447;
P.~Langacker and M.-x. Luo,
Phys.\ Rev.\ D {\bf 44} (1991) 817;
C.~Giunti, C.~W. Kim and U.~W. Lee,
Mod.\ Phys.\ Lett.\ A {\bf 6} (1991) 1745.

\bibitem{MP}
H.~Murayama and A.~Pierce,
  Phys.\ Rev.\ D {\bf 65} (2002) 055009
  [hep-ph/0108104].
  
\bibitem{EW}
S.~A.~R.~Ellis and J.~D.~Wells,
  Phys.\ Rev.\ D {\bf 91} (2015) 7,  075016
  [arXiv:1502.01362 [hep-ph]].
  
\bibitem{stau-co} J. Ellis, T. Falk, and K.A. Olive, Phys. Lett.  {\bf B444} (1998) 367
[arXiv:hep-ph/9810360];
J. Ellis, T. Falk, K.A. Olive, and M. Srednicki, {\it Astr. Part. Phys.}
{\bf 13} (2000) 181
[Erratum-ibid.\  {\bf 15} (2001) 413]
[arXiv:hep-ph/9905481];
R.~Arnowitt, B.~Dutta and Y.~Santoso,
Nucl.\ Phys.\ B {\bf 606} (2001) 59
[arXiv:hep-ph/0102181];
M.~E.~G\'omez, G.~Lazarides and C.~Pallis,
Phys. Rev. D {\bf D61} (2000) 123512
[arXiv:hep-ph/9907261];
  Phys.\ Lett. {\bf B487} (2000) 313
[arXiv:hep-ph/0004028];
  Nucl. Phys. B {\bf B638} (2002) 165
[arXiv:hep-ph/0203131];
T.~Nihei, L.~Roszkowski and R.~Ruiz de Austri,
  JHEP {\bf 0207} (2002) 024
[arXiv:hep-ph/0206266].

\bibitem{oldfp}
 J.~L.~Feng, K.~T.~Matchev and T.~Moroi,
  Phys.\ Rev.\ Lett.\  {\bf 84}, 2322 (2000)
  [arXiv:hep-ph/9908309];
  Phys.\ Rev.\  D {\bf 61}, 075005 (2000)
  [arXiv:hep-ph/9909334]; 
  J.~L.~Feng, K.~T.~Matchev and F.~Wilczek,
  Phys.\ Lett.\  B {\bf 482}, 388 (2000)
  [arXiv:hep-ph/0004043];
  H.~Baer, T.~Krupovnickas, S.~Profumo and P.~Ullio,
  JHEP {\bf 0510} (2005) 020
  [hep-ph/0507282];
  
  \bibitem{stop}  
  C.~Boehm, A.~Djouadi and M.~Drees,
  Phys.\ Rev.\  D {\bf 62}, 035012 (2000)
  [arXiv:hep-ph/9911496];
  J.~R.~Ellis, K.~A.~Olive and Y.~Santoso,
  Astropart.\ Phys.\  {\bf 18}, 395 (2003)
  [arXiv:hep-ph/0112113];
    J.~Edsjo, M.~Schelke, P.~Ullio and P.~Gondolo,
  JCAP {\bf 0304}, 001 (2003)
  [arXiv:hep-ph/0301106];
    J.~L.~Diaz-Cruz, J.~R.~Ellis, K.~A.~Olive and Y.~Santoso,
  JHEP {\bf 0705}, 003 (2007)
  [arXiv:hep-ph/0701229];
  I.~Gogoladze, S.~Raza and Q.~Shafi,
  Phys.\ Lett.\ B {\bf 706}, 345 (2012)
  [arXiv:1104.3566 [hep-ph]];
   M.~A.~Ajaib, T.~Li and Q.~Shafi,
  Phys.\ Rev.\ D {\bf 85}, 055021 (2012)
  [arXiv:1111.4467 [hep-ph]].

\bibitem{SLHA}
P.~Skands {\it et al.},
  JHEP {\bf 0407} (2004) 036
  [arXiv:hep-ph/0311123];
  B.~Allanach {\it et al.},
  Comput.\ Phys.\ Commun.\  {\bf 180} (2009) 8
  [arXiv:0801.0045 [hep-ph]].


\end{thebibliography}
\end{document}